\documentclass[12pt,preprint]{aastex}

\slugcomment{Astrophysical Journal}

\shorttitle{L-band spectroscopy of ULIRGs}
\shortauthors{Imanishi et al.}

\begin{document}


\title{Infrared 3--4 $\mu$m Spectroscopic Investigations of a Large
Sample of Nearby Ultraluminous Infrared Galaxies} 


\author{Masatoshi Imanishi\altaffilmark{1,2}}
\affil{National Astronomical Observatory, 2-21-1, Osawa, Mitaka, Tokyo
181-8588, Japan}
\email{masa.imanishi@nao.ac.jp}

\author{C. C. Dudley}
\affil{Naval Research Laboratory, Remote Sensing Division, 
Code 7213, Building 2, Room 240B, 4555 Overlook Ave SW, 
Washington DC 20375-5351, U.S.A.}
\email{dudley@vivaldi.nrl.navy.mil}

\and 

\author{Philip R. Maloney}
\affil{Center for Astrophysics and Space Astronomy, 
University of Colorado, Boulder, CO 80309-0389, U.S.A.}
\email{maloney@casa.colorado.edu}

\altaffiltext{1}{Based in part on data collected at Subaru Telescope,
which is operated by the National Astronomical Observatory of Japan.}

\altaffiltext{2}{Visiting Astronomer at the Infrared Telescope Facility,
which is operated by the University of Hawaii under
Cooperative Agreement no. NCC 5-538 with the National
Aeronautics and Space Administration, Office of Space
Science, Planetary Astronomy Program.}


\begin{abstract}
We present infrared $L$-band (3--4 $\mu$m) nuclear spectra of a large
sample of nearby ultraluminous infrared galaxies (ULIRGs). 
ULIRGs classified optically as non-Seyferts (LINERs, HII-regions, and
unclassified) are our main targets.
Using the 3.3 $\mu$m polycyclic aromatic hydrocarbon (PAH) emission and
absorption features at 3.1 $\mu$m due to ice-covered dust and at 3.4 $\mu$m
produced by bare carbonaceous dust, we search for signatures of
powerful active galactic nuclei (AGNs) deeply buried along virtually
all lines-of-sight. The 3.3 $\mu$m PAH emission, the signatures of
starbursts, is detected in all but two non-Seyfert ULIRGs, but the
estimated starburst magnitudes can account for only a small fraction of
the infrared luminosities. 
Three LINER ULIRGs show spectra typical of almost pure buried AGNs,
namely, strong absorption features with very small equivalent-width PAH
emission.  Besides these three sources, 14 LINER and 3 HII ULIRGs'
nuclei show strong absorption features whose absolute optical depths
suggest an energy source more centrally concentrated than the
surrounding dust, such as a buried AGN. 
In total, 17 out of 27 (63\%) LINER and 3 out of 13
(23\%) HII ULIRGs' nuclei show some degree of evidence for powerful
buried AGNs, suggesting that powerful buried AGNs may be more common in
LINER ULIRGs than in HII ULIRGs.  The evidence of AGNs is found in
non-Seyfert ULIRGs with both warm and cool far-infrared colors.
These spectra are compared with those of 15 ULIRGs' nuclei with optical
Seyfert signatures taken for comparison.
The overall spectral properties suggest that the total amount of dust
around buried AGNs in non-Seyfert ULIRGs is systematically larger than
that around AGNs in Seyfert 2 ULIRGs.
We argue that the optical (non-)detectability of Seyfert signatures in
ULIRGs is highly dependent on how deeply buried  the AGNs are, and that
it is essential to properly evaluate the energetic importance of buried
AGNs in non-Seyfert ULIRGs. 
\end{abstract}


\section{Introduction}

Ultraluminous infrared galaxies (ULIRGs), discovered by the {\it IRAS} 
all sky survey, radiate quasar-like luminosities ($>$$10^{12}L_{\odot}$)
as infrared dust emission \citep{sam96}.  Powerful energy sources,
starbursts and/or active galactic nuclei (AGNs), must be present
hidden behind dust. This ULIRG population dominates not only the
bright end of the local luminosity function \citep{soi87}, but also
the cosmic submillimeter background emission \citep{bla99}, so that
their energy sources are closely related to understanding the
connection between starbursts and AGNs.

It is now generally accepted that the majority of nearby ULIRGs are
energetically 
dominated by compact ($<$500 pc), highly dust-obscured nuclear cores,
rather than extended ($>$kpc), weakly-obscured starburst activity in
host galaxies \citep{soi00,fis00}. Determining whether the compact
cores are powered by AGNs and/or by very compact starbursts is
essential, but it is not easy to distinguish the energy sources of
ULIRGs' cores because the large amount of dust and molecular gas
concentrated in the nuclei \citep{sam96,spo02} can hide the signatures of
AGNs and make their detection difficult. If AGNs are
obscured by dust in a {\it torus} geometry, as is inferred for some 
types of nearby AGN populations \citep{ant93}, clouds
along the torus axis are photo-ionized by the AGN's hard radiation. 
At $\sim 100$ pc distances from the AGN, these clouds form the so-called
narrow line regions (NLRs; Robson 1996), and the NLRs produce optical
emission lines with flux ratios that are distinguishable from those of
normal starburst galaxies \citep{vei87}. Thus, AGNs obscured by
torus-shaped dust distributions can still be detectable through optical
spectroscopy (classified as Seyfert 2 or Sy 2; Veilleux et
al. 1999a). Such obscured AGNs with prominent NLRs are also recognizable
through strong high-excitation forbidden line emission, originating in
the NLRs, in high-resolution near- to mid-infrared spectra
\citep{gen98,vei99b,mur01,arm04}. However, when virtually all directions
around an AGN become opaque to the bulk of the AGN's ionizing photons
(mostly soft X-rays and UV), no significant NLRs develop.  In such
{\it buried} AGNs, X-ray dissociation regions (XDRs), dominated by
low-ionization species \citep{mal96} develop,
rather than NLRs, so that the signatures of buried AGNs are difficult
to find using the conventional searches for emission lines originating
in the NLRs.

To study deeply buried AGNs without significant NLRs, observations at
wavelengths of low dust extinction are necessary.  Infrared $L$-band
($\lambda =$ 2.8--4.1 $\mu$m) spectroscopy from the ground is one of
the most effective ways to find the signatures of such buried AGNs.
First, dust extinction in the $L$-band is substantially reduced
compared to shorter wavelengths (A$_{\rm L}$ $\sim$ 0.06 $\times$
A$_{\rm V}$; Rieke \& Lebofsky 1985; Lutz et al. 1996) and is as low
as that at 5--13 $\mu$m \citep{lut96}.  Second, the contribution from
AGN-powered dust emission to an observed flux, relative to stellar
emission, increases substantially compared with $\lambda <$ 2 $\mu$m.
Thus, the detection of a buried AGN becomes more feasible than at
$\lambda <$ 2 $\mu$m.  Third, normal starburst and AGN emission are
clearly distinguishable using $L$-band spectra \citep{moo86,imd00}.  
In a normal starburst galaxy, emission from polycyclic aromatic hydrocarbon
(PAH) molecules is excited by far UV photons in photo-dissociation
regions, the interfaces between molecular gas and HII-regions
\citep{sel81,nor95}.  Near an AGN, the PAH molecules can be destroyed
by strong X-ray radiation from the AGN \citep{voi92,sie04}.  If the PAH
molecules are sufficiently shielded by obscuration from the X-ray
emission of the central AGN, the PAHs can survive.  However, the far
UV emission from the AGN will also be attenuated by the obscuration,
so in a pure AGN without starburst activity, the PAH emission will be
very weak. Instead, a featureless continuum from hot ($\sim$1000 K),
submicron-sized dust grains \citep{dra84,mat89} heated by the AGN is
observed. If the AGN is deeply buried in dust, the 3.1 $\mu$m H$_{2}$O
ice absorption feature produced by ice-covered dust and/or the 3.4
$\mu$m absorption feature resulting from bare carbonaceous dust should
be observed \citep{imd00,idm01,im03}. In reality in ULIRGs, starbursts and
AGNs are likely to coexist.  Hence it is important to disentangle the
energy sources of such starburst/AGN composite ULIRGs. This can be
done using observations of PAH emission: because the column densities
and pressures in the interstellar medium of ULIRGs are very high, the
starburst regions in starburst/AGN composite ULIRGs should produce
strong PAH emission \citep{mal99}.  
In a pure normal starburst galaxy, the equivalent width ($\equiv$ a
line-to-continuum flux ratio) of the 3.3 $\mu$m PAH emission feature
should always be high, regardless of dust extinction of the starburst
\citep{moo86,imd00}, because both the PAH and continuum emission are
similarly flux-attenuated. 
If PAH emission is detected in a ULIRG, but its equivalent width is
substantially smaller than in normal starburst galaxies, it will 
strongly suggest that the PAH equivalent width is being reduced by
the contribution to the observed spectrum of a PAH-free featureless
continuum, 
produced by the AGN. Thus, the presence of a powerful AGN can be
inferred from the equivalent width of the 3.3 $\mu$m PAH emission
feature and dust absorption features, even if the AGN is buried.

In addition, using $L$-band spectra we have another independent way to
detect the presence of powerful buried AGNs in ULIRGs: the absolute
optical depth of dust absorption features.  In a normal starburst, the
energy sources (stars) and dust are spatially well mixed, while in a
buried AGN, the energy source is very compact and thus should be more
centrally concentrated than the surrounding dust \citep{soi00,im03,sie04}.
In the mixed dust/source geometry, the foreground less-obscured 
emission (which thus shows only weak absorption features) dominates
the observed fluxes, and therefore there are upper limits to the
absorption optical depths, unless unusual abundances are considered
and/or there is a very large amount of foreground dust in the host
galaxy \citep{im03}. On the other hand, a centrally-concentrated
energy source can show arbitrarily large optical depth, because a foreground
screen dust extinction model is applicable \citep{im03}. 
A few ULIRGs have been found to show substantially larger
optical depths than the upper limits estimated for mixed dust/source
geometry, and thus suggest centrally-concentrated energy sources
(e.g., UGC 5101 and IRAS 08572+3915; Imanishi \& Dudley 2000; Imanishi
et al. 2001; Imanishi \& Maloney 2003). These sources also 
display signatures of buried AGNs based on the equivalent width of the
3.3 $\mu$m PAH emission \citep{imd00,idm01} and observations at other
wavelengths \citep{dud97,soi00,ima03b}, demonstrating that this second
method can also be applied to investigate the energy sources of
ULIRGs.

A number of papers have presented $L$-band spectra of nearby ULIRGs
\citep{imd00,idm01,im03,ris03,ris05} and their energy sources have
been investigated.  However, the number of the observed sources is
still too small to draw statistically meaningful conclusions. In
this paper, we present ground-based $L$-band spectra of a large sample
of nearby ULIRGs.  Throughout this paper, $H_{0}$ $=$ 75 km s$^{-1}$
Mpc$^{-1}$, $\Omega_{\rm M}$ = 0.3, and $\Omega_{\rm \Lambda}$ = 0.7
are adopted.  The physical scale of 1$''$ is 0.34 kpc in the nearest
source at $z =$ 0.018, 0.91 kpc at $z =$ 0.05, and 2.44 kpc in the
furtherest source at $z =$ 0.15.

\section{Target}

ULIRGs in the {\it IRAS} 1 Jy sample \citep{kim98}, with no obvious
Seyfert signatures in the optical (non-Seyfert ULIRGs), are our main
targets to look for buried AGNs.  We restrict the sample to those at
$z \leqq 0.15$ so that the long wavelength side of the redshifted 3.4
$\mu$m dust absorption feature is at $\lambda <$ 4.1 $\mu$m; beyond
this wavelength the steeply increasing Earth's atmospheric background
signal degrades spectral quality drastically in the case of ground-based
observations. 
A declination cutoff $>$ $-$26$^{\circ}$ was also applied so that the
targets are observable with small airmasses from Mauna Kea, Hawaii,
our main observing site. These selection criteria result in 24 ULIRGs
classified optically as LINERs, 17 ULIRGs classified optically as
HII regions, and 4 optically unclassified ULIRGs, at $z <$ 0.15 in
this sample, where the optical classifications are based on Veilleux et
al. (1999a; their Table 2). 

We gave highest priority to LINER ULIRGs because a pure buried AGN is
expected to produce X-ray dissociation regions \citep{mal96} and to
show a LINER-type optical spectrum \citep{idm01}. 
All 24 LINER ULIRGs were observed, comprising a complete sample. 
For the HII and unclassified ULIRGs, 11 (out of 17)
and 2 (out of 4) sources were observed, respectively. One ULIRG, IRAS
13443+0822, consists of north-eastern and south-western nuclei, which
are optically classified as an HII region and a Sy 2,
respectively (Veilleux et al. 1999a).  Although this ULIRG is
classified as Sy 2 for the whole system (Veilleux et al. 1999a,
their Table 2), we observed only the north-eastern HII nucleus.  We
include this source in the category of HII ULIRGs, raising the number
of observed HII ULIRGs to 12. 
The observed 12 HII ULIRGs also constitute a complete sample, in the
right ascension range 10--21 hr.  
The details of the observed non-Seyfert ULIRGs are summarized in
Table 1.

In addition to these non-Seyfert ULIRGs, several ULIRGs with obvious
Seyfert signatures in the optical were also observed as a control
sample. There are 12 Sy 2 ULIRGs at $z <$0.15 and declination
$>$$-$26$^{\circ}$ in the {\it IRAS} 1 Jy sample, excluding the
above-mentioned IRAS 13443+0822. In total, 7 Sy 2 ULIRGs (Table 1)
were observed. The main aims of observing these Sy 2 ULIRGs are
twofold: (1) to check if our $L$-band spectroscopic method can
successfully find signatures of obscured AGNs in these ULIRGs which we
know possess obscured AGNs, and (2) to provide a control sample with
which the properties of buried AGNs in non-Seyfert ULIRGs can be
compared.

$L$-band spectra of several Sy 1 ULIRGs were also obtained.  For these
unobscured AGNs, since no significant 3.4 $\mu$m dust absorption
feature is expected, we extend the redshift restriction to $z < 0.17$
where the longer wavelength side of the 3.3 $\mu$m PAH emission
feature is covered in the ground-based $L$-band spectrum.  There are 7
Sy 1 ULIRGs at $z < 0.17$ with declination $>$$-$26$^{\circ}$ in the
IRAS 1 Jy sample, and all of these sources were observed (Table 1).

\section{Observations and Data Analysis}

Observations were made primarily using the IRCS near-infrared
spectrograph \citep{kob00} on the Subaru 8.2m telescope \citep{iye04}.
Several bright ULIRGs were observed with CGS4 \citep{moun90} on the
UKIRT 3.6m telescope, and with SpeX \citep{ray03} and NSFCAM
\citep{shu94} on the IRTF 3m telescope. Table 2 provides a detailed
observing log.

For the Subaru IRCS observing runs, the sky was clear during the
observations of all but one ULIRG (IRAS 16090$-$0139). The seeing at
$K$, measured in images taken before $L$-band spectroscopy, was
0$\farcs$4--0$\farcs$9 full-width at half maximum (FWHM). A
0$\farcs$9-wide slit and the $L$-grism were used with a 58-mas pixel
scale.  The achievable spectral resolution is R $\sim$ 140 at $\lambda
\sim$ 3.5 $\mu$m.  A standard telescope nodding technique (ABBA
pattern), with a throw of 7--10$''$ along the slit, was employed to
subtract background emission.  The optical guider of the Subaru
telescope was used to monitor the telescope tracking.  Exposure
time was 1.5--2.5 sec, and 15--40 coadds were made at each nod
position.
            
The spectra of two bright ULIRGs, IRAS 08572+3915 and Mrk 1014, were
taken using IRTF SpeX \citep{ray03}.  The 1.9--4.2 $\mu$m
cross-dispersed mode with a 1\farcs6 wide slit was employed.  This
mode enables $L$- (2.8--4.1 $\mu$m) and $K$-band (2--2.5 $\mu$m)
spectra to be obtained simultaneously, with a spectral resolution of R
$\sim$ 500.  The sky conditions were photometric throughout the
observations, and the seeing at $K$ was measured to be in the range
0$\farcs$5--0$\farcs$8 FWHM. A standard telescope nodding technique
(ABBA pattern) with a throw of 7$\farcs$5 was employed along the slit.
The telescope tracking was monitored with the infrared slit-viewer of
SpeX.  Each exposure was 15 sec, and 2 coadds were made at each
position.

For the IRTF NSFCAM observations of IRAS 07599+6508, 3C 273, and IRAS
15462$-$0450, we used the NSFCAM grism mode \citep{shu94}.  Sky
conditions were photometric throughout the observations, and the
seeing FWHM was measured to be 0$\farcs$5--0$\farcs$9 in the
$K$-band. The HKL grism and L blocker were used with the 4-pixel slit
(= 1$\farcs2$).  The resulting spectral resolution was R $\sim$ 150 at
$\lambda \sim$ 3.5 $\mu$m.  A standard 10$''$ telescope nodding
technique (ABBA pattern) was employed.  Each exposure was 1 sec, and
30 coadds were made at each nod position.  An optical guide star was
used for the telescope tracking, and the objects were observed in the
eastern sky, where the tracking performance was well confirmed.

Spectra of four ULIRGs (Arp 220, IRAS 05189$-$2524, Mrk 273, Mrk 231)
were taken with UKIRT CGS4 \citep{moun90}, using a 1$\farcs$2 wide
slit, and were published in \citet{imd00}.  The observing details are
found in that paper.

For all the observing runs, A-, F-, and G-type main sequence stars
(Table~\ref{tbl-2}) were observed as standard stars, with mean airmass
difference of $<$0.1 to the individual ULIRGs' nuclei, to correct for
the transmission of the Earth's atmosphere and to provide flux
calibration.  The $L$-band magnitudes of the standard stars were
estimated from their $V$-band ($\lambda =$ 0.6 $\mu$m) magnitudes,
adopting the $V-L$ colors appropriate to the stellar types of
individual standard stars \citep{tok00}.

Standard data analysis procedures were employed, using IRAF   
\footnote{ IRAF is distributed by the National Optical Astronomy
Observatories, which are operated by the Association of Universities
for Research in Astronomy, Inc. (AURA), under cooperative agreement
with the National Science Foundation.}.
Initially, frames taken with an A (or B) beam were subtracted from
frames subsequently taken with a B (or A) beam, and the resulting
subtracted frames were added and divided by a spectroscopic flat
image.  Then, bad pixels and pixels hit by cosmic rays were replaced
with the interpolated values of the surrounding pixels.  Finally the
spectra of ULIRGs' nuclei and standard stars were extracted, by
integrating signals over 0$\farcs$9--2$\farcs$0, depending on actual
signal profiles.  Wavelength calibration was performed taking into
account the wavelength-dependent transmission of the Earth's
atmosphere.  The spectra of ULIRGs' nuclei were divided by the
observed spectra of standard stars, multiplied by the spectra of
blackbodies with temperatures appropriate to individual standard stars
(Table~\ref{tbl-2}).

Flux calibration for all ULIRGs but one (IRAS 16090$-$0139) was done 
based on signals of ULIRGs and standard stars detected inside our slit
spectra. For the  
Subaru IRCS spectrum of IRAS 16090$-$0139, since its spectrum was
taken under non-photometric conditions, we calibrated its flux level
based on our own $L$-band photometry ($L$ = 13.3), made on a different
night using IRTF NSFCAM.  Seeing sizes
at $K$ (and $L$) were always smaller than the employed slit widths,
and good telescope tracking performances of Subaru, IRTF, and UKIRT
were established.  We thus expect possible slit loss to be
minimal. To estimate this ambiguity, we divided the whole data set
into sub-groups and compared their flux levels.  With the exception of
the highly time variable source 3C 273, the flux levels agree within
10\% for most objects and 30\% for the worst case, suggesting that the
flux uncertainty is at a similar level, which will not affect our main
conclusions significantly. For 3C 273, time variation was clearly
recognizable even during our IRTF NSFCAM observing run. For this
source, the resulting spectrum is a time averaged one.

To obtain an adequate signal-to-noise ratio in each element,
appropriate binning of spectral elements was performed, particularly
at $\lambda_{\rm obs} < 3.3$ $\mu$m in the observed frame, where the
scatter of data points is larger, due to poorer Earth atmospheric
transmission, than at $\lambda_{\rm obs} > 3.4$ $\mu$m.  The resulting
spectral resolution at $\lambda_{\rm obs} < 3.3$ $\mu$m is R
$\lessapprox 100$ in most cases.  The atmospheric transmission curve
at $\lambda_{\rm obs} =$ 2.8--3.3 $\mu$m is fairly smooth at this
spectral resolution. Thus, even if the net positions of the target
object and standard star on the slit differ slightly (on the sub-pixel
scale) along the wavelength direction, the standard data analysis
described above is expected to produce no significant spurious
features in spectra with R $\lessapprox 100$ (see Imanishi \& Maloney
2003). 

\section{Results}

\subsection{$L$-band spectra}

Figures 1, 2, 3, 4, and 5 present flux-calibrated $L$-band spectra of
the nuclei of ULIRGs optically classified as LINERs, HII-regions,
unclassified, Sy 2, and Sy 1, respectively. For ULIRGs
which have more than one nuclei brighter than $L$ $\sim$ 14.5 mag,
spectra of individual nuclei are shown separately.  The total number
of $L$-band spectra is 27 for LINER, 13 for HII, 2
for unclassified, 8 for Sy 2, and 7 for Sy 1 ULIRGs' {\it
nuclei}.  $L$-band spectra of several ULIRGs' nuclei, previously
presented by \citet{imd00} and \citet{im03}, are shown again here.

There are a variety of spectral shapes, but the majority of the
observed ULIRGs show emission features at $\lambda_{\rm obs}$ = (1 +
$z$) $\times$ 3.29 $\mu$m, the wavelength where the 3.3 $\mu$m PAH
emission feature has a peak. We thus identify these features as the 3.3
$\mu$m PAH emission. To estimate the strength of the 3.3 $\mu$m PAH
emission feature in the spectra where broad absorption features
coexist and overlap with the PAH emission feature in wavelength, we
make the reasonable assumption that the profiles of the 3.3 $\mu$m PAH
emission in these ULIRGs are similar to those of Galactic star-forming
regions and nearby starburst galaxies; the main emission profile
extends between $\lambda_{\rm rest}$ $=$ 3.24--3.35 $\mu$m in the rest
frame \citep{tok91,imd00}. 
Data points at slightly shorter than $\lambda_{\rm rest}$ $=$ 3.24
$\mu$m and slightly longer than $\lambda_{\rm rest}$ $=$ 3.35 $\mu$m,
unaffected by obvious absorption features, are adopted as the continuum
levels to estimate the 3.3 $\mu$m PAH emission strength.  
We adopt the spectral profile of type-1 sources \citep{tok91} as a
template for the 3.3 $\mu$m PAH emission. 
The adopted template reproduces the observed 3.3 $\mu$m PAH emission
features of the ULIRGs reasonably well, with our spectral resolution and
signal-to-noise ratios. 
Table~\ref{tbl-3} summarizes the fluxes, luminosities, and rest-frame
equivalent widths (EW$_{\rm 3.3PAH}$) of the 3.3 $\mu$m PAH emission
feature.  
The uncertainties of the 3.3 $\mu$m PAH emission fluxes, estimated from
the fittings, are $<$20\% in most cases, and $\sim$30\% even for ULIRGs
with large scatters at the wavelengths around the 3.3 $\mu$m PAH
emission. 
We also tried several continuum levels in a reasonable range, to
estimate possible systematic uncertainties coming from continuum
determination ambiguities.  
The estimated 3.3 $\mu$m PAH emission fluxes usually agree within
20\% for ULIRGs with smooth continuum emission. 
Even in ULIRGs with winding spectra, the estimated fluxes do not differ
by $>$30\%, as long as reasonable continuum levels are adopted. 
Thus, the total PAH flux uncertainties are unlikely to exceed $\sim$40\%
even in the worst case.  

In addition to the 3.3 $\mu$m PAH emission feature, clear absorption
features at $\lambda_{\rm rest}$ = 3.4 $\mu$m by bare carbonaceous
dust \citep{pen94} was found in three LINER ULIRGs, IRAS 08572+3915NW,
12127$-$1412NE, and 17044+6720 (Figure 1), and one Sy 2 ULIRG, IRAS
12072$-$0444N (Figure 4). The unobscured $L$-band spectral shapes of
AGNs are dominated by hot dust emission and almost linear
\citep{ima02,ima03,imw04}. Starburst galaxies also show $L$-band
spectra that are approximately linear, aside from the 3.3 $\mu$m PAH
emission feature \citep{imd00}.  To estimate the optical depths of the
3.4 $\mu$m carbonaceous dust absorption features, we therefore adopt a
linear continuum, which is shown as a dashed line in the spectra of
these four ULIRGs in Figures 1 and 4.  The estimated optical depths of
the 3.4 $\mu$m absorption feature ($\tau_{3.4}$) for these ULIRGs are
summarized in Table 4.

In the observed spectra of many ULIRGs, other important features are
present: (1) There is a spectral ``gap'', in that the flux densities
at the shorter wavelength side of the 3.3 $\mu$m PAH emission feature
are fainter than the extrapolation from the data points at the longer
wavelength side.  (2) The continuum is concave.  At the shorter
wavelength side of the 3.3 $\mu$m PAH emission, the continuum flux
level initially decreases with decreasing wavelength, but then begins
to increase again at $\lambda_{\rm rest} \sim 3.05$ $\mu$m, and then
the spectra become flat at the shortest wavelengths ($\lambda_{\rm
rest}$ $<$ 2.7 $\mu$m). This behavior is naturally explained by the
broad H$_{2}$O ice absorption feature caused by ice-covered dust
grains \citep{spo00,im03}. The signatures of this feature are found in
none of the Sy 1 ULIRGs, as expected from the unobscured view of the
AGN. The emission from stars is reduced in the $L$-band compared to
that in the $K$-band, whereas hot ($\sim$1000K) dust heated by an AGN
has strong emission at both $K$ and $L$. In a starburst/AGN composite
ULIRG, if the $L$-band emission were dominated by the AGN-powered hot
dust emission, while the longer wavelength tail of the stellar
emission extended to the shorter part of the $L$-band spectrum, then
the $L$-band spectrum may appear to be concave. However, this scenario
cannot produce the observed spectral gaps (point 1 above), so the
detected features are ascribed to H$_{2}$O ice absorption.

The shortest wavelength of the H$_{2}$O ice absorption feature is
$\lambda_{\rm rest}$ $\sim$ 2.7 $\mu$m \citep{whi88,smi93}. Due to
redshifting, for the majority of the observed ULIRGs at $z >$ 0.07, the
continuum on the short-wavelength side of the absorption feature is
included in our $L$-band spectra at
$\lambda_{\rm obs}$ $\sim$ 2.8--4.1 $\mu$m. 
Thus, continuum levels at both shorter and longer wavelength sides of
important features (broad 3.1 $\mu$m ice absorption, 3.3 $\mu$m PAH
emission, and 3.4 $\mu$m absorption) are covered in our spectra, making
continuum determination ambiguities small.
To estimate the optical depth of this H$_{2}$O absorption feature, we
adopt a linear continuum, as we did for the 3.4 $\mu$m absorption
feature.  
Although \citet{im03} adopted a concave quadratic continuum to
derive the lower limits, linear continua are arguably more plausible,
given that both the AGN and starburst components (outside the 3.3 $\mu$m
PAH emission feature) show linear continuum emission
\citep{imd00,ima02,ima03,imw04}. 
This H$_{2}$O ice absorption feature is spectrally very broad, and 
the continuum level can vary slightly, depending on the adopted data
points used for the continuum determination. 
We tried several plausible linear continuum levels, and finally adopt
the lowest plausible level to obtain a conservative lower limit for the
optical depth of the 3.1 $\mu$m H$_{2}$O ice absorption feature
($\tau_{3.1}$). 
The derived $\tau_{3.1}$ values for ULIRGs showing this feature clearly
are summarized in Table 4.

\subsection{$K$-band spectra}

Two ULIRGs, IRAS 08572+3915 and Mrk 1014, were observed with IRTF
SpeX, and so $K$-band ($\lambda$=2.0--2.5 $\mu$m) spectra were
simultaneously taken.  Figure 6 shows the $K$-band spectra of these
two sources.

The $K$-band spectrum of IRAS 08572+3915 (Figure 6, {\it Left}) is
similar to that presented by \citet{gol95}.  The spectrum is very red
and displays no detectable CO absorption features due to stars at
$\lambda_{\rm rest}$ $>$ 2.29 $\mu$m or $\lambda_{\rm obs}$ $>$ 2.43
$\mu$m, which suggests that featureless continuum emission from dust
heated by an AGN dominates the $K$-band spectrum.

The $K$-band spectrum of Mrk 1014 (Figure 6, {\it Right}) shows a
strong Pa$\alpha$ emission line. The emission line displays both broad
and narrow components, as expected from the optical classification of
this galaxy as Sy 1. The broad component has a line width of FWHM
$\sim$ 3100 km s$^{-1}$ and a flux of 5 $\times$ 10 $^{-14}$ ergs
s$^{-1}$ cm$^{-2}$.  The narrow component has a line width of FWHM
$\sim$ 600 km s$^{-1}$ and a flux of 1 $\times$ 10$^{-14}$ ergs
s$^{-1}$ cm$^{-2}$.

\section{Discussion}

In this section, we first focus on the investigations of the energy
sources of non-Seyfert ULIRGs based on our $L$-band spectra
($\S$5.1--5.5), and then combine these spectra with data at other
wavelengths and compare these sources with the properties of Seyfert
ULIRGs ($\S$5.6--5.8).

\subsection{The detected modestly obscured starbursts}

The detection of the 3.3 $\mu$m PAH emission clearly indicates the
presence of starbursts in the majority of ULIRGs. Since dust
extinction in the $L$-band is about 0.06 times as large as in the
optical $V$-band ($\lambda$ = 0.6 $\mu$m; Rieke \& Lebofsky 1985; Lutz
et al. 1996), the attenuation of 3.3 $\mu$m PAH emission with dust
extinction of A$_{\rm V}$ $\sim$ 15 mag is less than 1 mag. Thus, the
observed 3.3 $\mu$m PAH emission luminosities can be used to determine
the absolute magnitudes of modestly obscured (A$_{\rm V}$ $<$ 15 mag)
starburst activity.

At $\lambda_{\rm rest}$ $\sim$ 3.3 $\mu$m, Pf$\delta$ emission is
present, superposed on the 3.3 $\mu$m PAH emission.  The relative
contribution from this Pf$\delta$ emission line is expected to be the
highest in Sy 1 ULIRGs, because: (1) broad components from the AGNs
are unattenuated, and (2) 3.3 $\mu$m PAH emission is weak compared to
starburst-dominated ULIRGs. For the Sy 1 ULIRGs Mrk 1014 and IRAS
07599+6508, we can estimate the contamination by the Pf$\delta$ emission
using the measured Pa$\alpha$ emission line fluxes (This paper;
Taniguchi et al. 1994).   
For high-transition broad and narrow emission lines, such as Pa$\alpha$
and Pf$\delta$, case-B is applicable \citep{rhe00}, and so we adopt a
Pf$\delta$-to-Pa$\alpha$ flux ratio of 0.023 \citep{wyn84}. 
The estimated Pf$\delta$ fluxes, including both broad and narrow
components, are more than a factor of 20 smaller than the measured 3.3
$\mu$m PAH flux (Mrk 1014) or its upper limit (IRAS 07599+6508) in Table 3.
For another Sy 1 ULIRG Mrk 231, we find that the Pf$\delta$ flux,
estimated based on the Br$\gamma$ ($\lambda_{\rm rest}$ = 2.17 $\mu$m)
flux \citep{gol95} and case-B (Pf$\delta$/Br$\gamma$ = 0.26; Wynn-Williams
1984), is a factor of $>$25 smaller than the observed 3.3 $\mu$m PAH
flux (Table 3).  
For the remaining Sy 1 ULIRGs, we adopt a
median quasar spectrum at $\lambda_{\rm rest}$ = 1.8--2.0 $\mu$m
\citep{mur99} and an average spectral energy distribution of quasars at
$\lambda$ $>$ 1 $\mu$m 
(F$\nu$ $\propto$ $\nu^{-1.4}$; Neugebauer et al. 1987). 
Assuming case-B, we find that the equivalent width of the Pf$\delta$
line is $\sim$0.3 nm for a typical Sy 1 source.
For Sy 2 and non-Seyfert ULIRGs, the Pf$\delta$ contaminations are
expected to be much smaller than Sy 1 ULIRGs. 
Whenever Pa$\alpha$ fluxes are available in the literature
\citep{vei97,vei99b}, we confirm that the expected Pf$\delta$
fluxes never exceed 10\% of the observed 3.3 $\mu$m PAH fluxes or
their upper limits in Sy 2 and non-Seyfert ULIRGs.   
Thus, we can safely assume that Pf$\delta$ contamination is
insignificant for the excess component at $\lambda_{\rm rest}$ $\sim$
3.3 $\mu$m, which must be mostly ascribed to 3.3 $\mu$m PAH emission
($\S$4.1).  

The observed 3.3 $\mu$m PAH to infrared luminosity ratios 
(L$_{\rm 3.3PAH}$/L$_{\rm IR}$) are summarized in column 4 of Table 3. The
ratios are factors of 2 to more than 10 times smaller than those found
in lower luminosity, less-obscured starbursts 
($\sim$10$^{-3}$; Mouri et al. 1990; Imanishi 2002).  Since emission
from the ULIRGs' cores ($<$500 pc) is covered in our slit 
spectra, except for the nearest ULIRG Arp 220 ($z =$ 0.018), the small
L$_{\rm 3.3PAH}$/L$_{\rm IR}$ ratios suggest that the detected {\it
  nuclear} starbursts can account for only a small fraction of the
infrared luminosities of the observed ULIRGs.  \citet{fis00} have
previously found that the 6.2 $\mu$m PAH to infrared luminosity ratios
in ULIRGs are smaller by a similar factor than in less-obscured
starburst galaxies.  The 6.2 $\mu$m PAH luminosities were measured
with the large apertures of {\it ISO}, and yet they show a similar PAH
deficit: this indicates that PAH emission from the host galaxies,
outside our slit spectra, is not energetically important in the
infrared luminosities of ULIRGs \citep{soi00}. Both the detected 3.3
$\mu$m and 6.2 $\mu$m PAH emission probe modestly obscured (A$_{\rm
  V}$ $<$ 15 mag or so) starbursts, presumably at the outer regions of
the ULIRGs' cores. The energy sources responsible for the bulk of the
infrared emission of these ULIRGs must be deeply buried (A$_{\rm V}$
$>>$ 15 mag). 

\subsection{Buried AGNs with very weak starbursts}

The energetically dominant, deeply buried energy sources at the cores
of non-Seyfert ULIRGs can be either buried AGNs or very compact
starbursts or both.  Since buried AGNs can produce large infrared dust
emission luminosities, without PAH emission (see $\S$1), the small
observed L$_{\rm 3.3PAH}$/L$_{\rm IR}$ values are naturally
explained. Very highly obscured starbursts can also explain the small
observed L$_{\rm 3.3PAH}$/L$_{\rm IR}$ ratios, because flux
attenuation of the 3.3 $\mu$m PAH emission may be significant, while
that of longer wavelength infrared emission (8--1000 $\mu$m) may not.
These two scenarios are difficult to differentiate based on the
absolute luminosities of PAH emission, but can be distinguished by the
{\it equivalent width} of emission or absorption features.  
In a pure normal starburst, where HII regions, molecular gas, and
photo-dissociation regions are spatially mixed, 
the equivalent width of the 3.3 $\mu$m PAH emission feature is 
less affected by dust extinction ($\S$1), and so the observed value  
should be close to the intrinsic one (EW$_{\rm 3.3PAH}$ $\sim$ 100 nm;
Moorwood 1986; Imanishi \& Dudley 2000), regardless of dust
extinction of the starburst. 
On the other hand, an AGN produces a
PAH-free featureless continuum, with strong absorption features when the
AGN is highly dust-obscured.  
If this emission contributes significantly to an 
observed $L$-band flux, then the EW$_{\rm 3.3PAH}$ value should be
smaller than in normal starburst galaxies.  
The EW$_{\rm 3.3PAH}$ values in starbursts have an average value of 
EW$_{\rm 3.3PAH}$ $\sim$ 100 nm, with some scatter, but
never become lower than 40 nm (Moorwood 1986).
Thus, if we adopt EW$_{\rm 3.3PAH}$ 
${^{\displaystyle <}_{\displaystyle \sim}}$ 40 nm as strong signatures
of a significant contribution from the AGN's PAH-free continuum, then we
can argue that the three LINER ULIRGs IRAS 08572+3915NW, 12127$-$1412NE,
and 17044+6720 possess powerful AGNs. 
All of them display strong dust absorption features at 3.1 $\mu$m
and/or 3.4 $\mu$m, suggesting that the AGNs are deeply buried.

\citet{kim02} estimated the nuclear $K'$-band magnitudes within the
central 4 kpc diameter (K$^{'}_{4}$) for the ULIRGs in the IRAS 1 Jy
sample (their Table 3).  We combine these data with the $L$-band
magnitudes derived from our slit spectra (L$_{\rm spec}$), and find
that the K$^{'}_{4}$ $-$ L$_{\rm spec}$ colors of the three ULIRGs,
IRAS 08572+3915NW, 12127$-$1412NE, and 17044+6720, occupy the largest
(reddest) three values.  Hot dust emission heated by an AGN produces
redder $K - L$ colors than normal stellar emission and so the $K - L$
colors of AGNs are usually redder than starbursts
\citep{wil84,alo03,ia04,imw04}.  The red K$^{'}_{4}$ $-$ L$_{\rm
spec}$ colors again support the scenario that powerful buried AGNs are
present and produce strong hot dust emission in these ULIRGs.

\subsection{Buried AGNs with coexisting strong starbursts}

Based on the EW$_{\rm 3.3PAH}$ values, the presence of buried AGNs is
strongly indicated in the three LINER ULIRGs discussed above.  This
was possible primarily because the 3.3 $\mu$m PAH emission expected
from starbursts is exceptionally weak in these ULIRGs.  Hence these
ULIRGs may be classified as almost pure buried AGNs.  However, the
nuclei of ULIRGs are generally extremely rich in molecular gas
\citep{sam96}, which not only fuels the existing supermassive black
holes and thereby increases the luminosities of AGNs, but is also likely to
fuel star formation.  Hence it is expected that, if powerful
AGNs are present in the majority of non-Seyfert ULIRGs, they will
coexist with starbursts. It is therefore important to look
for the signatures of powerful buried AGNs in the cores of
starburst-containing ULIRGs. To detect the AGN signatures in these
composite cores requires more careful analysis of the spectra than for
the almost pure buried AGNs \citep{idm01,soi02,im03,spo04}. Even if the
intrinsic luminosities of a buried AGN and surrounding less-obscured
starbursts are similar, the flux from the buried AGN 
will be more highly attenuated by dust extinction. In addition to this
extinction, when the broad 3.1 $\mu$m H$_{2}$O ice absorption feature
is present \citep{spo00,im03}, the AGN flux is suppressed even more
severely. In this case, the 3.3 $\mu$m PAH emission is not diluted
significantly by the PAH-free continuum produced by the AGN, and so
the small EW$_{\rm 3.3PAH}$ diagnostic can no longer be used.

To determine whether the remaining ULIRGs' cores consist of (a)
modestly obscured starbursts and deeply buried {\it starbursts}, or of
(b) modestly obscured starbursts and deeply buried {\it AGNs}, we use
the absolute optical depth values of dust absorption features found in
the $L$-band spectra. As mentioned in $\S$1, these values can be used
to distinguish whether the energy sources are spatially well mixed
with dust, or are more centrally concentrated than the dust. Here we
discuss this point more quantitatively.

In a mixed dust/source geometry, flux attenuation as a
function of optical depth is
\begin{eqnarray}
I(\tau_{\lambda}) & = & I_{0} \times \frac{1 - e^{-\tau_{\lambda}}}{\tau_{\lambda}},  
\end{eqnarray}
where $I_{0}$ is the unattenuated intrinsic flux and $\tau_{\lambda}$ is the
optical depth at each wavelength; this takes different values inside
and outside the absorption features.  
The 3.1 $\mu$m H$_{2}$O ice absorption feature is observed if dust grains
are located deep inside molecular gas, are shielded from ambient UV
radiation, and are covered with an ice mantle \citep{whi88}.
The feature is not observed in the diffuse interstellar medium outside
molecular clouds, since ice mantles cannot survive exposure to the
ambient UV radiation.  
Thus, its optical depth reflects the total column density of ice-covered
dust grains in molecular gas, in front of the continuum-emitting energy
sources. 
The observed optical depth of the 3.1 $\mu$m H$_{2}$O ice absorption
feature is, by definition, dependent on the difference of flux
attenuation between the ice absorption feature and the 3--4 $\mu$m  
continuum outside the feature. 
\citet{im03} have shown that, in a mixed dust/source geometry,   
\begin{eqnarray}
\tau_{3.1} & \equiv & \ln [\frac{1 - e^{-\tau_{cont}}}{\tau_{cont}} 
\times \frac{\tau_{ice}}{1 - e^{-\tau_{ice}}}] \\
& = & \ln [(1+f) \frac{1 - e^{-0.06 \times A_{\rm V}}}{1 - e^{-0.06 \times
A_{\rm V} \times (1+f)}}],  
\end{eqnarray}
where $f$ is the fraction of dust that is covered with an ice mantle,
and it is assumed that the $\tau_{3.1}$/A$_{\rm V}$ ratio (the
abundance of H$_{2}$O ice relative to dust grains in the regions 
where an ice mantle can survive) is similar to that in Galactic
molecular clouds \citep{smi93,tan90,mur00}.

Stars are formed within the high density cores of molecular clouds.
If we observe such molecular clouds individually, then the energy
sources should be more centrally concentrated than the gas and
accompanying dust.  However, starburst galaxies consist of many such
stars and the energy sources will be distributed throughout the volume
of the starburst \citep{soi00,sie04}. Hence for observations of entire
starburst nuclei, the stellar energy sources and dust will be
relatively well mixed spatially (Figure 7a).  In fact, detailed
studies of the prototypical starburst galaxy M82 strongly suggest that
the stars of the burst and the obscuring dust are spatially well mixed
\citep{pux91,mcl93,for01}. Thus, it seems quite reasonable to regard
that starbursts can be represented by this mixed dust/source
geometry model.

The observed $\tau_{3.1}$ depends on A$_{\rm V}$ and the $f$-value,
the fraction of dust that is covered with an ice mantle. If we assume
that the $f$ values at the cores of ULIRGs are similar to that
in the well-studied prototype starburst galaxy M82 ($f$ $\sim$ 30\%;
Imanishi \& Maloney 2003), then $\tau_{3.1}$ will be $<$0.3 for any
A$_{\rm V}$ value in a mixed dust/source geometry \citep{im03}. In
reality, the cores of ULIRGs are estimated to have much higher surface
brightnesses than M82 \citep{soi00}, implying a much larger ambient
radiation field. In a mixed dust/source geometry, the radiation
environment in the cores of ULIRGs will therefore be much harsher and
it will be much more difficult for ice mantles to survive than in
M82. In this geometry, the $f$-values in ULIRGs' cores must be smaller
than the assumed value of $f$ = 30\%, and thus the upper limit of
$\tau_{3.1} =$ 0.3 should be robust.

On the other hand, in a buried AGN, the energy source is very compact
and is more centrally concentrated than the surrounding dust (Figure 7b,
{\it Left}).
In this geometry, a simple foreground screen dust model is 
applicable and the flux attenuation is 
\begin{eqnarray}
I(\tau_{\lambda}) & = &  I_{0} \times e^{-\tau_{\lambda}}.
\end{eqnarray}
The $\tau_{3.1}$ value is expressed as
\begin{eqnarray}
\tau_{3.1} & \equiv & \ln \frac{e^{-\tau_{cont}}}{e^{-\tau_{ice}}} \\
& = & \ln \frac{e^{-0.06 \times A_{\rm V}}}
{e^{-0.06 \times A_{\rm V} \times (1+f)}} \\
& = & 0.06 \times A_{\rm V} \times f. \\
\end{eqnarray}
The $\tau_{3.1}$ value can become arbitrarily large with increasing
A$_{\rm V}$ provided $f$ is non-zero.  In this centrally-concentrated
energy-source geometry, once the shielding column density of dust and
molecular gas around the central energy source becomes sufficiently
large, dust grains at a given distance from the center can be covered
with ice mantles ($f$ $>$ 0), and so a large $\tau_{3.1}$ value can be
produced.  Therefore, observing larger $\tau_{3.1}$ values than the limit
for a mixed dust/source geometry (0.3) is another
independent means to determine the geometry of ULIRGs' cores, and
therefore the nature of the central energy sources.

Aside from the three LINER ULIRGs with strong buried AGN signatures
discussed above (IRAS 08572+3915NW, 12127$-$1412NE, and 17044+6720),
14 LINER, 3 HII, and 2 optically-unclassified ULIRGs' nuclei have
observed $\tau_{3.1}$ values larger than the threshold (Table 4). 
Note that these observed values are contaminated by weakly obscured
starburst emission.  Hence these observed optical depths provide only
lower limits to the true $\tau_{3.1}$ values for the central buried 
energy sources.  For these ULIRGs' nuclei, a centrally-concentrated
energy-source geometry provides a natural explanation for the large
observed $\tau_{3.1}$ values.

There is one case in which a mixed dust/source geometry for the cores
can exceed the threshold value of 0.3. If the host galaxies are viewed
close to edge-on and a large column density of ice-covered dust is
present in the interstellar medium of the host galaxies, exterior to the
starburst core, then a foreground screen model is applicable and
$\tau_{3.1}$ can be large, just as for a centrally-concentrated energy
source. However, the
presence of ice-covered dust requires that the dust be located deep
inside molecular gas \citep{whi88}.  Strong 3.1 $\mu$m H$_{2}$O ice
absorption is found in the dense cores of molecular gas \citep{mur00},
and such dense molecular gas, probed by e.g., HCN (J = 
1--0) emission at $\nu$ = 88.632 GHz ($\lambda$ = 3.3848 mm) in the
millimeter wavelength range, is normally seen only in the nuclei of
ULIRGs (Imanishi, Nakanishi, \& Kohno 2005, in preparation). It is
thus very likely that the strong observed H$_{2}$O ice absorption
largely arises in the cores of the ULIRGs rather than in the extended
interstellar medium of the host galaxies. The large observed H$_{2}$O ice
absorption optical depths can thus be taken as some evidence of a
centrally-concentrated energy source geometry at the cores of these
ULIRGs.  

\citet{soi00} studied the cores of seven nearby ULIRGs and estimated
that, in the majority (six out of seven) of these ULIRGs' cores, the
emission surface brightnesses are very high ($\sim$10$^{13}$
L$_{\odot}$ kpc$^{-2}$) and close to the maximum observed values
seen in the cores of Galactic HII regions, even if the energy sources
are uniformly distributed over the cores ($<$500 pc in size).  
Provided that the emission properties of the six ULIRGs' cores
\citep{soi00} are representative of the majority of nearby ULIRGs, 
surface brightnesses of the {\it centrally-concentrated} energy sources
would be even higher, making {\it exceptionally centrally-concentrated
starbursts} less likely, if not completely ruled out.
AGNs are more natural explanations for the centrally-concentrated high
surface brightness energy sources at the cores of ULIRGs.  
We thus take the centrally-concentrated energy sources as {\it tentative}
evidence for buried AGNs.  

We can apply the same argument concerning the geometry of the energy
sources and dust to the absolute optical depth of the 3.4 $\mu$m
carbonaceous dust absorption feature ($\tau_{3.4}$).  This absorption
feature is found if a source is obscured by bare dust grains
\citep{pen94,ima96,raw03}, but absent if the absorbing dust is
ice-covered \citep{men01}.  Thus, the optical depth of the 3.4 $\mu$m
absorption feature traces the column density of bare carbonaceous dust
grains.  \citet{im03} have quantitatively demonstrated that
$\tau_{3.4}$ cannot be larger than 0.2 in a mixed dust/source
geometry.  Two LINER ULIRGs, IRAS 08572+3915NW and IRAS
12127$-$1412NE, show $\tau_{3.4}$ $>$ 0.2 (Table 4).  These ULIRGs
have already been classified as buried AGNs based on their EW$_{\rm
3.3PAH}$ values ($\S$5.2).  For them, additional support for powerful
buried AGNs is provided here from the geometry implied by the depth of
their 3.4 $\mu$m absorption features.

For the remaining ULIRGs where (1) the EW$_{\rm 3.3PAH}$ values are in
the possible scattered range of starburst galaxies ($>$40 nm), and (2) no strong
dust absorption features are observed, there are no obvious signatures
of buried AGNs. Their spectra can be explained by either of the
following scenarios: (1) the cores are dominated by starbursts with a
mixed dust/source geometry, or (2) the cores do possess powerful
buried AGNs, but the emission from the AGNs is so highly attenuated
that its contribution to the observed $L$-band fluxes is not
significant.  We have no way to distinguish between these two
scenarios, and their energy sources remain undetermined by our
$L$-band spectroscopic method. There are some known examples (e.g.,
NGC 4418) which show no obvious buried AGN signatures in the $L$-band
spectra \citep{ima04}, but almost certainly contain powerful AGNs
based on longer wavelength data \citep{dud97,spo01}.  Observations at
wavelengths with lower dust extinction are necessary to unveil
possible buried AGNs which are undetectable even in the $L$-band.

\subsection{Dust Extinction and Dereddened AGN luminosities}

In a centrally-concentrated energy source geometry, dust at a
temperature of 1000 K, which is close to the innermost dust
sublimation radius, produces emission with a peak at $\lambda \sim$ 3
$\mu$m (Figure 7b, {\it Right}), assuming approximately blackbody
emission.  Dust grains at larger radii, with lower temperatures, show
emission peaks at longer infrared wavelengths. The dust extinction
toward the 3 $\mu$m continuum-emitting region, A$_{\rm V}$(3$\mu$m),
can be estimated from the $\tau_{3.4}$ (bare dust) and $\tau_{3.1}$
(ice-covered dust) values, and can provide a better estimate for dust
extinction toward the buried AGN itself than the estimate using longer
infrared wavelengths, which only probes dust extinction toward the
outer, cooler emission regions (Figure 7b, {\it Right}).  If we assume
that the optical depth of these dust absorption features, relative to
A$_{\rm V}$, is similar to the Galactic diffuse interstellar medium
($\tau_{3.4}$/A$_{\rm V}$ = 0.004--0.007; Pendleton et al. 1994,
$\tau_{3.1}$/A$_{\rm V}$ = 0.06; Smith et al. 1993; Tanaka et
al. 1990; Murakawa et al. 2000), and if we adopt a foreground screen
model which should be applicable to the centrally-concentrated energy
source geometry, then the estimated A$_{\rm V}$ values are summarized
in column 4 of Table 4.

For the three ULIRGs with very small EW$_{\rm 3.3PAH}$ values, IRAS
08572+3915NW, 12127$-$1412NE, and IRAS 17044+6720, the $L$-band
emission is concluded to predominantly come from hot dust emission
heated by the buried AGN.  We can estimate the dereddened luminosity
of the AGN-heated dust emission fairly straightforwardly for these
sources.  Assuming that dust extinction at $L$, relative to $V$, is
similar to the Galactic diffuse interstellar medium (A$_{\rm
L}$/A$_{\rm V}$ $\sim$ 0.06; Rieke \& Lebofsky 1985; Lutz et
al. 1996), the dereddened AGN luminosities at $\lambda_{\rm rest}$ =
3--4 $\mu$m are $>$50, $>$10, and 0.7--2 $\times$ 10$^{45}$ ergs
s$^{-1}$, respectively, for IRAS 08572+3915NW, 12127$-$1412NE, and
17044+6720.  The dereddened AGN luminosities could quantitatively
account for the bulk of the infrared luminosities of these ULIRGs
(L$_{\rm IR}$ $\sim$ 5 $\times$ 10$^{45}$ ergs s$^{-1}$).  For IRAS
08572+3915NW and 12127$-$1412NE, the dereddened AGN luminosities
significantly exceed the observed infrared luminosities.  A similar
excess was found in the ULIRG Superantennae \citep{ris03}, and the
excess implies that the $\tau_{3.4}$/A$_{\rm L}$ ratios in some
ULIRGs' cores are larger than the ratios found in the Galactic
interstellar medium.

For the remaining starburst/buried-AGN composite ULIRGs (large
EW$_{\rm 3.3PAH}$ values with $>$40 nm and strong 3.1 $\mu$m H$_{2}$O
ice absorption with $\tau_{3.1}$ $>$ 0.3), no detectable 3.4 $\mu$m
absorption features are found.  The optical depths of the 3.1 $\mu$m
and 3.4 $\mu$m absorption features probe the column density of
ice-covered and bare dust grains, respectively, but it seems strange
that there are no bare dust grains in the nuclear regions of these
ULIRGs.  The 3.3 $\mu$m PAH emission feature is often accompanied by a
sub-peak at 3.4 $\mu$m \citep{tok91,imd00} and this sub-peak may
dilute the 3.4 $\mu$m dust absorption feature at the same wavelength.
In fact, some ULIRGs' spectra with strong 3.3 $\mu$m PAH emission show
sub-peaks at $\lambda_{\rm rest}$ = 3.4 $\mu$m in Figures 1, 2, and 3
(e.g., IRAS 09039+0503, 13509+0442, and 14197+0813).  For those ULIRGs
with EW$_{\rm 3.3PAH}$ $>$ 40 nm, the estimated A$_{\rm V}$ values
(Table 4, column 4) should be taken as only lower limits for dust
extinction toward the inner 3 $\mu$m continuum emitting regions at the
cores.  Although quantitative values are highly uncertain for these
sources, the cores are usually assumed to dominate the infrared
luminosities of nearby ULIRGs \citep{soi00}, and the cores of these
composite ULIRGs are suggested to have centrally-concentrated energy
sources, most likely buried AGNs. Thus, the buried AGNs in these ULIRGs
could account for the bulk of their observed infrared luminosities.

\subsection{Dependence on optical spectral type}

In total, $L$-band spectra of 27 LINER and 13 HII ULIRGs' nuclei have
been obtained.  3 LINER ULIRGs' and no HII ULIRGs' nuclei show clear
evidence for almost pure buried AGNs ($\S$5.2).  Additionally, 14
LINER and 3 HII ULIRGs' nuclei have tentative evidence for powerful
buried AGNs, in addition to modestly obscured starbursts ($\S$5.3).
Combining these values, 17 out of 27 LINER ULIRGs' nuclei (63\%) and 3
out of 13 HII ULIRGs (23\%) display some degree of evidence for
powerful buried AGNs.  The observed LINER ULIRGs comprise a complete
sample ($\S$2).  The observed HII ULIRGs also constitute a complete
sample in the right ascension range 10--22 hr ($\S$2).  The
fraction of sources with signatures of buried AGNs is significantly
higher in LINER ULIRGs than HII ULIRGs.  This suggests that a larger
fraction of LINER ULIRGs may have powerful buried AGNs than HII
ULIRGs.

Some ULIRGs have spectra of poor signal-to-noise ratios, and the
presence of strong dust absorption features at 3.1 $\mu$m
($\tau_{3.1}$ $>$ 0.3) or 3.4 $\mu$m ($\tau_{3.4}$ $>$ 0.2) may be
missed.  Strong 3.1 $\mu$m H$_{2}$O ice absorption may be present, but
be missed due to poor signal-to-noise ratios, in four LINER ULIRGs'
nuclei (IRAS 12112+0305NE, 12112+0305SW, 14252$-$1550W, and
16468+5200E) in Figure 1, and one HII ULIRG's nucleus (IRAS
21208$-$0519N) in Figure 2.  The two LINER ULIRGs (IRAS 
10485$-$1447 and 16090$-$0139) may have large $\tau_{3.4}$ values.
Even if only the HII ULIRG IRAS
21208$-$0519N is included as sources with tentative buried AGN
signatures, the AGN fraction in HII ULIRGs' nuclei is only 31\% (4/13),
significantly below the fraction for LINER ULIRGs' nuclei.

In a pure buried AGN, X-ray dissociation regions (XDRs), dominated by
low-ionization species \citep{mal96}, should develop and a LINER-type
optical spectrum is expected \citep{idm01}, rather than a typical
Seyfert spectrum originating in the narrow line regions (NLRs). 
The discovery of pure powerful buried AGNs with no detectable starbursts
in the two LINER ULIRGs (IRAS 08572+3915NW and 12127$-$1412NE) may be
explained by this scenario.   

For the remaining LINER ULIRGs with coexisting modestly-obscured
starbursts, the LINER optical spectra can be due to XDRs and/or 
shocks driven by the starbursts \citep{tan99,lut99}.   
In the case of XDRs, even though the central AGN itself is deeply
buried, emission from the surrounding XDRs can be relatively unobscured. 
Although the origin of the LINER optical spectra in these ULIRGs is
unclear, the higher detection rate of powerful buried AGNs in LINER
ULIRGs suggests that emission from XDRs may be important in the optical. 

\subsection{Comparison with VLA radio observations}

\citet{nag03} performed VLA observations at $\nu =$ 15 GHz ($\lambda
=$ 2 cm), with 150 mas spatial resolution, of a large sample of ULIRGs
in the same IRAS 1 Jy sample, and searched for compact radio core
emission, another good indicator of an AGN.  
They concluded that the detection rate of the radio cores in
LINER ULIRGs ($\sim$74\%) was substantially higher than in HII ULIRGs
($\sim$14\%).  Our higher detection rate of buried AGN signatures in
LINER ULIRGs than in HII ULIRGs is both qualitatively and
quantitatively similar to their conclusion.

Table 3 (columns 6 and 7) summarizes information on which individual 
ULIRGs have detectable (or non-detectable) radio cores \citep{nag03}, in
comparison with the buried AGN signatures from our $L$-band spectroscopic
method.  
Among the three ULIRGs with clear buried AGN signatures, two sources,
IRAS 08572+3915NW and 17044+6720, were observed with VLA and both 
(100\% = 2/2) show core radio emission.  For ULIRGs with
tentative evidence for buried AGNs, 8 LINER, 3 HII, and 1
optically-unclassified ULIRGs' nuclei have VLA data, and 5 LINER, 2 HII,
and 1 unclassified ULIRGs' nuclei have detectable VLA radio cores.
Hence the detection rate is 67 \% \{= (5+2+1)/(8+3+1)\}.  For ULIRGs
without buried AGN signatures in the $L$-band spectra, VLA cores were
detected in 8 out of 17 sources (47 \%).  
The higher detection rate of the VLA cores in ULIRGs
that show a higher level of buried AGN evidence in the $L$-band
suggest that our $L$-band spectroscopic method is a reliable, if not
perfect, tracer of elusive buried AGNs in non-Seyfert ULIRGs.

Our infrared $L$-band spectra can probe dust emission heated by an AGN
and thus quantitatively determine the dust emission luminosity, the
dominant emission mechanism of ULIRGs. However, the spectra may miss the
AGN signatures 
if the AGN emission is highly attenuated by dust extinction.  On the
other hand, the radio observations are negligibly affected by dust
extinction, and so could detect highly dust-obscured AGNs that may be
missed in our $L$-band spectroscopic method.  However, the radio data
trace synchrotron emission and may not be proportional to ULIRGs' dust
emission luminosities \citep{nag03}. Thus, the infrared $L$-band and
radio observations are complementary to each other.  The fact that
both independent methods provide a similar result suggests that the
higher detection rate of powerful buried AGNs in LINER ULIRGs than HII
ULIRGs is probably real.

\subsection{Comparison with Seyfert ULIRGs} 

We compare the $L$-band spectral properties of non-Seyfert ULIRGs
showing buried AGN signatures, with those of Seyfert ULIRGs (i.e.,
known AGN-possessing ULIRGs).  The apparent main difference between
them is that the ionizing radiation from the putative AGNs in
non-Seyfert ULIRGs is obscured by the surrounding dust along virtually
all lines-of-sight, while dust around the AGNs in Seyfert ULIRGs is
distributed in a ``torus'', and ionizing radiation from the AGNs can
escape along the torus axis, allowing the Seyfert signatures to be
detectable in the optical.  In Table 3, all Seyfert ULIRGs' nuclei show
EW$_{\rm 3.3PAH}$ smaller than $\sim$40 nm.  
The EW$_{\rm 3.3PAH}$ values of Sy 1 ULIRGs are systematically
smaller than Sy 2 ULIRGs, which is reasonable because unattenuated
emission from unobscured AGNs in Sy 1 ULIRGs can reduce the EW$_{\rm
3.3PAH}$ values more strongly than attenuated emission from obscured
AGNs in Sy 2 ULIRGs.  In Sy 2 ULIRGs, the significantly smaller
EW$_{\rm 3.3PAH}$ values than seen in starburst galaxies suggest that
the strong dilution of the PAH emission by the PAH-free continuum
produced by the AGN is present.  Our $L$-band spectroscopic method
has succeeded in detecting obscured AGN signatures in all of the known
obscured-AGN-possessing Sy 2 ULIRGs observed, demonstrating that this
method is very effective.

Among the eight Sy 2 ULIRGs' nuclei, \citet{nag03} presented VLA data
for four sources, and three sources show core emission (Table 3).  The
VLA core detection rate is 75\% (3/4). 
For the seven Sy 1 ULIRGs' nuclei observed by us, five sources have VLA
data and all show radio core emission (100\% = 5/5; Table 3). 
These detection rates are again clearly higher than those for ULIRGs
with no obvious AGN signatures in our $L$-band spectra.

Figure 8(a) plots the L$_{\rm 3.3PAH}$/L$_{\rm IR}$ ratios of ULIRGs,
separated by optical spectral types.  Both non-Seyfert and Seyfert
ULIRGs are plotted.  The median L$_{\rm 3.3PAH}$/L$_{\rm IR}$ values
are 0.1 (LINER), 0.2 (HII), 0.09 (Sy 2) and $<$0.15 (Sy 1).  The
L$_{\rm 3.3PAH}$/L$_{\rm IR}$ ratios do not differ significantly
between non-Seyfert and Seyfert ULIRGs, suggesting that both types of
ULIRGs possess detectable modestly-obscured starburst activity with
similar absolute luminosities.

Figure 8(b) plots the rest-frame equivalent widths of the 3.3 $\mu$m
PAH emission (EW$_{\rm 3.3PAH}$), separated by optical spectral types.
In this plot, Sy 2 ULIRGs tend to show smaller EW$_{\rm 3.3PAH}$
values than non-Seyfert ULIRGs, while Sy 1 ULIRGs show the smallest
values.  The median EW$_{\rm 3.3PAH}$ values are 75 nm (LINER), 85 nm
(HII), 15 nm (Sy 2) and $<$4 nm (Sy 1).  Even though the absolute
luminosities of the detectable modestly-obscured starbursts are
similar, less-obscured AGNs can have a larger contribution to the
observed $L$-band flux, and thereby decrease the EW$_{\rm 3.3PAH}$
values.  The trends in Figure 8(a) and (b) can be explained by a
scenario where AGNs in Sy 2 ULIRGs tend to be less obscured by dust
along our line-of-sight than AGNs in non-Seyfert ULIRGs.  In fact,
while many non-Seyfert ULIRGs show clear absorption features at 3.1
$\mu$m by H$_{2}$O ice and at 3.4 $\mu$m by bare carbonaceous dust
grains (Figure 1, 2, 3), no such absorption features are found in the
$L$-band spectra of Sy 2 ULIRGs, with the exception of IRAS
12072$-$0444N (Figure 4), supporting the less-obscuration scenario for
AGNs in Sy 2 ULIRGs.  We argue that buried AGNs in non-Seyfert ULIRGs
and optically visible AGNs (obscured by torus-shaped dust) in Sy 2
ULIRGs are different not only in the dust geometry, but also in the
dust column density along our line-of-sight, in that the dust columns
toward the AGNs in non-Seyfert ULIRGs are much higher.
Since the dust covering factor around buried AGNs in non-Seyfert ULIRGs 
(almost all directions) is also larger than AGNs in Sy 2 ULIRGs
(torus-shaped), the total amount of nuclear dust must be larger in the
former. 
The optical (non-)detectability of Seyfert signatures in ULIRGs depends
not on the presence or absence of powerful AGNs, but largely on the mass
of gas and dust in the nuclei. 

\subsection{Dependence on far-infrared colors}

Based on the {\it IRAS} 25 $\mu$m to 60 $\mu$m flux ratio 
($f_{\rm 25}$/$f_{\rm 60}$), ULIRGs are divided into warm ($>$ 0.2) and
cool ($<$ 0.2) sources (Sanders et al. 1988).  The majority of
non-Seyfert ULIRGs show cool far-infrared 
colors (Table 1).  Optical spectroscopy of galaxies with warm
far-infrared colors often reveals Seyfert signatures \citep{deg87,kee05},
while starburst galaxies usually show cool far-infrared colors.  The
warm colors of galaxies showing optical Seyfert signatures are usually
interpreted as emission from dust in the close vicinity of an AGN,
whose temperature can be higher than that in starbursts, because of
the higher radiation density in the former.  
It is sometimes argued that cool components of the far-infrared emission
from ULIRGs must be starburst-dominated (e.g., Downes \& Solomon 1998). 
However, the far-infrared color from AGN-heated dust is highly dependent
on the column density of dust around an AGN, in that a larger column
density makes the far-infrared color cooler 
($f_{\rm 25}$/$f_{\rm 60}$ ratio smaller).  
Our $L$-band spectra suggest that buried AGNs in non-Seyfert ULIRGs are
more highly dust obscured, and thus produce cooler far-infrared colors
than AGNs in Sy 2 ULIRGs. 

Figure 9(a) compares the observed 3.3 $\mu$m PAH to infrared
luminosity ratios, with {\it IRAS} 25 $\mu$m to 60 $\mu$m flux ratios
(i.e., far-infrared color), for non-Seyfert ULIRGs.  There is no clear
trend.  If the detected 3.3 $\mu$m PAH emission only probes the
modestly-obscured energetically-insignificant surface starbursts
($\S$5.1), and buried energy sources are responsible for the infrared
emission at 25 $\mu$m and 60 $\mu$m, no trend is expected.

Figure 9(b) compares the EW$_{\rm 3.3PAH}$ and far-infrared colors for
non-Seyfert ULIRGs.  The three ULIRGs with almost pure buried AGNs,
IRAS 08572+3915NW, 12127$-$1412NE, and 17044+6720, are distributed on
the warm side, but IRAS 12127$-$1412NE has a cool far-infrared color.
Thus, there is at least one example of a pure buried AGN that produces
a cool far-infrared color.  The bulk of ULIRGs with tentative evidence
of buried AGNs are also classified as cool sources, and their
far-infrared color distribution overlaps with the colors of ULIRGs with
no AGN signatures. 

We next compare the properties of the two pure buried AGNs, IRAS
08572+3915NW and 12127$-$1412NE.  IRAS 08572+3915NW shows only 3.4
$\mu$m dust absorption feature, with no detectable 3.1 $\mu$m H$_{2}$O
ice absorption.  Assuming Galactic dust properties, A$_{\rm V}$ $>$
110 mag is estimated toward the inner 3 $\mu$m continuum emitting
regions, close to the central AGN.  On the other hand, the bare dust
column of IRAS 12127$-$1412NE ($\tau_{3.4}$ $=$ 0.35) is estimated to
be A$_{\rm V}$ $>$ 50 mag.  This source displays ice absorption
($\tau_{3.1}$), and the column of ice-covered dust is estimated to be
A$_{\rm V}$ $>$ 6 mag.  The total dust column in IRAS 08572+3915NW may
be higher than IRAS 12127$-$1412NE, and yet the former source shows a
warmer infrared color.  This apparently contradicts the simple
prediction that the far-infrared color is cooler when a larger amount
of dust obscures an AGN, assuming the dust properties are similar.
However, IRAS 08572+3915NW and 12127$-$1412NE are different in that
strong ice absorption is found only in the latter source.  Ice-covered
dust grains can make a cooler far-infrared color 
(smaller $f_{\rm 25}$/$f_{\rm 60}$ ratio) than do bare dust grains,
because of the change of optical constants in the infrared
\citep{dud03}.  The cooler color of IRAS 12127$-$1412NE than
08572+3915NW may be due to the presence of a significant amount of
ice.

\section{Summary}

We presented the results of infrared $L$-band (3--4 $\mu$m) slit
spectroscopy of the nuclei of a large sample of nearby ULIRGs in the
IRAS 1 Jy sample.   
ULIRGs with no obvious Seyfert signatures in the optical (i.e., LINER,
HII, and unclassified) were mainly observed and spectra of 27 LINER, 13
HII, and 2 optically-unclassified ULIRGs' nuclei were obtained. 
Using the 3.3 $\mu$m PAH emission, 3.1 $\mu$m H$_{2}$O ice absorption
and 3.4 $\mu$m carbonaceous dust absorption features, we investigated
whether the cores of these ULIRGs can be explained by
very compact starbursts only, or show signatures of buried AGNs. 
Since our spectra covered both the shorter and longer wavelength sides
of these features, we were able to distinguish between
absorption-dominated and emission-dominated sources with small ambiguities. 
These spectra were compared with known AGN-possessing ULIRGs' nuclei (8 
Sy 2 and 7 Sy 1) and we found the following main conclusions. 

\begin{enumerate}
\item The 3.3 $\mu$m PAH emission, the probe of starbursts, was 
      detected in all but two non-Seyfert ULIRGs. 
      It is likely that modestly-obscured (A$_{\rm V}$ $<$ 15 mag)
      starbursts at the surface of ULIRGs' cores are responsible for
      the detected PAH emission.  
      The observed 3.3 $\mu$m PAH to infrared luminosity ratios were
      smaller by a factor of 2 to $>$10 than those found in
      less infrared-luminous starbursts, suggesting
      that the detected modestly-obscured starbursts are energetically
      insignificant. 
\item Three LINER ULIRGs, IRAS 08572+3915NW, 12127$-$1412NE, and 17044+6720,
      showed 3.3 $\mu$m PAH emission equivalent widths substantially
      smaller than typical values found in starburst galaxies. 
      These ULIRGs were classified as sources dominated by almost pure
      buried AGNs, with no detectable or very weak starburst activity.
\item Besides these three ULIRGs, the absolute optical depths of 3.1
      $\mu$m H$_{2}$O ice absorption 
      feature were used to distinguish whether the energy sources at the
      cores of ULIRGs are more centrally concentrated than the
      surrounding dust (as is expected for a buried AGN), or are
      spatially well mixed with dust (a normal starburst).   
      The large optical depths in 14 LINER, 3 HII, and 2 unclassified 
      ULIRGs' nuclei suggested that the energy sources are more
      centrally concentrated than the surrounding dust, unless a large 
      amount of foreground ice-covered dust is present in the
      interstellar medium of the host galaxies, which we consider unlikely.
      These ULIRGs were classified as showing tentative 
      evidence for powerful buried AGNs. 
\item Based on the optical depths of the 3.4 $\mu$m carbonaceous dust
      absorption feature, centrally-concentrated energy sources were
      suggested in the two LINER ULIRGs, IRAS 08572+3915NW and 12127$-$1412NE.
      Both of them had already been classified as buried AGNs from the
      3.3 $\mu$m PAH equivalent widths, and thus powerful buried AGNs
      were consistently identified with two independent methods.
\item In total, 17 out of 27 (63\%) LINER ULIRGs' nuclei and 3 out of 13
      HII ULIRGs' nuclei (23\%) showed some evidence for buried AGNs.
      The higher detection rate of buried AGNs in LINER ULIRGs than in
      HII ULIRGs was similar to the higher detection rate of compact
      radio cores (also a signature of AGNs) in the former. 
\item $L$-band spectra of 8 Sy 2 and 7 Sy 1 ULIRGs' nuclei
      were obtained, and the 3.3 $\mu$m PAH emission was detected in 7
      (out of 8) and 5 (out of 7) sources, respectively, suggesting that
      these Seyfert ULIRGs also contain detectable
      starburst activity.  
\item Based on the equivalent widths of the 3.3 $\mu$m PAH emission, 
      our $L$-band spectra successfully detected AGN signatures in all
      of the observed Seyfert 2 ULIRGs (known-obscured-AGN-possessing ULIRGs),
      demonstrating that our $L$-band spectroscopic method is indeed
      effective in revealing signatures of obscured AGNs.
\item The detection rate of VLA radio cores was higher in ULIRGs that
      display evidence of buried AGNs in the $L$-band.
      This suggested that our $L$-band spectroscopic method is 
      reliable, if not perfect.
\item Among 8 observed Sy 2 ULIRGs, only IRAS 12072$-$0444N showed a
      clear 3.4 $\mu$m carbonaceous dust absorption feature. 
      The remaining Sy 2 ULIRGs showed neither
      detectable 3.4 $\mu$m nor 3.1 $\mu$m absorption features.
\item There was no systematic difference in the 3.3 $\mu$m PAH to
      infrared luminosity ratios between Seyfert and non-Seyfert ULIRGs.
      However, Sy 1 and 2 ULIRGs tended to show smaller rest-frame
      equivalent widths of the 3.3 $\mu$m PAH emission than non-Seyfert
      ULIRGs, with Sy 1 ULIRGs (unobscured AGNs) exhibiting the
      smallest values.  
      The optical depths of absorption features in the $L$-band were 
      also generally smaller in Sy 2 ULIRGs than in non-Seyfert
      ULIRGs.
      These results were naturally explained if AGNs in non-Seyfert
      ULIRGs are more highly obscured by dust along our sightline than
      the AGNs in Sy 2 ULIRGs.
\item Evidence for buried AGNs was found in non-Seyfert ULIRGs with both
      warm and cool far-infrared colors. 
      The LINER ULIRG dominated by a pure buried AGN, IRAS 12127$-$1412NE,
      has a cool far-infrared color. 
      The cool far-infrared color in this pure buried AGN could be the
      consequences of a large dust column density around an AGN,
      including ice-covered dust, as is implied from strong dust
      absorption features at 3.1 $\mu$m and 3.4 $\mu$m.
\item Our $L$-band spectroscopy revealed the presence of AGNs in many
      non-Seyfert ULIRGs, and these AGNs are thought to be 
      obscured (at least opaque to the AGN's ionizing radiation) 
      along virtually all sightlines so as to be optically elusive 
      \citep{mai03}.
      This dust geometry is fundamentally different from the
      AGNs in Sy 2 ULIRGs which are obscured by dust in a torus
      geometry. 
      Together with the above finding that dust column along our
      line-of-sight is higher in AGNs in non-Seyfert ULIRGs,  
      it was implied that a larger amount of dust is concentrated
      surrounding the AGNs at the nuclei of non-Seyfert ULIRGs than
      Seyfert ULIRGs.  
      We argued that the (non-)detection of Seyfert signatures in the
      optical is highly dependent on the total amount of dust and its
      covering factor around a central AGN, and that many non-Seyfert
      ULIRGs may contain powerful buried AGNs.
      Other types of careful investigations at the wavelengths of lower
      dust extinction are important to find new buried AGNs which may
      even be undetectable in our $L$-band spectroscopic method. 
\end{enumerate}

\acknowledgments

We are grateful to H. Terada, M. Ishii, R. Potter, E. Pickett,
A. Hatakeyama, B. Golisch, D. Griep, P. Sears for their support during
our Subaru and IRTF observing runs.   
We thank the anonymous referee for his/her useful comments.
M.I. is supported by Grants-in-Aid for Scientific Research (16740117).
Research in Infrared Astronomy at the Naval Research Laboratory is supported
by the Office of Naval Research (USA).
Some part of the data analysis was made using a computer system operated
by the Astronomical Data Analysis Center (ADAC) and the Subaru Telescope
of the National Astronomical Observatory, Japan.
This research has made use of the SIMBAD database, operated at CDS,
Strasbourg, France, and of the NASA/IPAC Extragalactic Database
(NED) which is operated by the Jet Propulsion Laboratory, California
Institute of Technology, under contract with the National Aeronautics
and Space Administration.

\clearpage

\begin{deluxetable}{lcrrrrcrc}
\tabletypesize{\scriptsize}
\tablecaption{Observed ULIRGs and their far-infrared emission
properties.
\label{tbl-1}}
\tablewidth{0pt}
\tablehead{
\colhead{Object} & \colhead{Redshift}   & 
\colhead{f$_{\rm 12}$}   & 
\colhead{f$_{\rm 25}$}   & 
\colhead{f$_{\rm 60}$}   & 
\colhead{f$_{\rm 100}$}  & 
\colhead{log L$_{\rm IR}$} & 
\colhead{f$_{25}$/f$_{60}$} & 
\colhead{Optical}   \\
\colhead{} & \colhead{}   & \colhead{(Jy)} & \colhead{(Jy)} 
& \colhead{(Jy)} & \colhead{(Jy)}  & \colhead{L$_{\odot}$} & \colhead{}
& \colhead{Class}   \\
\colhead{(1)} & \colhead{(2)} & \colhead{(3)} & \colhead{(4)} & 
\colhead{(5)} & \colhead{(6)} & \colhead{(7)} & \colhead{(8)} & 
\colhead{(9)}
}
\startdata
IRAS 00188$-$0856 & 0.128 & $<$0.12 & 0.37 & 2.59 & 3.40 & 12.3 & 0.14 (C) & LINER \\  
IRAS 03250+1606 & 0.129 & $<$0.10 & $<$0.15 & 1.38 & 1.77 & 12.1 & $<$0.11 (C) & LINER \\  
IRAS 08572+3915 & 0.058 & 0.32 & 1.70 & 7.43 & 4.59 & 12.1 & 0.23 (W) & LINER \\  
IRAS 09039+0503 & 0.125 & 0.07 & 0.12 & 1.48 & 2.06 & 12.1 & 0.08 (C) &LINER \\  
IRAS 09116+0334 & 0.146 & $<$0.09 & $<$0.14 & 1.09 & 1.82 & 12.2 & $<$0.13 (C) &LINER \\  
IRAS 09539+0857 & 0.129 & $<$0.15 & $<$0.15 & 1.44 & 1.04 & 12.0 & $<$0.10 (C) &LINER \\  
IRAS 10378+1108 & 0.136 & $<$0.11 & 0.24 & 2.28 & 1.82 & 12.3 & 0.11 (C) &LINER \\  
IRAS 10485$-$1447 & 0.133 & $<$0.11 & 0.25 & 1.73 & 1.66 & 12.2 & 0.14 (C)& LINER \\  
IRAS 10494+4424 & 0.092 & $<$0.12 & 0.16 & 3.53 & 5.41 & 12.2 & 0.05 (C) &LINER \\  
IRAS 11095$-$0238 & 0.106 & 0.06 & 0.42 & 3.25 & 2.53 & 12.2 & 0.13 (C)& LINER \\  
IRAS 12112+0305 & 0.073 & 0.12 & 0.51 & 8.50 & 9.98 & 12.3 & 0.06 (C) &LINER \\  
IRAS 12127$-$1412 & 0.133 & $<$0.13 & 0.24 & 1.54 & 1.13 & 12.2 & 0.16 (C)& LINER \\  
IRAS 12359$-$0725 & 0.138 & 0.09 & 0.15 & 1.33 & 1.12 & 12.1 & 0.11 (C)& LINER \\  
IRAS 14252$-$1550 & 0.149 & $<$0.09 & $<$0.23 & 1.15 & 1.86 & 12.3 & $<$0.20 (C)& LINER \\  
IRAS 14348$-$1447 & 0.083 & 0.07 & 0.49 & 6.87 & 7.07 & 12.3 & 0.07 (C)& LINER \\  
IRAS 15327+2340 (Arp 220) & 0.018 & 0.48 & 7.92 & 103.33 & 112.40 & 12.1 & 0.08 (C) & LINER \\  
IRAS 16090$-$0139 & 0.134 & 0.09 & 0.26 & 3.61 & 4.87 & 12.5 & 0.07 (C)& LINER \\  
IRAS 16468+5200   & 0.150 & $<$0.06 & 0.10 & 1.01 & 1.04 & 12.1 & 0.10 (C)& LINER \\  
IRAS 16487+5447 & 0.104 & $<$0.07 & 0.20 & 2.88 & 3.07 & 12.2 & 0.07 (C) &LINER \\ 
IRAS 17028+5817   & 0.106 & $<$0.06 & 0.10 & 2.43 & 3.91 & 12.1 & 0.04 (C)& LINER \\  
IRAS 17044+6720   & 0.135 & $<$0.07 & 0.36 & 1.28 & 0.98 & 12.1 & 0.28 (W)& LINER \\  
IRAS 21329$-$2346 & 0.125 & 0.05 & 0.12 & 1.65 & 2.22 & 12.1 & 0.07 (C)& LINER \\  
IRAS 23234+0946 & 0.128 & $<$0.06 & 0.08 & 1.56 & 2.11 & 12.1 & 0.05 (C) &LINER \\  
IRAS 23327+2913 & 0.107 & $<$0.06 & 0.22 & 2.10 & 2.81 & 12.1 & 0.10 (C) &LINER \\ \hline  
IRAS 10190+1322 & 0.077 & $<$0.07 & 0.38 & 3.33 & 5.57 & 12.0 & 0.11 (C) &HII \\  
IRAS 11387+4116 & 0.149 & 0.12 & $<$0.14 & 1.02 & 1.51 & 12.2 & $<$0.14 (C) &HII \\  
IRAS 11506+1331 & 0.127 & $<$0.10 & 0.19 & 2.58 & 3.32 & 12.3 & 0.07 (C) &HII \\ 
IRAS 13443+0802  & 0.135 & $<$0.12 & $<$0.11 & 1.50 & 1.99 & 12.2 & $<$0.07 (C) &HII \tablenotemark{a} \\ 
IRAS 13509+0442 & 0.136 & 0.10 & $<$0.23 & 1.56 & 2.53 & 12.3 & $<$0.15 (C) &HII \\ 
IRAS 13539+2920 & 0.108 & $<$0.09 & 0.12 & 1.83 & 2.73 & 12.1 & 0.07 (C) &HII \\  
IRAS 14060+2919 & 0.117 & $<$0.10 & 0.14 & 1.61 & 2.42 & 12.1 & 0.09 (C) &HII \\ 
IRAS 15206+3342 & 0.125 & 0.08 & 0.35 & 1.77 & 1.89 & 12.2 & 0.19 (C) &HII \\  
IRAS 15225+2350 & 0.139 & $<$0.07 & 0.18 & 1.30 & 1.48 & 12.2 & 0.14 (C) &HII \\  
IRAS 16474+3430 & 0.111 & $<$0.13 & 0.20 & 2.27 & 2.88 & 12.2 & 0.09 (C) &HII \\ 
IRAS 20414$-$1651 & 0.086 & $<$0.65 & 0.35 & 4.36 & 5.25 & 12.3 & 0.08 (C) & HII \\  
IRAS 21208$-$0519 & 0.130 & $<$0.09 & $<$0.15 & 1.17 & 1.66 & 12.1 & $<$0.13 (C) & HII \\ \hline 
IRAS 14197+0813   & 0.131 & $<$0.17 & $<$0.19 & 1.10 & 1.66 & 12.2 & $<$0.17 (C) & unclassified \\ 
IRAS 14485$-$2434 & 0.148 & $<$0.11 & $<$0.15 & 1.02 & 1.05 & 12.2 & $<$0.15 (C) & unclassified \\  \hline 
IRAS 05189$-$2524 & 0.042 & 0.73 & 3.44 & 13.67 & 11.36 & 12.1 & 0.25 (W) & Sy2 \\ 
IRAS 08559+1053 & 0.148 & $<$0.10 & 0.19 & 1.12 & 1.95 & 12.2 & 0.17 (C) &Sy2 \\ 
IRAS 12072$-$0444 & 0.129 & $<$0.12 & 0.54 & 2.46 & 2.47 & 12.4 & 0.22 (W) & Sy2 \\ 
IRAS 13428+5608 (Mrk 273) & 0.037 & 0.24 & 2.28 & 21.74 & 21.38 & 12.1 & 0.10 (C) & Sy2 \\ 
IRAS 13451+1232 (PKS 1345+12) & 0.122 & 0.14 & 0.67 & 1.92 & 2.06 & 12.3 & 0.35 (W) & Sy2 \\ 
IRAS 15130$-$1958 & 0.109 & $<$0.14 & 0.39 & 1.92 & 2.30 & 12.1 & 0.20 (W) & Sy2 \\ 
IRAS 17179+5444  & 0.147 & $<$0.08 & 0.20 & 1.36 & 1.91 & 12.3 & 0.15 (C) & Sy2 \\ 
\hline 
IRAS 01572+0009 (Mrk 1014) & 0.163 & 0.12 & 0.54 & 2.22 & 2.16 & 12.6 &0.24 (W) & Sy1 \\ 
IRAS 07599+6508 & 0.149 & 0.26 & 0.53 & 1.69 & 1.73 & 12.5 & 0.31 (W) &Sy1 \\  
IRAS 11598$-$0112  & 0.151 & $<$0.14 & $<$0.38 & 2.39 & 2.63 & 12.5 & $<$0.16 (C) & Sy1 \\ 
IRAS 12265+0219 (3C 273) & 0.159 & 0.55 & 0.90 & 2.06 & 2.89 & 12.8 & 0.44 (W) & Sy1 \\ 
IRAS 12540+5708 (Mrk 231) & 0.042 & 1.87 & 8.66 & 31.99 & 30.29 & 12.5 & 0.27 (W) & Sy1 \\ 
IRAS 15462$-$0450 & 0.100 & 0.10 & 0.45 & 2.92 & 3.00 & 12.2 & 0.15 (C)& Sy1 \\ 
IRAS 21219$-$1757 & 0.112 & 0.21 & 0.45 & 1.07 & 1.18 & 12.1 & 0.42 (W) & Sy1 \\ 
\hline  
\enddata

\tablenotetext{a}{The north-eastern nucleus \citep{kim02} classified
optically as an HII-region \citep{vei99a} is observed.}

\tablecomments{
Col.(1): Object name.  
Col.(2): Redshift.
Col.(3)--(6): f$_{12}$, f$_{25}$, f$_{60}$, and f$_{100}$ are 
{\it IRAS} fluxes at 12 $\mu$m, 25 $\mu$m, 60 $\mu$m, and 100 $\mu$m,
respectively, taken from \citet{kim98}.
Col.(7): Decimal logarithm of infrared (8$-$1000 $\mu$m) luminosity
in units of solar luminosity (L$_{\odot}$), calculated with
$L_{\rm IR} = 2.1 \times 10^{39} \times$ D(Mpc)$^{2}$
$\times$ (13.48 $\times$ $f_{12}$ + 5.16 $\times$ $f_{25}$ +
$2.58 \times f_{60} + f_{100}$) ergs s$^{-1}$ \citep{sam96}.
Since the calculation is based on our adopted cosmology, the infrared
luminosities slightly ($<$10\%) differ from the values shown in Kim \&
Sanders (1998, their Table 1, column 15).  
For sources that have upper limits in some {\it IRAS} bands, we assume
that the actual flux is the upper limit. 
Even though zero values are adopted, the difference of the infrared
luminosities is very small, less than 0.2 dex, which will not affect our
main conclusions in this paper. 
Col.(8): ULIRGs with f$_{25}$/f$_{60}$ $<$ 0.2 and $>$ 0.2 are
classified as cool and warm (denoted as ``C'' and ``W''), respectively
\citep{san88}.
Col.(9): Optical spectral classification by \citet{vei99a}.
}

\end{deluxetable}

\clearpage

\begin{deluxetable}{llccclccc}
\rotate
\tabletypesize{\scriptsize}
\tablecaption{Observing log \label{tbl-2}}
\tablewidth{0pt}
\tablehead{
\colhead{Object} & 
\colhead{Date} & 
\colhead{Telescope} & 
\colhead{Integration} & 
\colhead{P.A. \tablenotemark{a}}  & 
\multicolumn{4}{c}{Standard Stars} \\
\colhead{} & 
\colhead{(UT)} & 
\colhead{Instrument} & 
\colhead{(Min)} &
\colhead{($^{\circ}$)} &
\colhead{Name} &  
\colhead{$L$-mag} &  
\colhead{Type} &  
\colhead{$T_{\rm eff}$ (K)}  \\
\colhead{(1)} & \colhead{(2)} & \colhead{(3)} & \colhead{(4)}
& \colhead{(5)} & \colhead{(6)}  & \colhead{(7)} & \colhead{(8)} &
\colhead{(9)}
}
\startdata
IRAS 00188$-$0856 & 2002 August 19 & Subaru IRCS & 25 & 90 & HR 72 & 5.0 & G0V & 5930 \\ 
IRAS 03250+1606   & 2002 August 19 &Subaru IRCS & 23 & 0 & HR 763 & 4.3 & F7V & 6240 \\
IRAS 08572+3915 NW \tablenotemark{b} & 2003 March 18 & IRTF SpeX & 30 & 0 &
HR 3451 & 5.0 & F7V & 6240 \\  
IRAS 09039+0503 & 2005 February 21 & Subaru IRCS & 48 & 0 & HR 3492 & 4.4 & A0V & 9480 \\ 
IRAS 09116+0334 W & 2005 February 20 & Subaru IRCS & 40 & 0 & HR 3651 & 6.1 & A0V & 9480 \\ 
IRAS 09539+0857 & 2005 February 22 & Subaru IRCS & 64 & 0 & HR 3651 & 6.1 & A0V & 9480 \\ 
IRAS 10378+1108 & 2005 February 21 & Subaru IRCS & 32 & 0 & HR 4079 & 5.3 & F6V & 6400 \\ 
IRAS 10485$-$1447 & 2005 February 21 & Subaru IRCS & 32 & 0 & HR 4158 & 4.3 & F7V & 6240 \\ 
IRAS 10494+4424 & 2005 February 20 & Subaru IRCS & 32 & 0 & HR 4098 & 5.0 & F9V & 6000 \\ 
IRAS 11095$-$0238 & 2005 February 21 & Subaru IRCS & 32 & 40
\tablenotemark{c} & HR 4356 & 5.4 & A0V & 9480 \\  
IRAS 12112+0305 NE, SW & 2004 February 8 & Subaru IRCS & 32 & 38 \tablenotemark{d} &
HR 4533 & 4.8 & F7V & 6240 \\  
IRAS 12127$-$1412 NE & 2005 February 21 & Subaru IRCS &
16 & 0 \tablenotemark{e} & HR 4529 & 4.9 & F7V & 6240 \\ 
IRAS 12359$-$0725 N & 2005 February 23 & Subaru IRCS &
40 & 0 \tablenotemark{f} & HR 4533 & 4.8 & F7V & 6240 \\ 
IRAS 14252$-$1550 E, W & 2005 February 22 & Subaru IRCS & 80 & 125 \tablenotemark{g} &
HR 5355 & 5.9 & A0 & 9480 \\  
IRAS 14348$-$1447 NE, SW & 2005 May 28 & Subaru IRCS & 32 & 30
\tablenotemark{h} & HR 5652 & 4.5 &A0V & 9480 \\  
IRAS 15327+2340 (Arp 220) \tablenotemark{i} & 2000 February 20 & UKIRT CGS4 & 64 & 90 & 
HR 5634 & 3.8 & F5V & 6500 \\  
IRAS 16090$-$0139 & 2003 May 18 & Subaru IRCS & 14 & 0 & HR 5859 & 5.6 & A0V & 9480 \\  
IRAS 16468+5200 E  & 2005 February 23 & Subaru IRCS & 72 & 85
\tablenotemark{j} & HR 5949 & 6.3 & A0V & 9480 \\ 
IRAS 16487+5447 W  & 2002 March 28 & Subaru IRCS & 36 & 70 \tablenotemark{k}& HR 5949 &
6.3 & A0V & 9480\\ 
IRAS 17028+5817 W & 2004 June 12 & Subaru IRCS & 20 & 175 \tablenotemark{l} 
& HR 5949 & 6.3 & A0V & 9480 \\  
IRAS 17044+6720   & 2005 June 13 & Subaru IRCS & 26 & 90 & HR 6025 & 5.4 & A0V & 9480 \\ 
IRAS 21329$-$2346 & 2002 October 24 & Subaru IRCS & 32 & 90 & HR 7898 & 4.5 & G8V & 5400\\ 
IRAS 23234+0946 W & 2002 October 24 & Subaru IRCS & 32 & 120 \tablenotemark{m}& HR 8653 &
4.6 & G8IV & 5400\\ 
IRAS 23327+2913 S & 2002 October 24 & Subaru IRCS & 24 & 90 \tablenotemark{n} & HR 8955 &
5.1 & F6V & 6400 \\ \hline 
IRAS 10190+1322 E,W & 2005 February 22 & Subaru IRCS & 32 & 65
\tablenotemark{o} & HR 3998 & 5.1 & F7V & 6240 \\  
IRAS 11387+4116 & 2005 February 23 & Subaru IRCS & 32 & 0 & HR 4345 & 5.0 & G0V & 5930 \\ 
IRAS 11506+1331 & 2005 February 22 & Subaru IRCS & 32 & 0 & HR 4437 & 4.7 & G0V & 5930 \\ 
IRAS 13443+0802 NE & 2005 May 28 & Subaru IRCS & 24 & 170
\tablenotemark{p} & HR 5243 & 4.9 & F6V & 6400 \\  
IRAS 13509+0442 & 2005 February 22 & Subaru IRCS & 32 & 0 & HR 5011 & 3.8 & G0V & 5930 \\ 
IRAS 13539+2920 NW & 2004 June 12 & Subaru IRCS & 20 & 0 \tablenotemark{q} 
& HR 5346 & 4.8 & F8V & 6000 \\ 
IRAS 14060+2919 & 2005 June 12 & Subaru IRCS & 18 & 0 & HR 5346 & 4.8 & F8V & 6000 \\ 
IRAS 15206+3342 & 2005 June 13 & Subaru IRCS & 18 & 0 & HR 5728 & 4.5 & G3V & 5800 \\ 
IRAS 15225+2350 & 2005 February 20 & Subaru IRCS & 32 & 0 & HR 5630 & 5.0 & F8V & 6000 \\ 
IRAS 16474+3430 S & 2005 February 21 & Subaru IRCS & 32
& 161 \tablenotemark{r} & HR 6064 & 5.2 & G1V & 5900 \\
IRAS 20414$-$1651 & 2005 May 28 & Subaru IRCS & 24 & 0 & HR 7855 & 4.9 & F6V& 6400\\
IRAS 21208$-$0519 N & 2005 May 28 & Subaru IRCS & 20 & 17
\tablenotemark{s} & HR 8041 & 4.7 & G1V & 5900\\ \hline 
IRAS 14197+0813 & 2005 May 28 & Subaru IRCS & 32 & 0 & HR 5307 & 5.1& F7V&6240 \\
IRAS 14485$-$2434 & 2005 May 28 & Subaru IRCS & 24 & 0 & HR 5504 & 5.0& F7V& 6240\\
\hline
IRAS 05189$-$2524 \tablenotemark{i} & 1999 September 9 & UKIRT CGS4 & 53
& 19 & HR 1762 & 4.7 & A0V & 9480 \\ 
IRAS 08559+1053 & 2005 February 23 & Subaru IRCS & 24 & 0 & HR 3510 &
4.9 & G1V & 5900 \\ 
IRAS 12072$-$0444 N, S & 2005 May 29 & Subaru IRCS & 16 & 0
\tablenotemark{t} & HR 4533 & 4.8 & F7V & 6240\\  
IRAS 13428+5608 (Mrk 273) \tablenotemark{i} & 2000 February 20 & UKIRT CGS4 & 43 & 35 &
HR 4761 & 4.8 & F6-8V & 6200 \\ 
IRAS 13451+1232 W (PKS 1345+12) & 2005 February 23 & Subaru IRCS & 32 &
105 \tablenotemark{u} & HR 5011 & 3.8 &G0V & 5930 \\ 
IRAS 15130$-$1958 & 2005 May 28 & Subaru IRCS & 12 & 0 & HR 5504 & 5.0 & F7V & 6240\\ 
IRAS 17179+5444   & 2005 May 28 & Subaru IRCS & 16 & 0 & HR 6360 & 4.5 & G5V & 5700\\ \hline
IRAS 01572+0009 (Mrk 1014) & 2003 September 8 & IRTF SpeX & 160 & 0 &
HR 650 & 4.1 & F8V & 6000  \\
IRAS 07599+6508 & 2004 April 4 & IRTF NSFCAM & 40 & 0 & HR 3028 & 4.8 & F6V & 6400 \\
IRAS 11598$-$0112 & 2005 May 29 & Subaru IRCS & 8 & 0 & HR 4533 & 4.8 & F7V & 6240 \\ 
IRAS 12265+0219 (3C 273) & 2004 April 5 & IRTF NSFCAM & 28 & 0 & HR 4708 & 5.0 & F8V & 6000 \\
IRAS 12540+5708 (Mrk 231) \tablenotemark{v} & 2000 February 20 & UKIRT CGS4 & 21 & 0 &
HR 4761 & 4.8 & F6-8V & 6200 \\
IRAS 15462$-$0450 & 2004 April 5 & IRTF NSFCAM & 120 & 0 & HR 5779 & 5.2 & F7V & 6240  \\ 
IRAS 21219$-$1757 & 2005 May 29 & Subaru IRCS & 8 & 0 & HR 7994 & 4.9 & G1V & 5900\\ 
\hline  
\enddata

\tablecomments{Column (1): Object name.
Col. (2): Observing date in UT.
Col. (3): Telescope and instrument. 
Col. (4): Net on-source integration time in min.
Col. (5): Position angle of the slit.
Col. (6): Standard star name.
Col. (7): Adopted $L$-band magnitude.
Col. (8): Stellar spectral type.
Col. (9): Effective temperature.}

\tablenotetext{a}{
0$^{\circ}$ corresponds to the north-south direction.
Position angle increases with counter-clockwise on the sky plane.}
\tablenotetext{b}{A spectrum at $\lambda$ = 3.2--3.8 $\mu$m was
shown by \citet{imd00}. 
We present a new spectrum with a wider wavelength coverage of $\lambda$
= 2.8--4.1 $\mu$m, to investigate a broad 3.1 $\mu$m H$_{2}$O ice
absorption feature.
We observed the north-western nucleus which is much brighter in the infrared
$L$-band than the south-eastern nucleus \citep{zho93}.
}
\tablenotetext{c}{To simultaneously observe double nuclei with a
separation of $\sim$0$\farcs$5 \citep{bus02}. 
In our spectrum, a double peak along the slit direction is
recognizable, but signals from the double nuclei are not 
clearly resolvable. 
A spectrum combining both nuclei is presented in this paper.}
\tablenotetext{d}{To simultaneously observe double nuclei with a
separation of $\sim$3$''$ \citep{sco00,kim02}.}
\tablenotetext{e}{Only the brighter north-eastern nucleus \citep{kim02}
was observed.} 
\tablenotetext{f}{Only the northern primary nucleus \citep{tru01} was
observed.} 
\tablenotetext{g}{The south-western nucleus \citep{kim02} was resolved
into double nuclei (eastern and western) with a separation of
$\sim$1$\farcs$1 in our $K$-band image. 
Both the eastern and western nuclei were observed.} 
\tablenotetext{h}{To simultaneously observe double nuclei with a
separation of $\sim$3$''$ \citep{sco00,kim02,cha02}.} 
\tablenotetext{i}{The same spectrum as previously shown by \citet{imd00}
is presented again.
The wavelength coverage is narrower than the other ULIRGs.} 
\tablenotetext{j}{To simultaneously observe double nuclei with a
separation of $\sim$3$\farcs$5 \citep{kim02}. 
Only the eastern nucleus was bright enough to extract a meaningful
$L$-band spectrum.} 
\tablenotetext{k}{To simultaneously observe double nuclei with a
separation of $\sim$3$''$ \citep{mur96,kim02}.   
Only the western primary nucleus \citep{mur01} was 
bright enough to extract a meaningful $L$-band spectrum.}  
\tablenotetext{l}{Only the western primary nucleus \citep{tru01} was
observed.}
\tablenotetext{m}{To simultaneously observe double nuclei with a
separation of $\sim$4$''$ in the $K$-band image \citep{kim02}.   
Only the western nucleus was bright enough to extract a meaningful
$L$-band spectrum.}
\tablenotetext{n}{Only the southern primary nucleus \citep{tru01} was
observed.}
\tablenotetext{o}{To simultaneously observe double nuclei with a
separation of $\sim$4$''$ \citep{kim02}.}
\tablenotetext{p}{To simultaneously observe north-eastern and eastern
nuclei with a separation of $\sim$5$''$ \citep{kim02}.
Both nuclei were detected, but only the north-eastern nucleus was bright
enough to extract a meaningful $L$-band spectrum.} 
\tablenotetext{q}{Only the north-western primary nucleus \citep{kim02} was
observed.}
\tablenotetext{r}{To simultaneously observe double nuclei with a
separation of $\sim$3$\farcs$5 \citep{kim02}.
Only the southern nucleus was bright enough to extract a meaningful
$L$-band spectrum.}
\tablenotetext{s}{To simultaneously observe double nuclei with a
separation of $\sim$7$''$ \citep{kim02}.
Only the northern nucleus was bright enough to extract a meaningful
$L$-band spectrum.}
\tablenotetext{t}{The nucleus \citep{kim02} was resolved
into double nuclei (northern and southern) with a separation of 
$\sim$0$\farcs$9 in our $K$-band image. Both nuclei were observed.} 
\tablenotetext{u}{To simultaneously observe double nuclei with a
separation of $\sim$2$''$ \citep{sco00}.
Only the western AGN-harboring main nucleus \citep{imt04} was bright
enough to extract a meaningful $L$-band spectrum.} 
\tablenotetext{v}{$L$-band spectra have previously been presented by
\citet{ima98} and \citet{imd00}.
The spectrum by \citet{imd00} is shown again. 
The wavelength coverage is narrower than the other ULIRGs.} 

\end{deluxetable}

\begin{deluxetable}{lcccccc}
\tabletypesize{\scriptsize}
\rotate
\tablecaption{Properties of the nuclear 3.3 $\mu$m PAH emission and 
buried AGN signatures. 
\label{tbl-3}}
\tablewidth{0pt}
\tablehead{
\colhead{Object} & 
\colhead{f$_{3.3 \rm PAH}$} & 
\colhead{L$_{3.3 \rm PAH}$}  & 
\colhead{L$_{3.3 \rm PAH}$/L$_{\rm IR}$} & 
\colhead{rest EW$_{3.3 \rm PAH}$} & \multicolumn{2}{c}{AGN evidence} \\
\colhead{} & \colhead{($\times$10$^{-14}$ ergs s$^{-1}$ cm$^{-2}$)} & 
\colhead{($\times$10$^{41}$ergs s$^{-1}$)}  & 
\colhead{($\times$10$^{-3}$)} & 
\colhead{(nm)} & \colhead{L-band} & \colhead{Radio} \\  
\colhead{(1)} & \colhead{(2)} & \colhead{(3)} & \colhead{(4)} & 
\colhead{(5)} & \colhead{(6)} & \colhead{(7)}  
}
\startdata
IRAS 00188$-$0856   & 3.5 & 13  & 0.15 & 50 & $\bigcirc$ & $\bigcirc$ (0.6) \\
IRAS 03250+1606     & 3.0 & 12  & 0.25 & 80 & $\bigcirc$ & X \\
IRAS 08572+3915 NW  & $<$3.5\tablenotemark{a} & $<$2.5\tablenotemark{a} &
$<$0.05\tablenotemark{a} & $<$5\tablenotemark{a} & 
{\large $\circledcirc$} & $\bigcirc$ (0.8) \\  
IRAS 09039+0503     & 1.5 & 5.5 & 0.10 & 95 & $\bigcirc$ & X \\
IRAS 09116+0334 W   & 3.0 & 15  & 0.25 & 75 & $\bigcirc$ & X \\
IRAS 09539+0857     & 0.9 & 3.5 & 0.07 & 65 & X & $\bigcirc$ (1.0) \\
IRAS 10378+1108     & 1.0 & 4.5 & 0.06 & 40 & $\bigcirc$ & $\bigcirc$ (0.8) \\
IRAS 10485$-$1447   & 2.0 & 7.0 & 0.10 & 80 & $\bigcirc$ & --- \\
IRAS 10494+4424     & 3.5 & 6.5 & 0.10 & 110 & $\bigcirc$ & $\bigcirc$ (0.4) \\ 
IRAS 11095$-$0238   & 2.5 & 6.0 & 0.10 & 150 & X & --- \\
IRAS 12112+0305 NE  & 3.5 & 4.0 & 0.05 & 100 & X & $\bigcirc$ (0.8) \\
IRAS 12112+0305 SW  & 3.5 & 4.0 & 0.05 & 160 & X & X \\
IRAS 12127$-$1412 NE & 0\tablenotemark{b} & 0\tablenotemark{b} &
0\tablenotemark{b} & 0\tablenotemark{b} & {\large $\circledcirc$} & --- \\ 
IRAS 12359$-$0725 N & 2.0 & 9.5 & 0.20 & 75 & $\bigcirc$ & --- \\
IRAS 14252$-$1550 E & 0.7 & 3.5 & 0.06 & 60 & $\bigcirc$ & --- \\
IRAS 14252$-$1550 W & 0.7 & 3.5 & 0.06 & 60 & X &--- \\
IRAS 14348$-$1447 NE & 1.5 & 2.0 & 0.03 & 70 & $\bigcirc$ & --- \\
IRAS 14348$-$1447 SW & 3.5 & 5.0 & 0.07 & 110 & X & --- \\
IRAS 15327+2340 (Arp 220) & 25 & 1.5 & 0.03 & 80 & X & $\bigcirc$ (0.2) \\  
IRAS 16090$-$0139   & 3.0 & 13  & 0.10 & 75 & $\bigcirc$ & --- \\
IRAS 16468+5200 E   & 1.0 & 5.5 & 0.10 & 120 & X & X \\
IRAS 16487+5447 W   & 2.5 & 6.0 & 0.10 & 70 & $\bigcirc$ & $\bigcirc$ (0.7) \\
IRAS 17028+5817 W   & 4.0 & 10  & 0.20 & 120 & $\bigcirc$ & $\bigcirc$ (0.9) \\
IRAS 17044+6720     & 2.0 & 8.5 & 0.15 & 10 & {\large $\circledcirc$} & $\bigcirc$ (1.3)\\
IRAS 21329$-$2346   & 1.0 & 4.0 & 0.08 & 50 & $\bigcirc$ & --- \\
IRAS 23234+0946 W   & 2.5 & 10 & 0.20 & 75 & X & $\bigcirc$ (0.4) \\
IRAS 23327+2913 S   & 2.0 & 5.0 & 0.10 & 45 & X & X \\ \hline
IRAS 10190+1322 E   & 5.5 & 7.0 & 0.20 & 95 & X & $\bigcirc$ (1.0) \\
IRAS 10190+1322 W   & 1.0 & 1.5 & 0.04 & 50 & X & X \\
IRAS 11387+4116     & 1.5 & 8.0 & 0.15 & 75 & X & X \\
IRAS 11506+1331     & 5.5 & 21  & 0.25 & 95 & $\bigcirc$ & $\bigcirc$ (0.3) \\
IRAS 13443+0802 NE  & 1.0 & 5.0 & 0.08 & 75 & X & $\bigcirc$ (0.7) \\ 
IRAS 13509+0442     & 4.0 & 18  & 0.25 & 135 & X & X \\
IRAS 13539+2920 NW  & 4.5 & 11  & 0.25 & 85 & X & $\bigcirc$ (0.2) \\
IRAS 14060+2919     & 7.5 & 23  & 0.45 & 150 & X & X \\
IRAS 15206+3342     & 4.0 & 14  & 0.25 & 55 & X & X \\
IRAS 15225+2350     & 1.5 & 7.0 & 0.15 & 40 & $\bigcirc$ & $\bigcirc$ (1.0) \\
IRAS 16474+3430 S   & 5.0 & 14  & 0.25 & 105 & $\bigcirc$ & X \\ 
IRAS 20414$-$1651   & 2.5 & 4.0 & 0.05 & 75 & X & $\bigcirc$ (0.7) \\ 
IRAS 21208$-$0519 N & 1.5 & 6.0 & 0.10 & 100 & X & X \\ \hline
IRAS 14197+0813     & 2.0 & 7.5 & 0.15 & 90 & $\bigcirc$ & $\bigcirc$\tablenotemark{c} (?) \\ 
IRAS 14485$-$2434   & 2.5 & 13.5 & 0.25 & 70 & $\bigcirc$ & --- \\ \hline 
IRAS 05189$-$2524   & 11  & 4.0 & 0.09 &  4 & {\large $\circledcirc$} & --- \\
IRAS 08559+1053     & 3.5 & 17  & 0.25 & 10 & {\large $\circledcirc$} & X \\
IRAS 12072$-$0444 S & 1.0 & 3.5 & 0.04 & 45 & {\large $\circledcirc$} & --- \\ 
IRAS 12072$-$0444 N & 2.0 & 8.0 & 0.09 & 20 & {\large $\circledcirc$} & --- \\ 
IRAS 13428+5608 (Mrk 273) & 14  & 4.0 & 0.09 & 35 & {\large
$\circledcirc$} & $\bigcirc$ (0.4) \\
IRAS 13451+1232 W (PKS 1345+12) & $<$2.5 & $<$7.0 & $<$0.10 & $<$5 & 
{\large $\circledcirc$} & $\bigcirc$ (0.9) \\ 
IRAS 15130$-$1958   & 3.5 & 9.0 & 0.15 & 10 & {\large $\circledcirc$} & --- \\ 
IRAS 17179+5444     & 1.0 & 6.0 & 0.09 & 20 & {\large $\circledcirc$} &
$\bigcirc$ (0.9)\\ \hline 
IRAS 01572+0009 (Mrk 1014) & 3.0 & 21 & 0.15 & 7 & {\large
$\circledcirc$} & $\bigcirc$ (1.0) \\ 
IRAS 07599+6508     & $<$7.5 & $<$40 & $<$0.35 & $<$4 & {\large $\circledcirc$} &
$\bigcirc$ (0.8) \\ 
IRAS 11598$-$0112   & 1.5 & 7.0 & 0.06 &  4 & {\large $\circledcirc$} & --- \\ 
IRAS 12265+0219 (3C 273)  & $<$7.0 & $<$40 & $<$0.20 & $<$2 & {\large $\circledcirc$}
& $\bigcirc$ (1.0) \\
IRAS 12540+5708 (Mrk 231) & 18  & 6.0  & 0.05 & 2 & {\large
$\circledcirc$} & $\bigcirc$ (0.9) \\
IRAS 15462$-$0450   & 3.5 & 7.0 & 0.15 & 10 & {\large $\circledcirc$} & --- \\ 
IRAS 21219$-$1757   & 1.0 & 2.5 & 0.05 & 1 & {\large $\circledcirc$} &
$\bigcirc$ (1.0) \\ \hline
\enddata

\tablenotetext{a}{
A conservative upper limit of the 3.3 $\mu$m PAH emission was 
estimated based on our new spectrum, and is larger than our original
estimate \citep{imd00}.}

\tablenotetext{b}{
The upper limit of the 3.3 $\mu$m PAH emission is difficult to estimate,
because of the winding spectral shape.
The observed profiles are explained by absorption features, and so 
we adopt the PAH flux as zero.} 

\tablenotetext{c}{We classify IRAS 14197+0813 as a ULIRG with detectable
core emission, based on column 9 of Table 1 in Nagar et al. (2003).
}

\tablecomments{
Col. (1): Object name. 
Col. (2): Observed flux of the 3.3 $\mu$m PAH emission. 
The second effective digit, if smaller than unity, is shown in units of 0.5. 
Col. (3): Observed luminosity of the 3.3 $\mu$m PAH emission.  
Col. (4): Observed 3.3 $\mu$m PAH-to-infrared luminosity ratio in units
          of 10$^{-3}$, a typical value for less infrared luminous
          starbursts \citep{mou90,ima02}.   
Col. (5): Rest-frame equivalent width of the 3.3 $\mu$m PAH emission.  
          Those for starbursts are $\sim$100 nm \citep{moo86,imd00}.  
Col. (6): Evidence of an AGN from our $L$-band spectroscopic method.
          $\circledcirc$: strong evidence. $\bigcirc$:
          tentative evidence. X: no evidence.
Col. (7): VLA radio data by \citet{nag03}.
          $\bigcirc$: core detected. X: undetected. 
          ---: no data.
          The values in parentheses are the core-to-total radio flux
          ratio in mJy, taken from \citet{nag03}.
}

\end{deluxetable}

\begin{deluxetable}{lccc}
\tablecaption{Absorption features in the $L$-band 
\label{tbl-4}}
\tablewidth{0pt}
\tablehead{
\colhead{Object} & 
\colhead{observed $\tau_{3.1}$} & 
\colhead{observed $\tau_{3.4}$} &
\colhead{A$_{\rm V}$ [mag]} \\
\colhead{(1)} & \colhead{(2)} & \colhead{(3)} & \colhead{(4)} 
}
\startdata
IRAS 00188$-$0856   & 1.8     & \nodata & $>$30 \\
IRAS 03250+1606     & 0.8     & \nodata & $>$13 \\
IRAS 08572+3915 NW  & \nodata & 0.8     & 110--200\\  
IRAS 09039+0503     & 0.8     & \nodata & $>$13 \\
IRAS 09116+0334 W   & 0.4     & \nodata & $>$6 \\
IRAS 10378+1108     & 0.6     & \nodata & $>$10 \\
IRAS 10485+1108     & 0.8     & \nodata & $>$13 \\
IRAS 10494+4424     & 1.0     & \nodata & $>$16 \\
IRAS 12127$-$1412 NE & 0.4     & 0.35    & 56--94 \\ 
IRAS 12359$-$0725 N  & 0.5     & \nodata & $>$8 \\
IRAS 14252$-$1550 E  & 0.7     & \nodata & $>$11 \\
IRAS 14348$-$1447 NE & 0.5     & \nodata & $>$8 \\
IRAS 16090$-$0139   & 0.8     & \nodata & $>$13 \\
IRAS 16487+5447 W   & 1.2     & \nodata & $>$20 \\
IRAS 17028+5817 W   & 0.6     & \nodata & $>$10 \\
IRAS 17044+6720     & \nodata & 0.15   & 21--38 \\ 
IRAS 21329$-$2346   & 1.0     & \nodata & $>$16 \\  \hline
IRAS 11506+1331     & 1.0     & \nodata & $>$16 \\
IRAS 15225+2350     & 0.4     & \nodata & $>$6 \\
IRAS 16474+3430 S   & 0.8     & \nodata & $>$13 \\ \hline
IRAS 14197+0813     & 0.4     & \nodata & $>$6 \\
IRAS 14485$-$2434   & 0.4     & \nodata & $>$6 \\ \hline
IRAS 12072$-$0444 N & \nodata & 0.4 & 57--100 \\
\enddata

\tablecomments{
Col. (1): Object name. Only ULIRGs which show clear absorption features
are listed.
LINER, HII, optically-unclassified, and Seyfert-2 ULIRGs
(from the top to the bottom) are separated by the horizontal solid lines. 
Col. (2): Observed optical depth of the 3.1 $\mu$m H$_{2}$O ice
absorption feature. 
Since relatively low plausible continuum levels are adopted,
these values should be taken as conservative lower limits. 
The $\tau_{3.1}$ values are derived by combining a few data
points around the ice absorption peak, and the statistical
errors are at a level of $<$0.1. 
Col. (3): Observed optical depth of the 3.4 $\mu$m bare carbonaceous dust
absorption feature. 
Col. (4): Dust extinction in A$_{\rm V}$ toward the 3--4 $\mu$m
continuum emission regions, derived from the absorption optical depths,
assuming the Galactic dust extinction curve. 
For those with detected 3.1 $\mu$m absorption, but with no measurable
$\tau_{3.4}$ values, the derived A$_{\rm V}$ values should be taken as
lower limits (see text in $\S$5.4).  
}

\end{deluxetable}

\clearpage

\begin{figure}
\begin{center} {\bf \Large LINER (1)} \end{center}
\includegraphics[angle=-90,scale=.35]{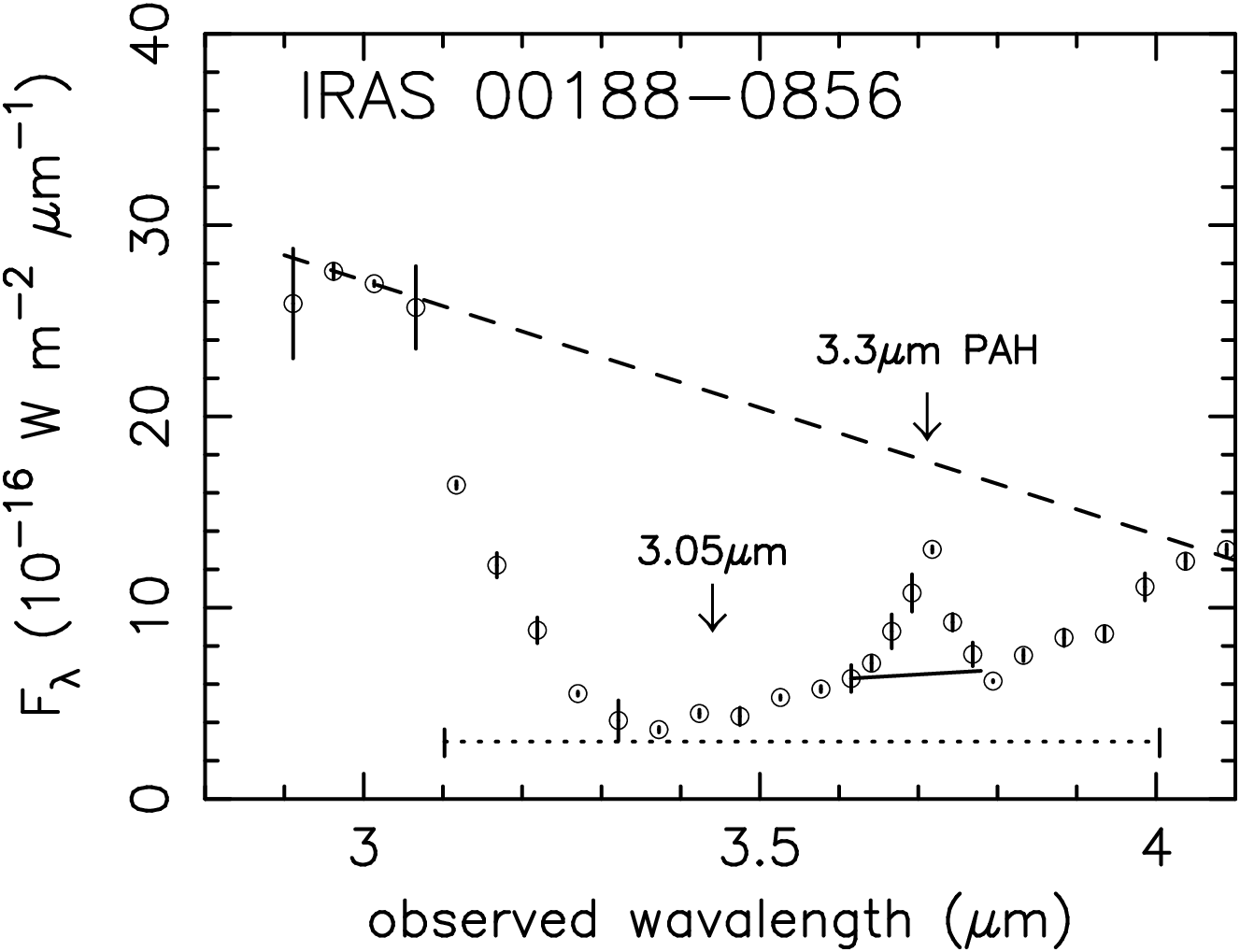} \hspace{0.3cm}
\includegraphics[angle=-90,scale=.35]{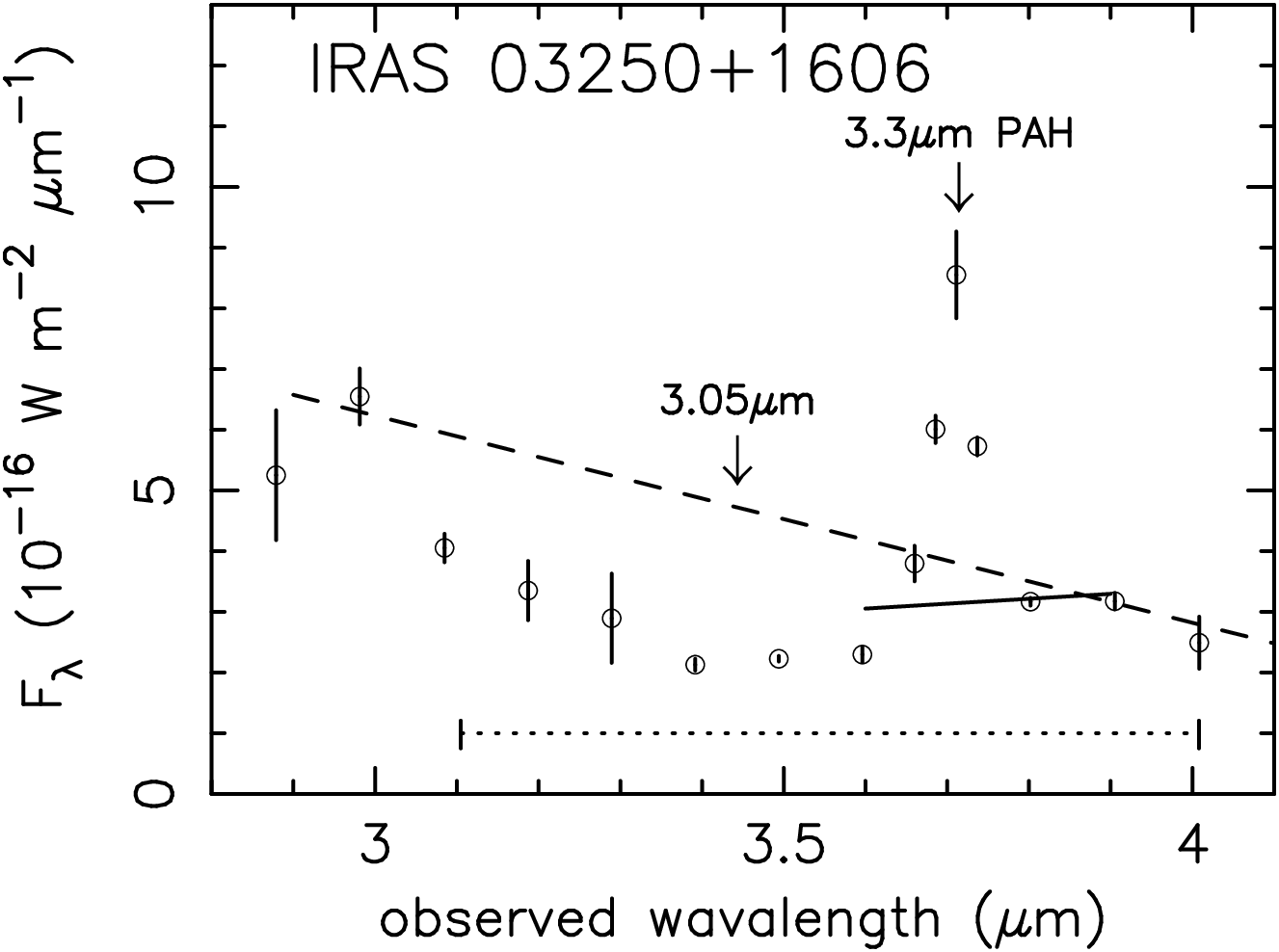} \\
\includegraphics[angle=-90,scale=.35]{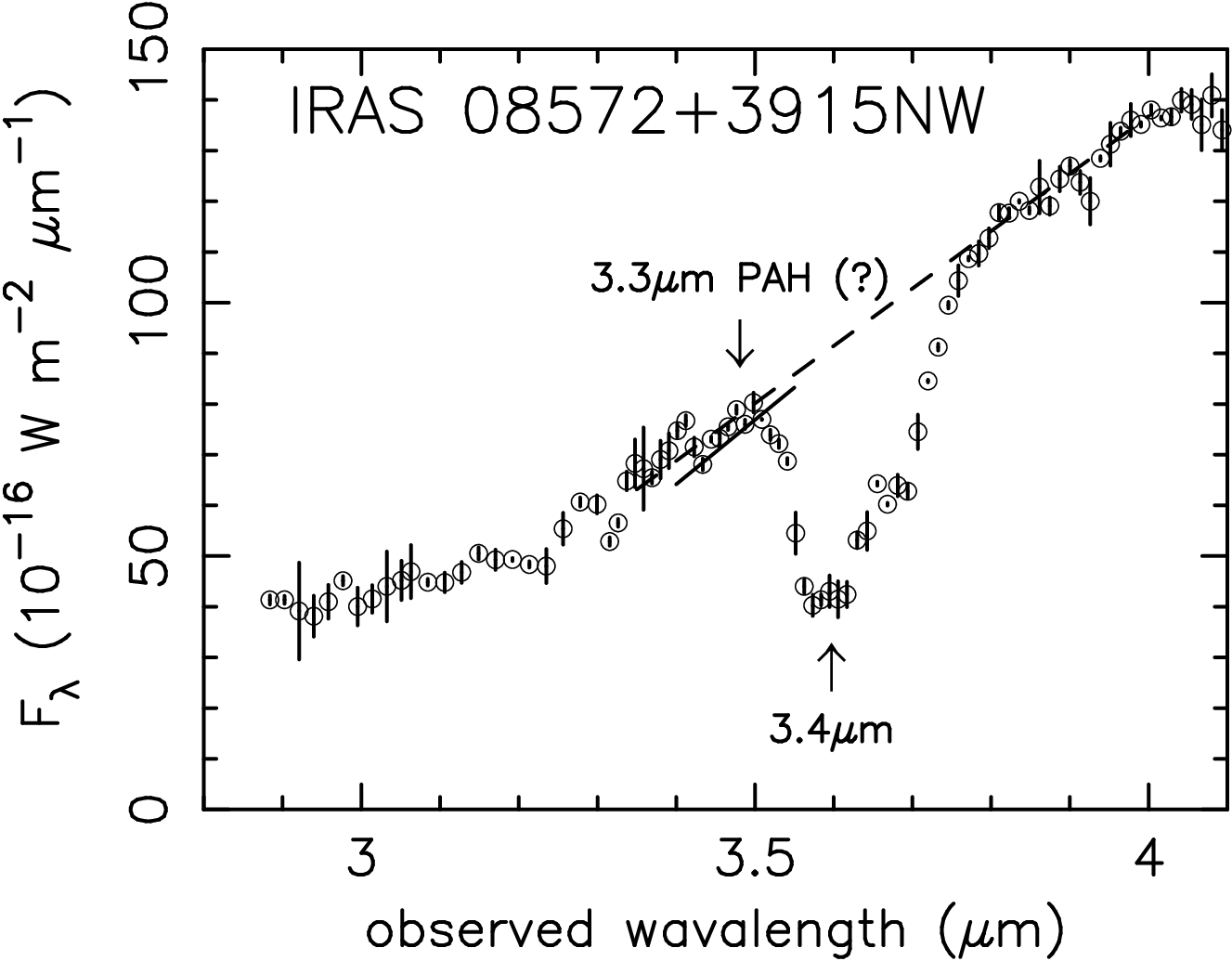} \hspace{0.3cm}
\includegraphics[angle=-90,scale=.35]{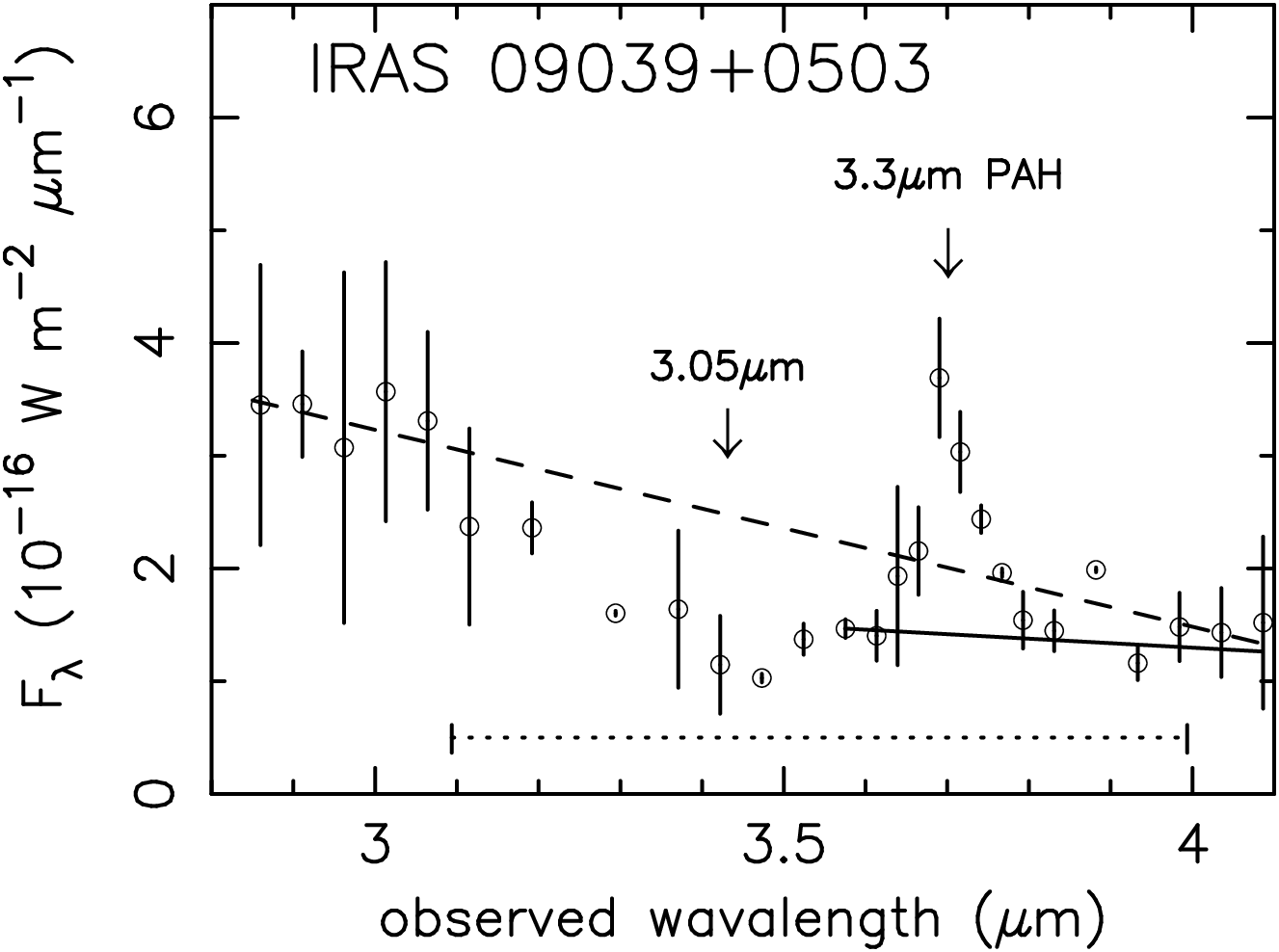} \\
\includegraphics[angle=-90,scale=.35]{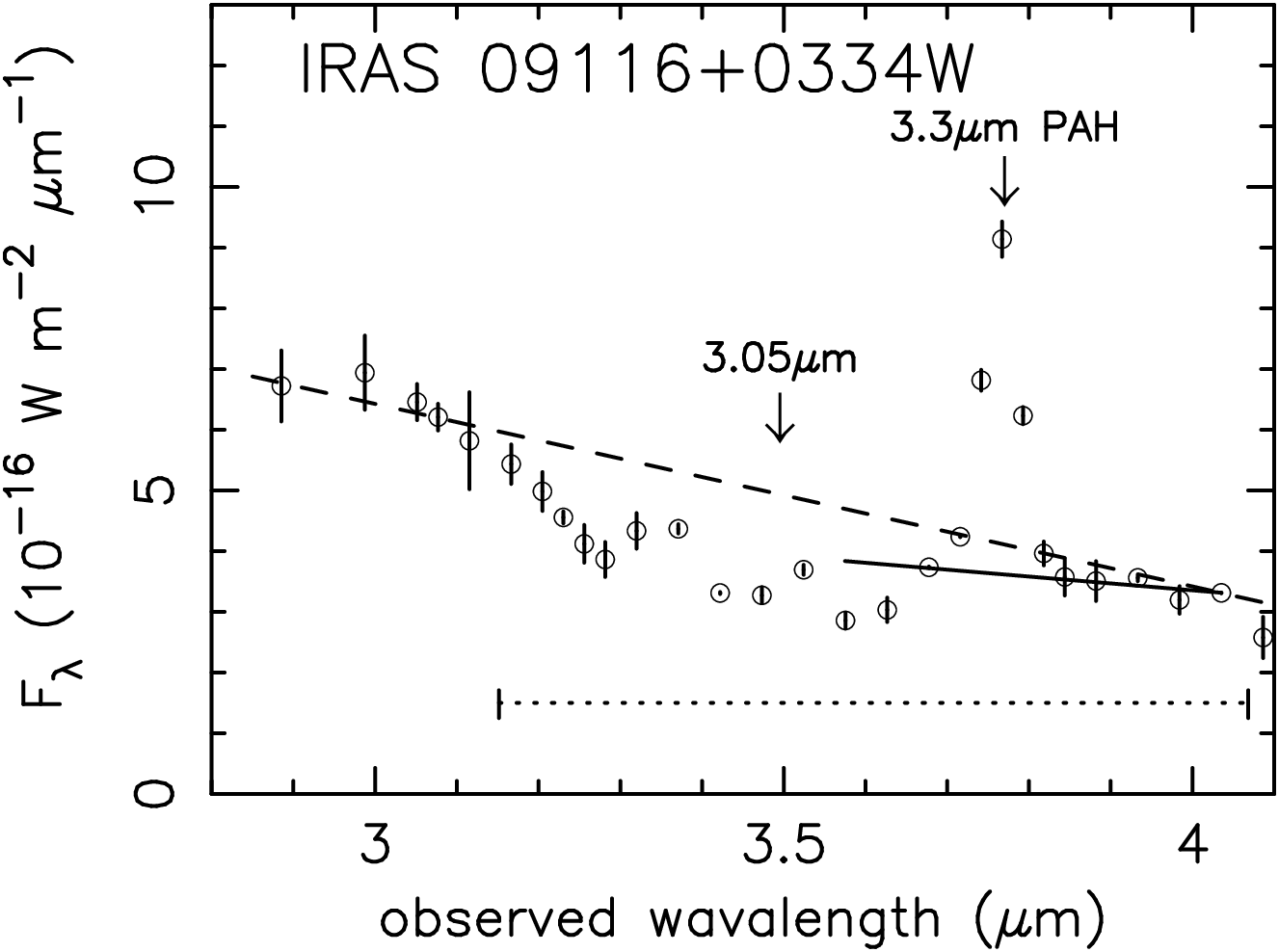} \hspace{0.3cm} 
\includegraphics[angle=-90,scale=.35]{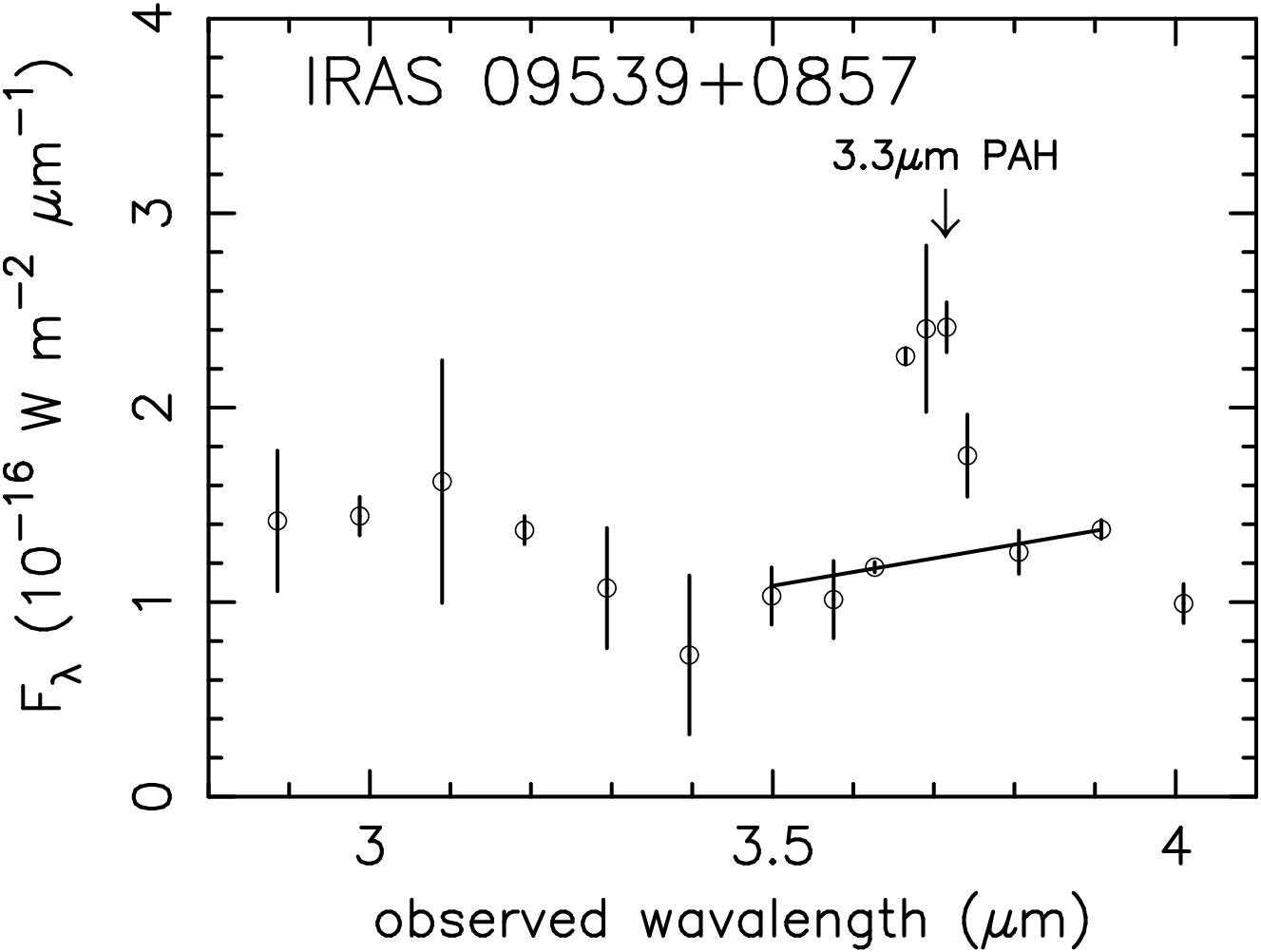} \\
\end{figure}

\clearpage

\begin{figure}
\begin{center} {\bf \Large LINER (2)} \end{center}
\includegraphics[angle=-90,scale=.35]{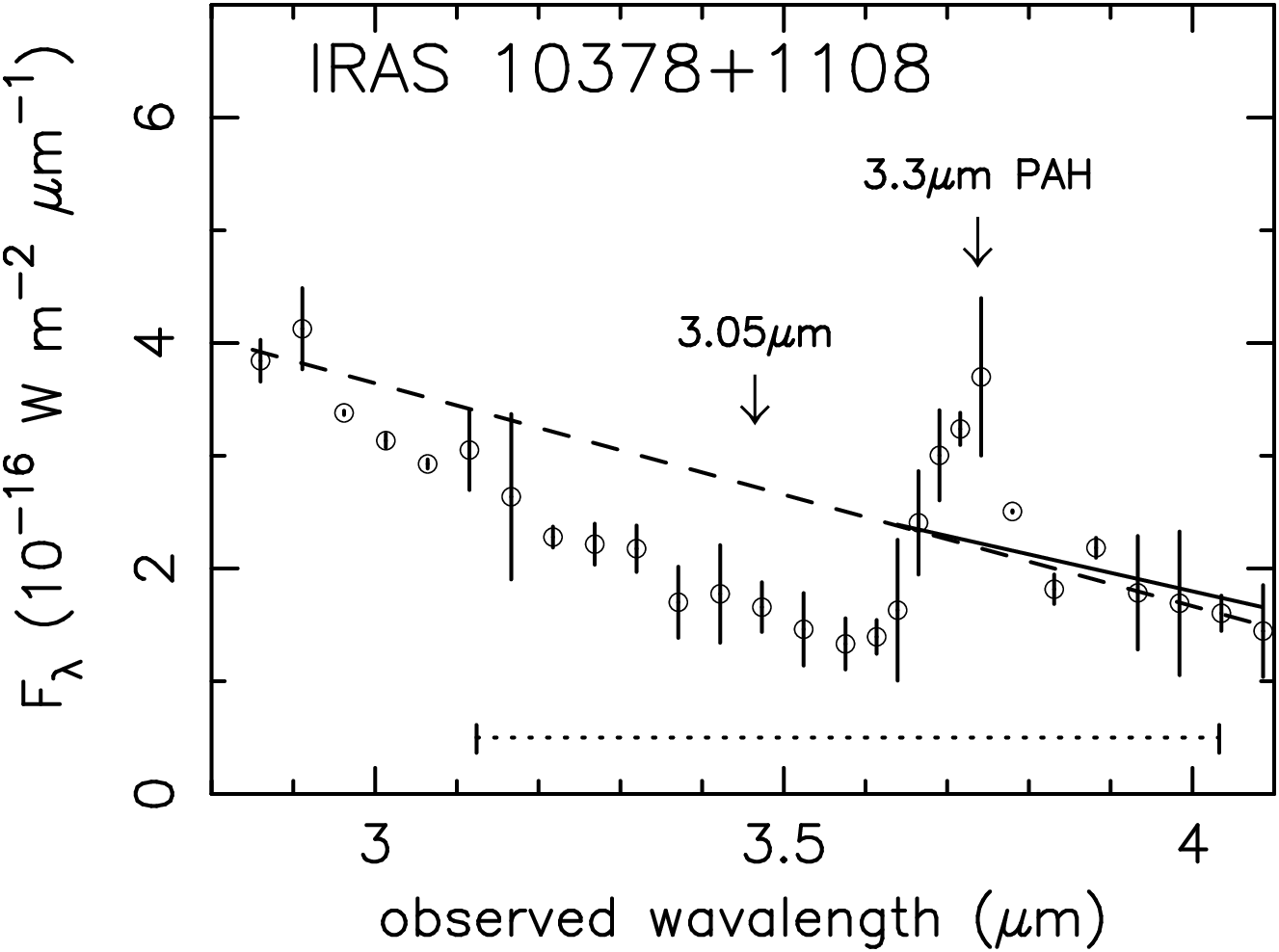} \hspace{0.3cm} 
\includegraphics[angle=-90,scale=.35]{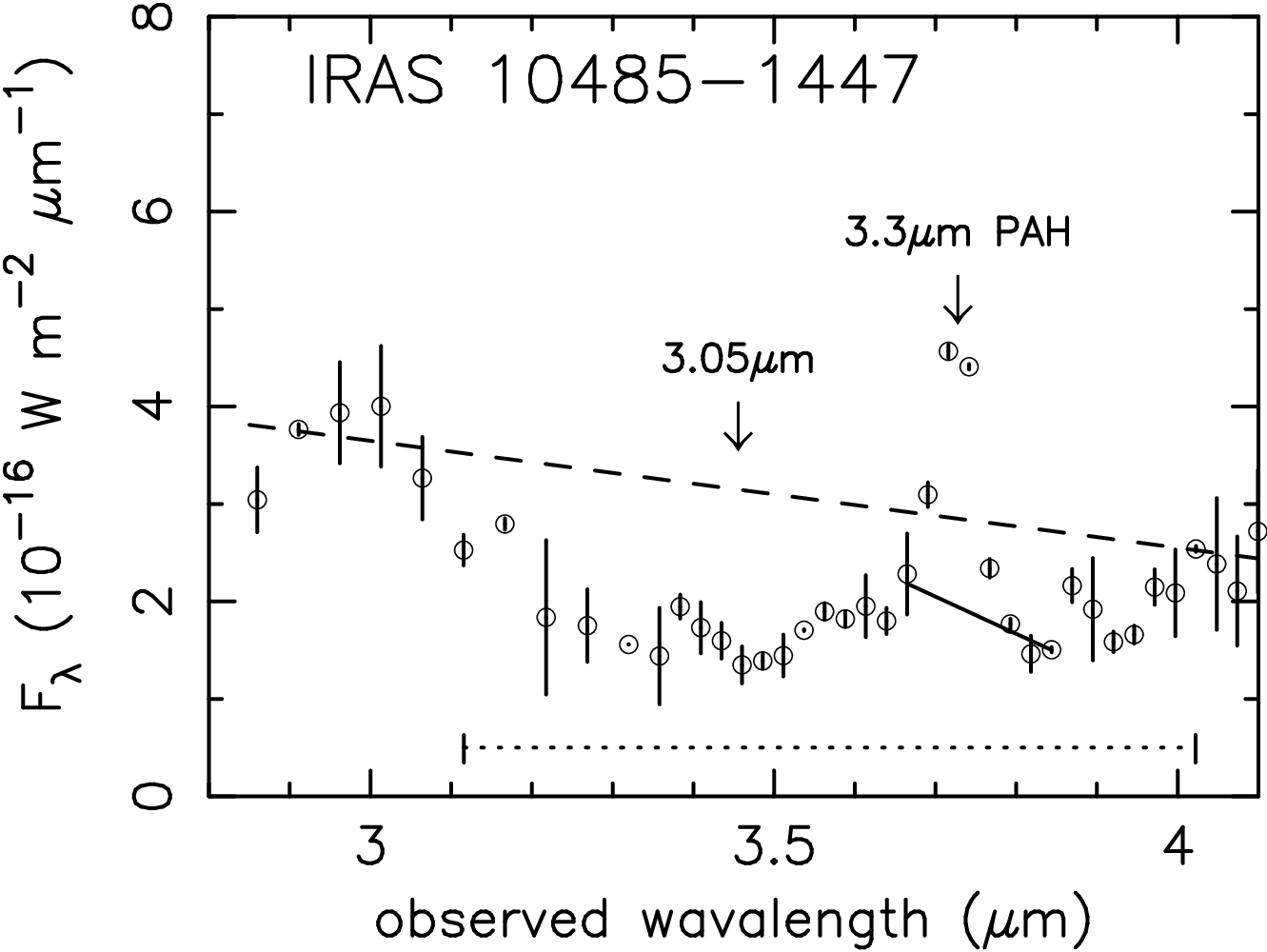} \\
\includegraphics[angle=-90,scale=.35]{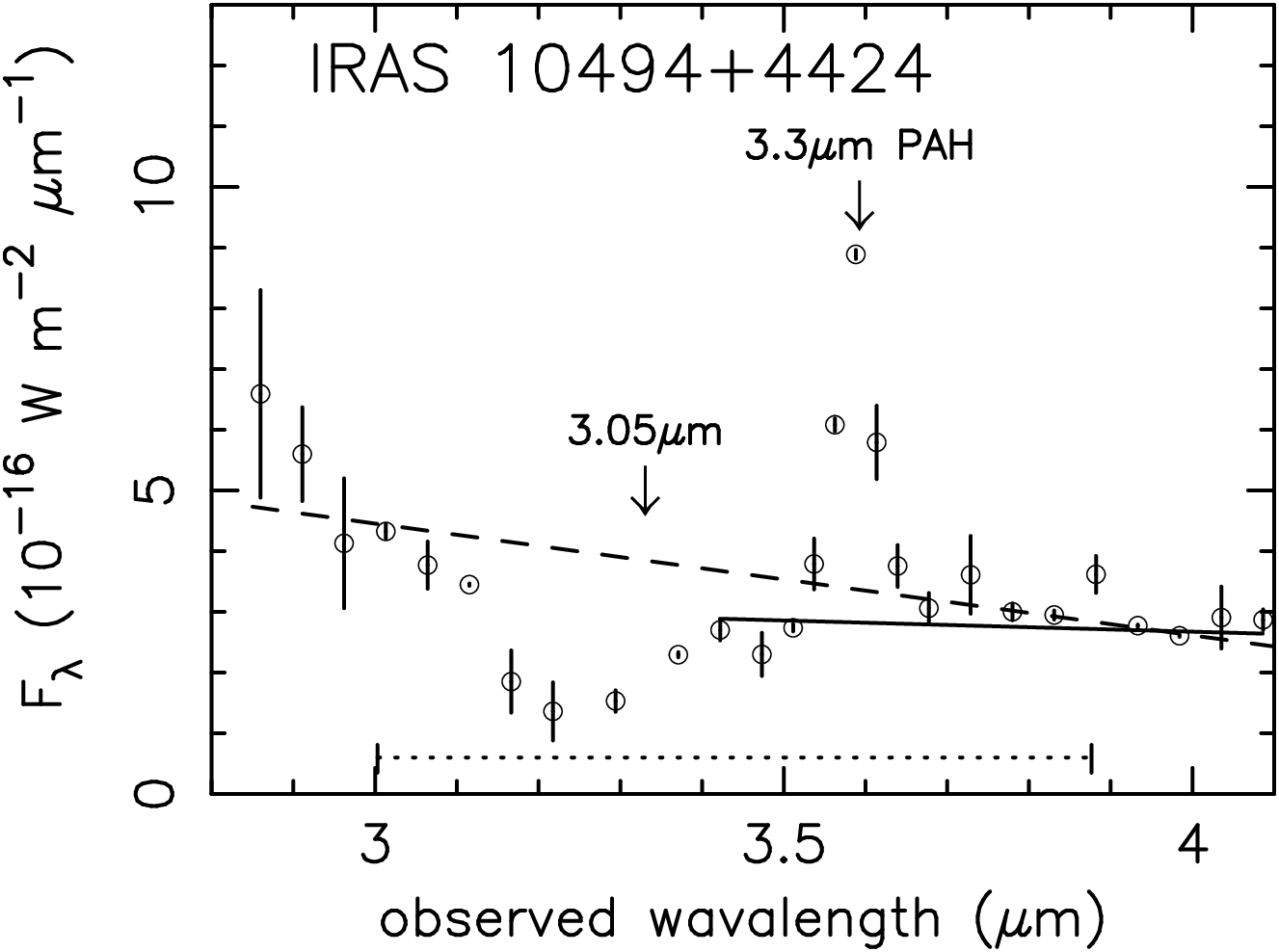} \hspace{0.3cm} 
\includegraphics[angle=-90,scale=.35]{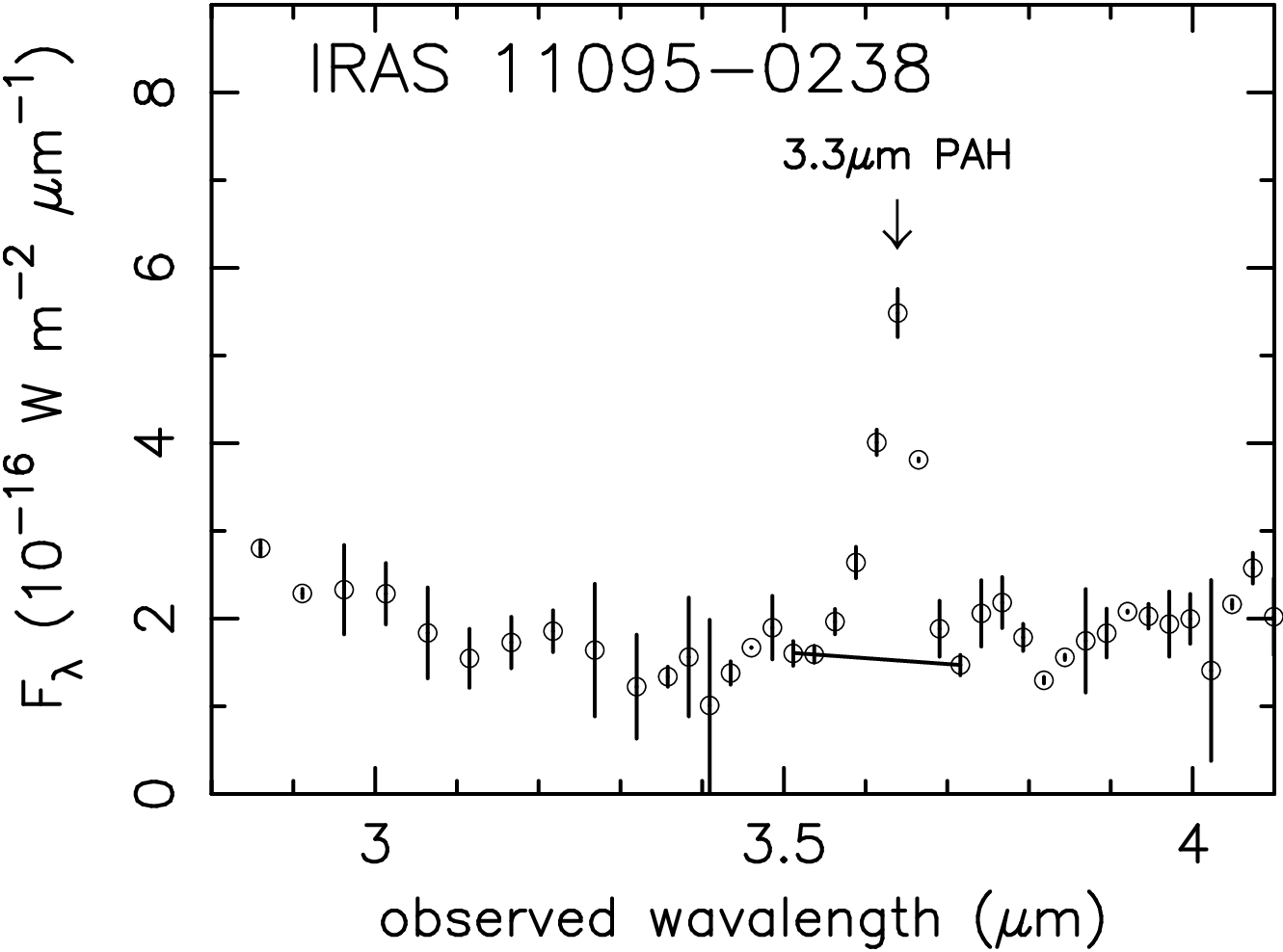} \\
\includegraphics[angle=-90,scale=.35]{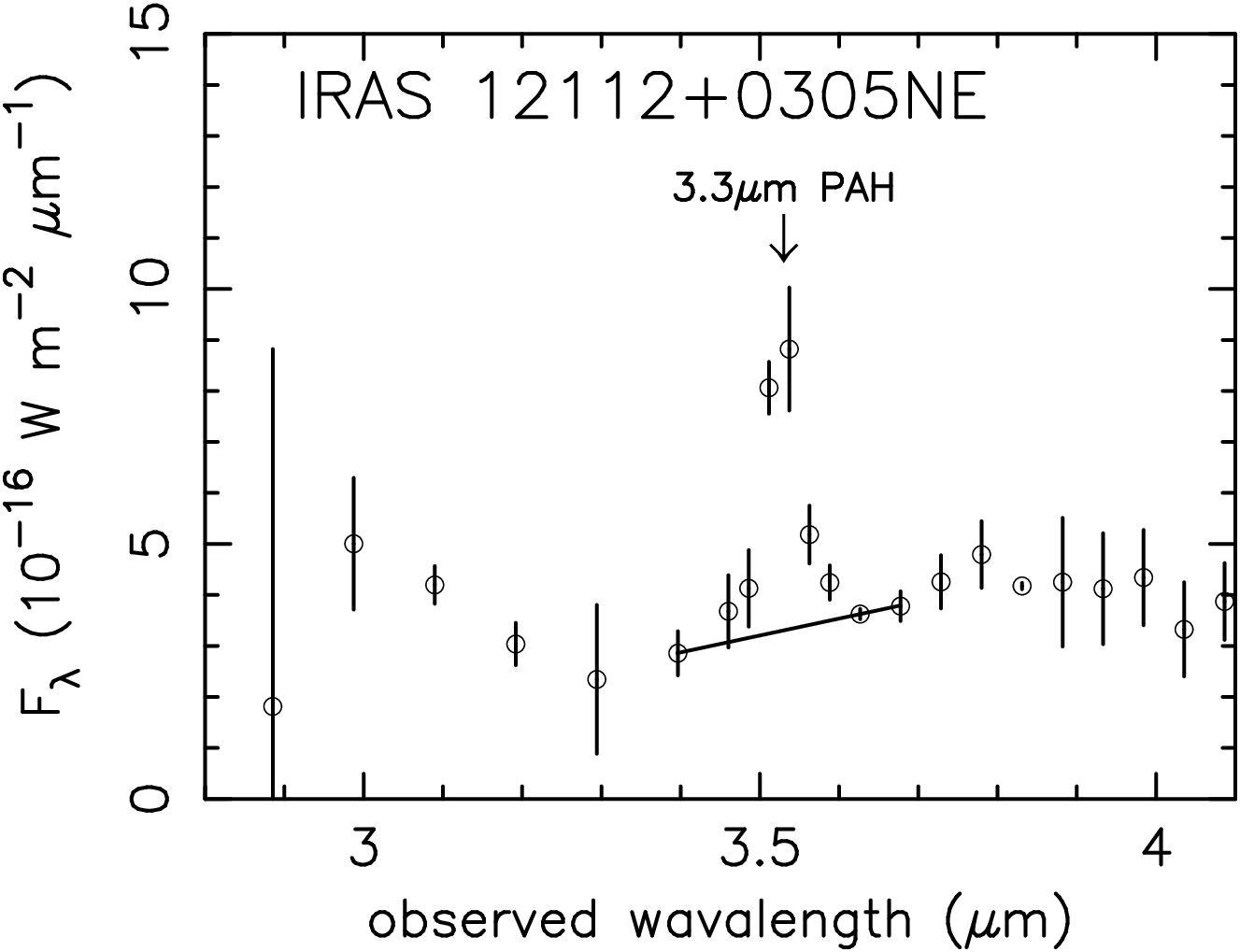} \hspace{0.3cm} 
\includegraphics[angle=-90,scale=.35]{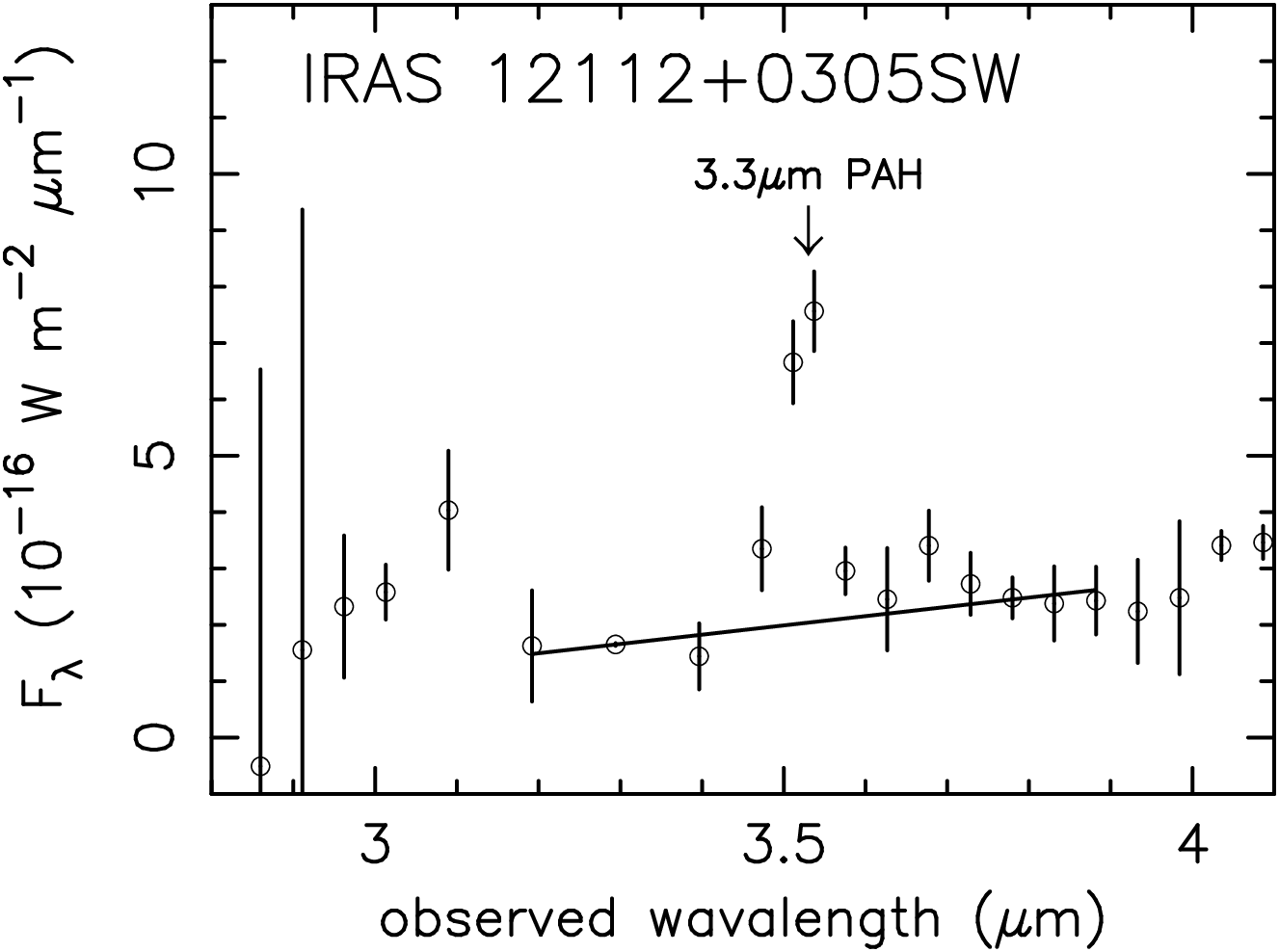} \\
\end{figure}

\clearpage

\begin{figure}
\begin{center} {\bf \Large LINER (3)} \end{center}
\includegraphics[angle=-90,scale=.35]{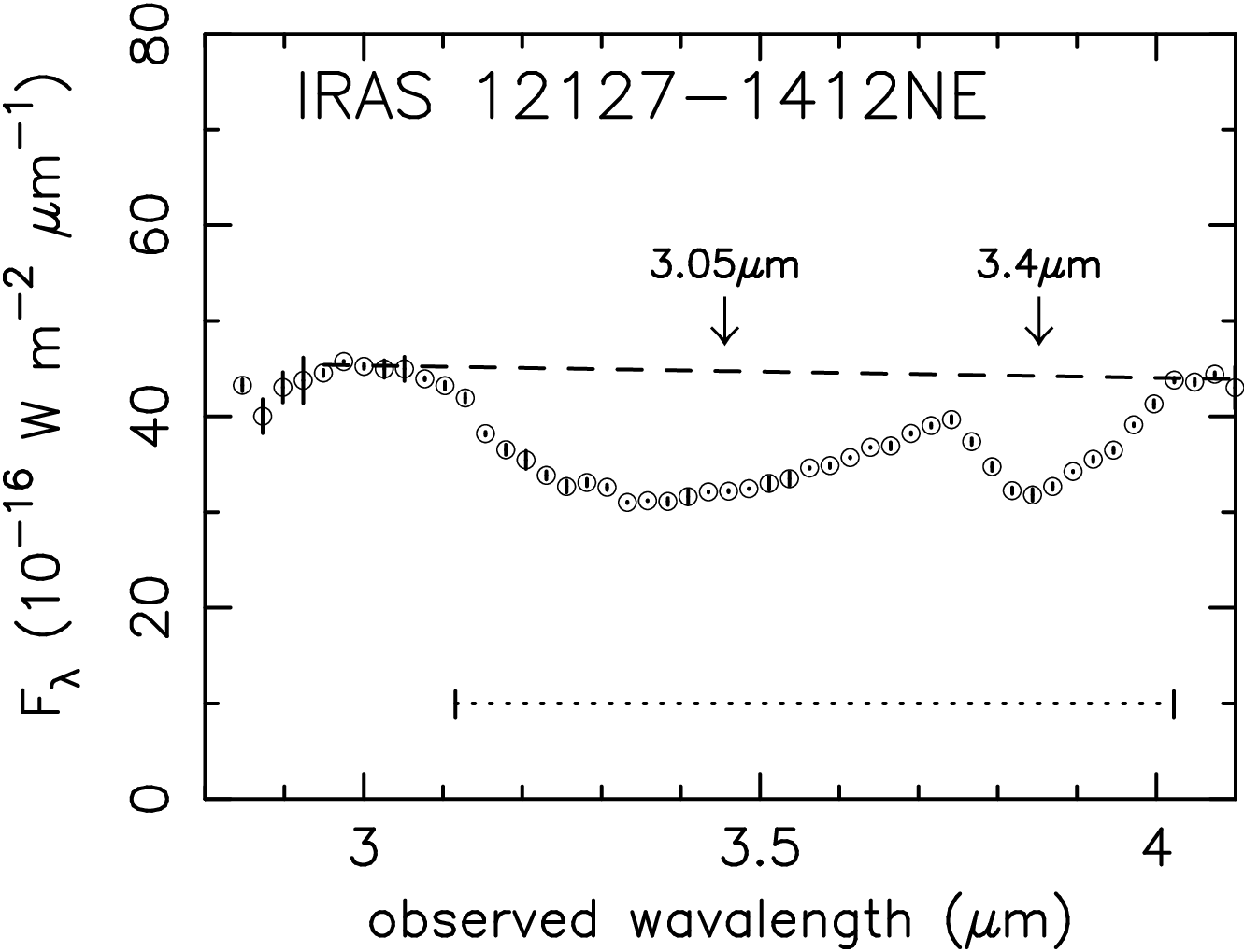} \hspace{0.3cm} 
\includegraphics[angle=-90,scale=.35]{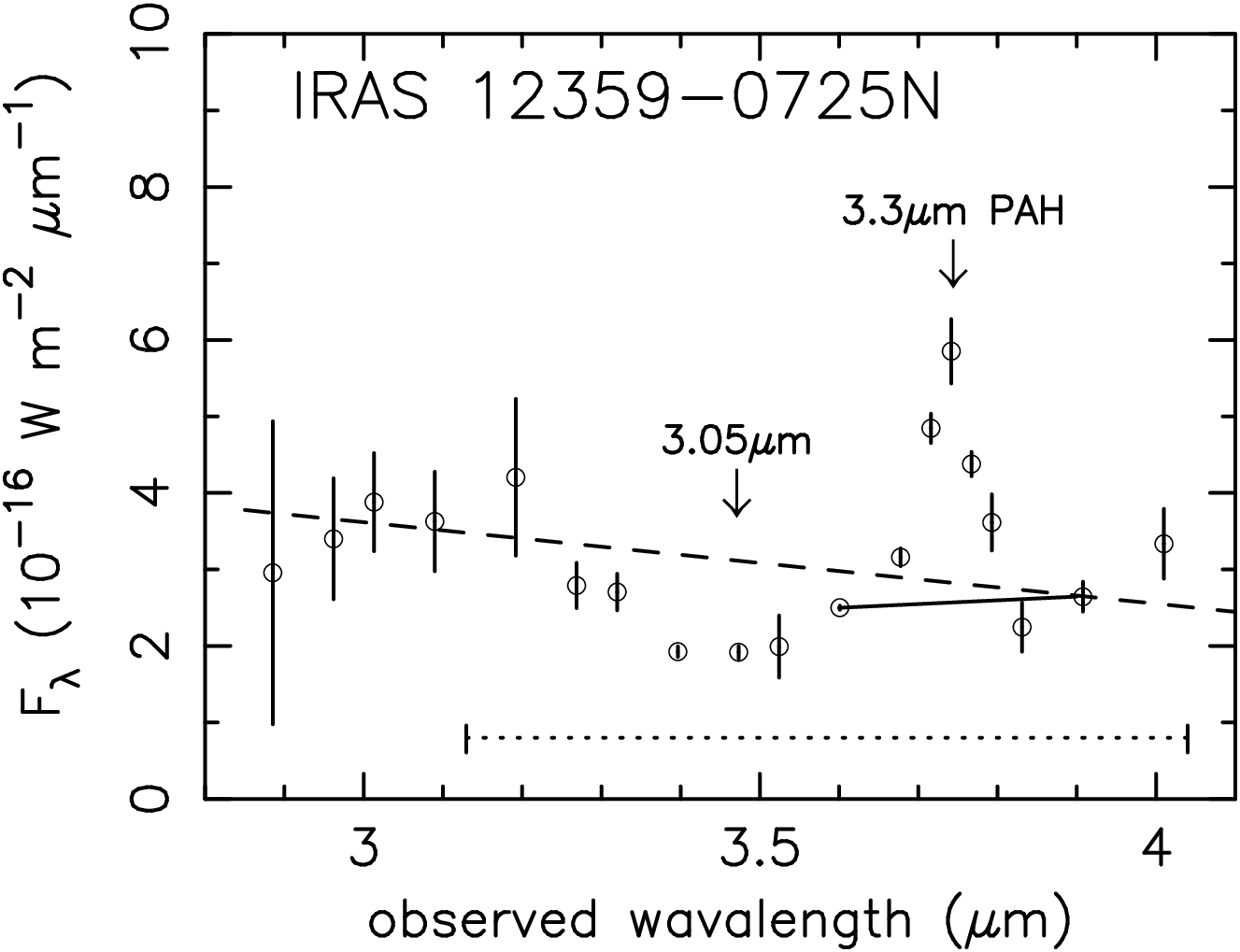} \\
\includegraphics[angle=-90,scale=.35]{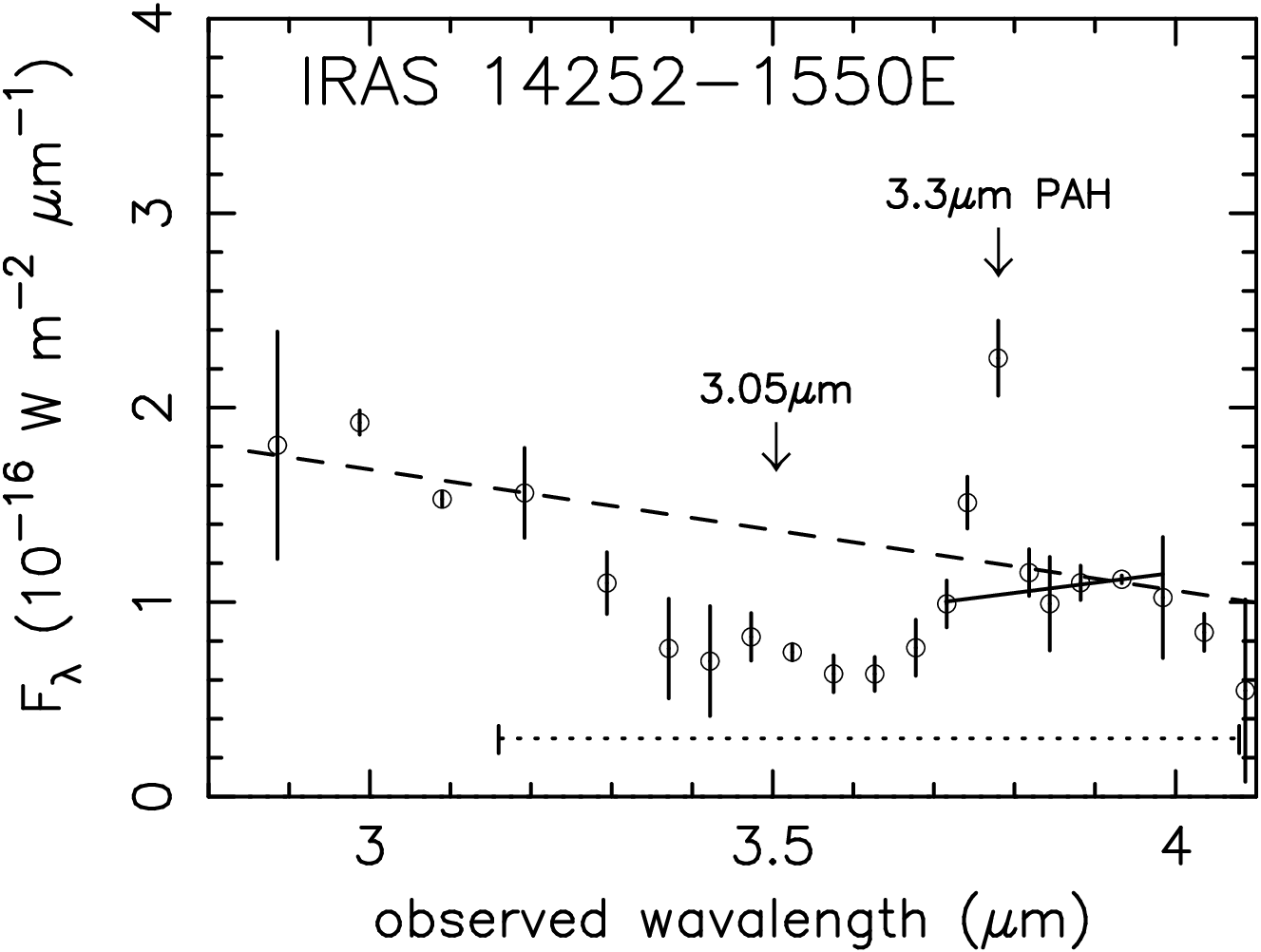} \hspace{0.3cm}
\includegraphics[angle=-90,scale=.35]{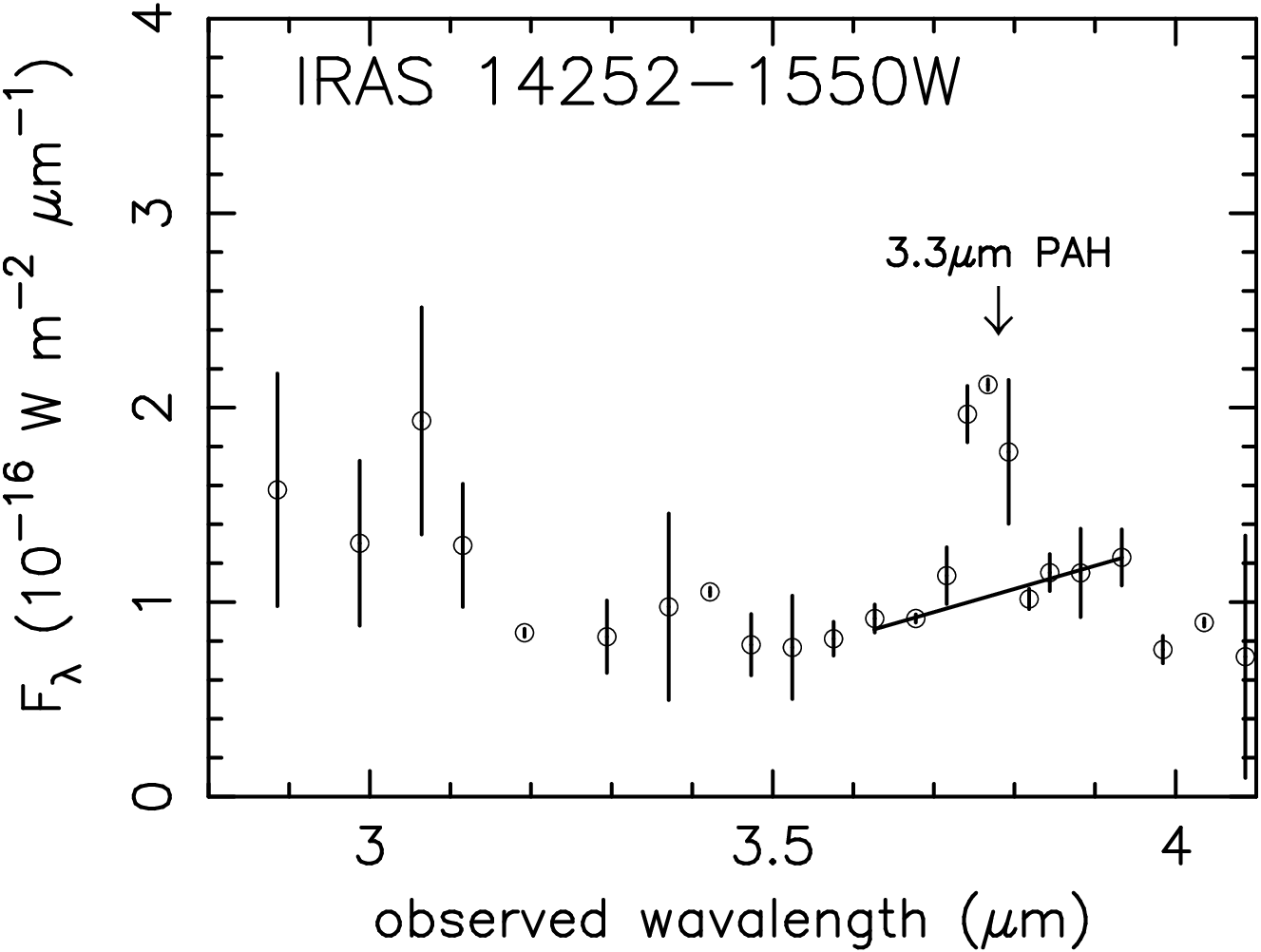} \\
\includegraphics[angle=-90,scale=.35]{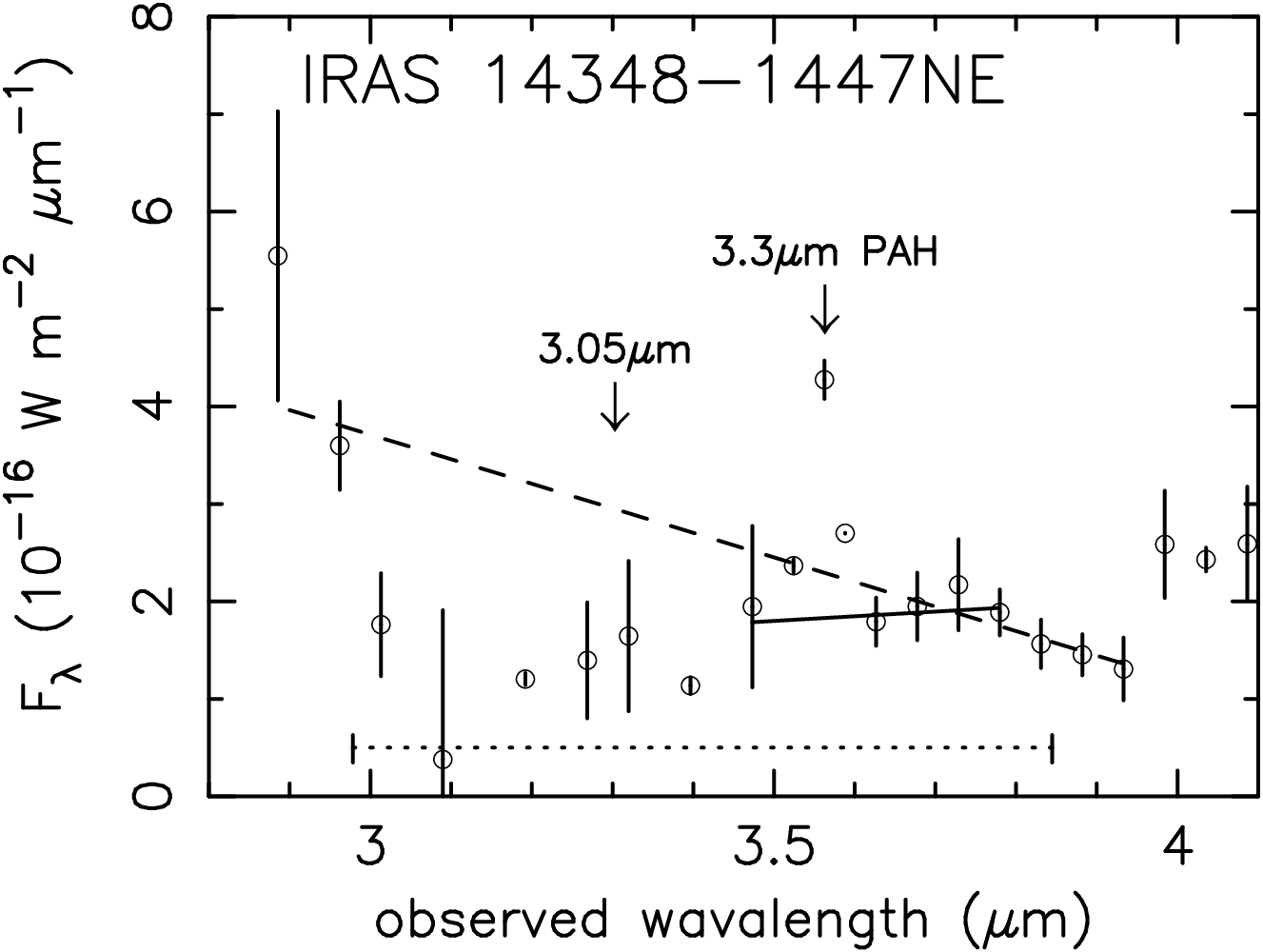} \hspace{0.3cm}
\includegraphics[angle=-90,scale=.35]{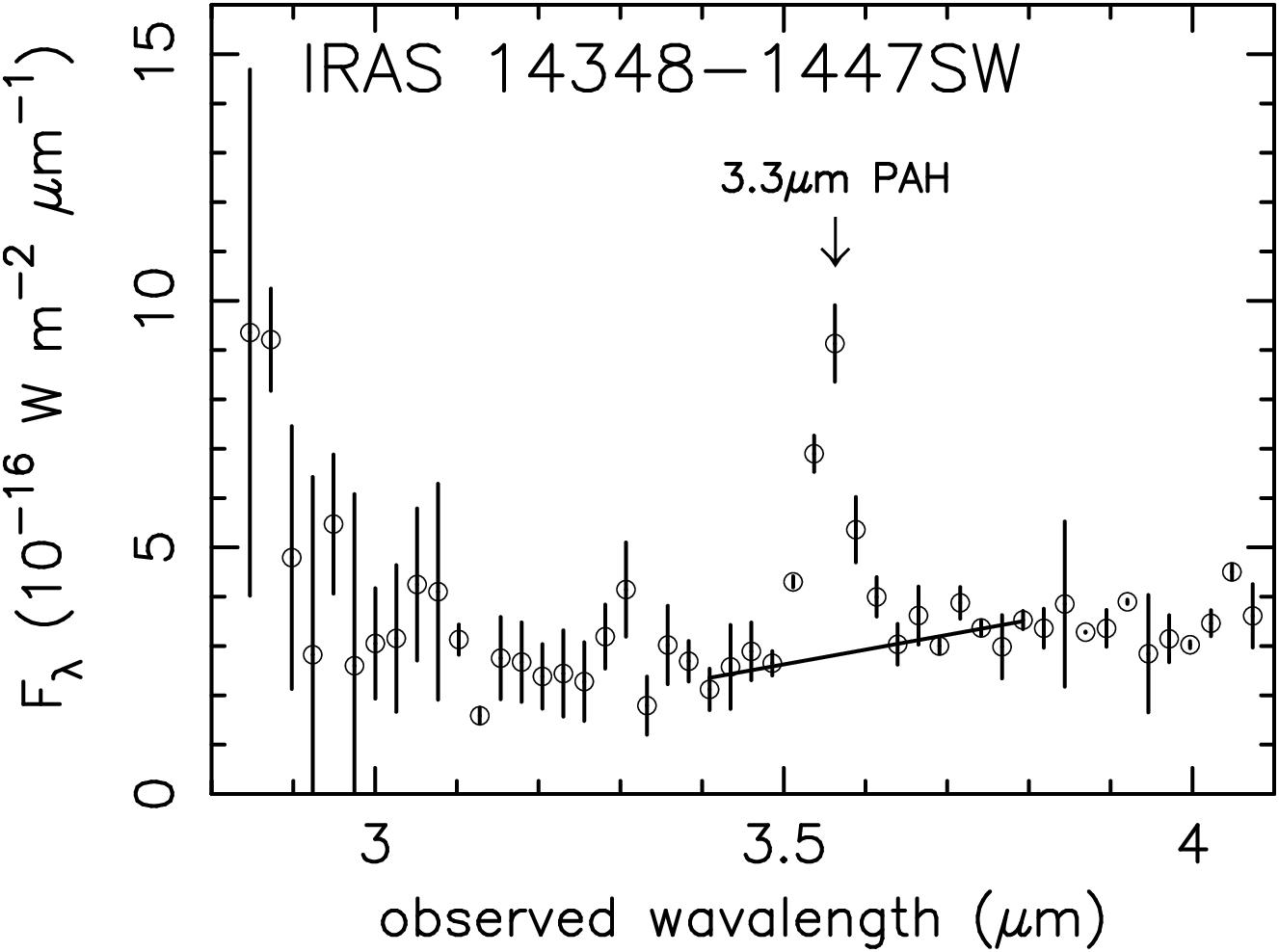} \\
\end{figure}

\clearpage

\begin{figure}
\begin{center} {\bf \Large LINER (4)} \end{center}
\includegraphics[angle=-90,scale=.35]{f1s.eps} \hspace{0.3cm} 
\includegraphics[angle=-90,scale=.35]{f1t.eps} \\
\includegraphics[angle=-90,scale=.35]{f1u.eps} \hspace{0.3cm}  
\includegraphics[angle=-90,scale=.35]{f1v.eps} \\
\includegraphics[angle=-90,scale=.35]{f1w.eps} \hspace{0.3cm} 
\includegraphics[angle=-90,scale=.35]{f1x.eps} \\ 
\end{figure}

\clearpage

\begin{figure}
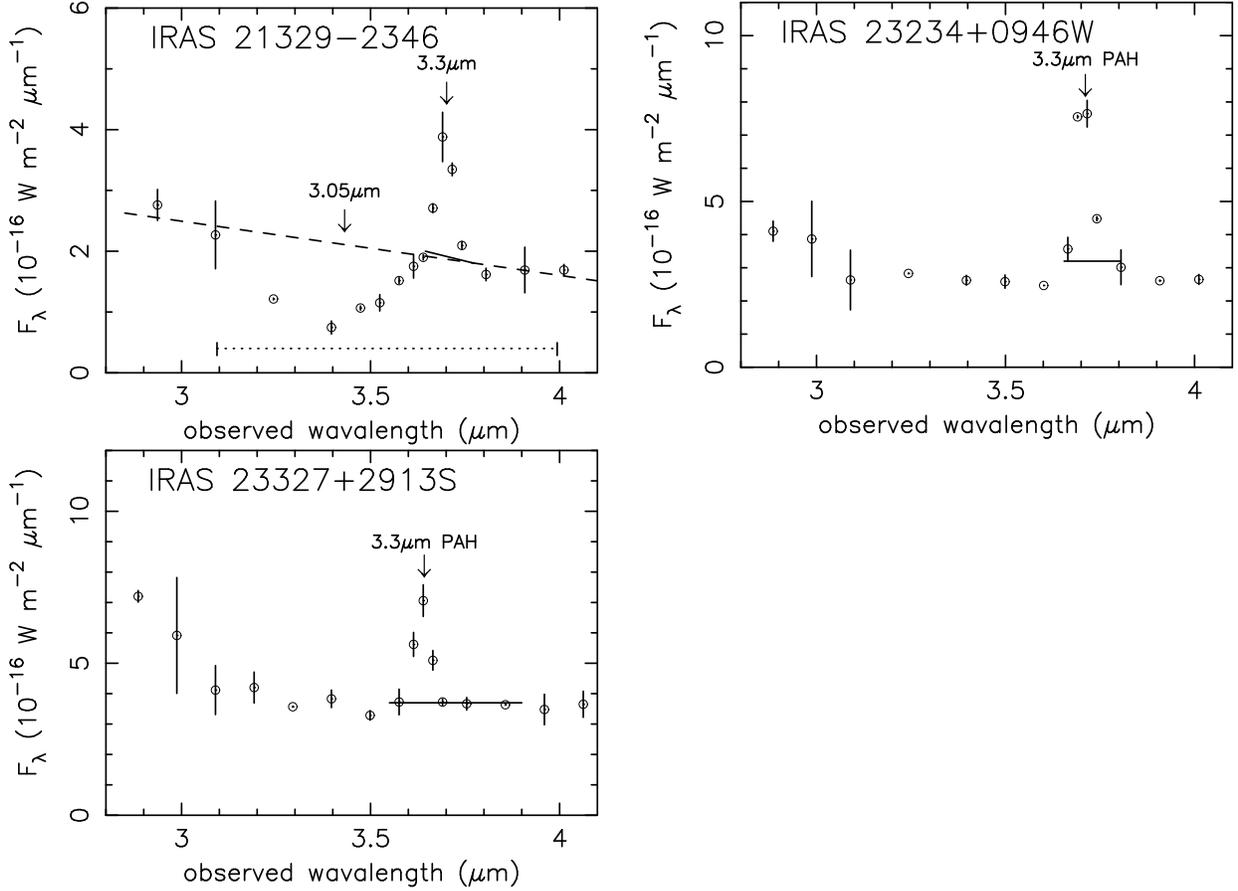

\begin{center} {\bf \Large LINER (5)} \end{center}
\includegraphics[angle=-90,scale=.35]{f1y.eps} \hspace{0.3cm}  
\includegraphics[angle=-90,scale=.35]{f1z.eps} \\
\includegraphics[angle=-90,scale=.35]{f1aa.eps} \hspace{0.3cm}
\caption{
Infrared $L$-band ($\lambda_{\rm obs}$ = 2.8--4.1 $\mu$m) spectra of ULIRGs'
nuclei classified optically as LINERs. 
The abscissa and ordinate are the observed wavelength in $\mu$m and
F$_{\lambda}$ in 10$^{-16}$ W m$^{-2}$ $\mu$m$^{-1}$, respectively.
The lower arrows with ``3.3 $\mu$m PAH'' indicate the expected wavelength
of the 3.3 $\mu$m PAH emission ($\lambda_{\rm rest}$ = 3.29 $\mu$m).
The solid lines are the adopted continuum to estimate the 3.3 $\mu$m PAH
emission fluxes.
The arrows with ``3.05 $\mu$m'' or ``3.4 $\mu$m'' indicate the 
expected absorption peak at $\lambda_{\rm rest}$ = 3.05 $\mu$m and 3.4
$\mu$m, respectively.   
For ULIRGs with obvious signatures of these absorption features, dashed
straight lines are plotted to show the adopted continuum levels used to
estimate the absorption optical depths.
The dotted lines indicate the wavelength range where the 
effects of the broad 3.05 $\mu$m ice absorption feature 
can be significant ($\lambda_{\rm rest}$ = 2.75--3.55 $\mu$m 
adopted from the spectrum of Elias 16; Smith et al. 1989).  
The ice absorption feature is usually very strong at 
$\lambda_{\rm rest}$ = 2.9--3.2 $\mu$m, but the profiles and wavelength
range of weaker absorption wings are found to vary among different
Galactic objects (Smith et al. 1989).    
}
\end{figure}

\clearpage

\begin{figure}
\begin{center} {\bf \Large HII (1)} \end{center}
\includegraphics[angle=-90,scale=.35]{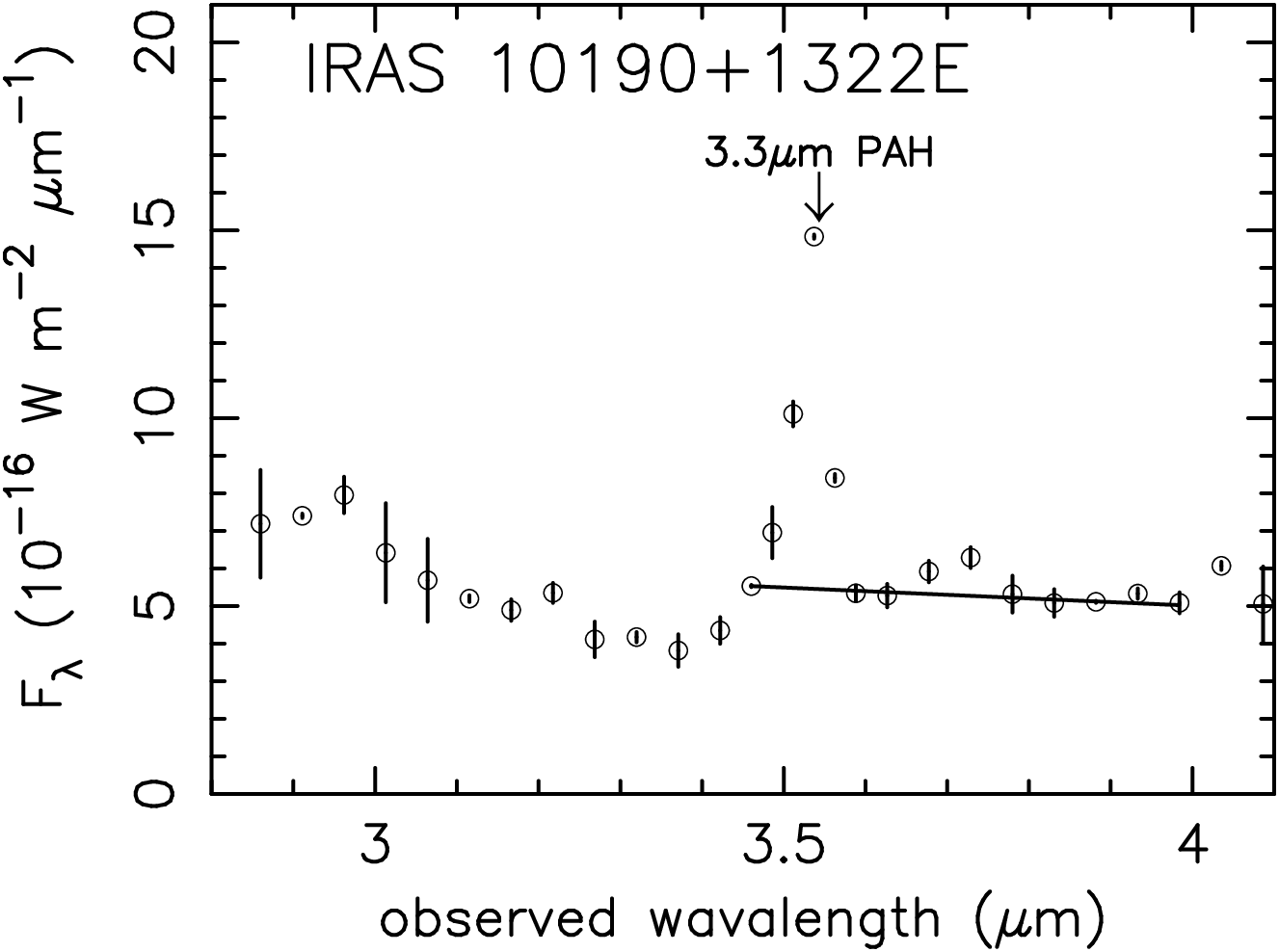} \hspace{0.3cm}
\includegraphics[angle=-90,scale=.35]{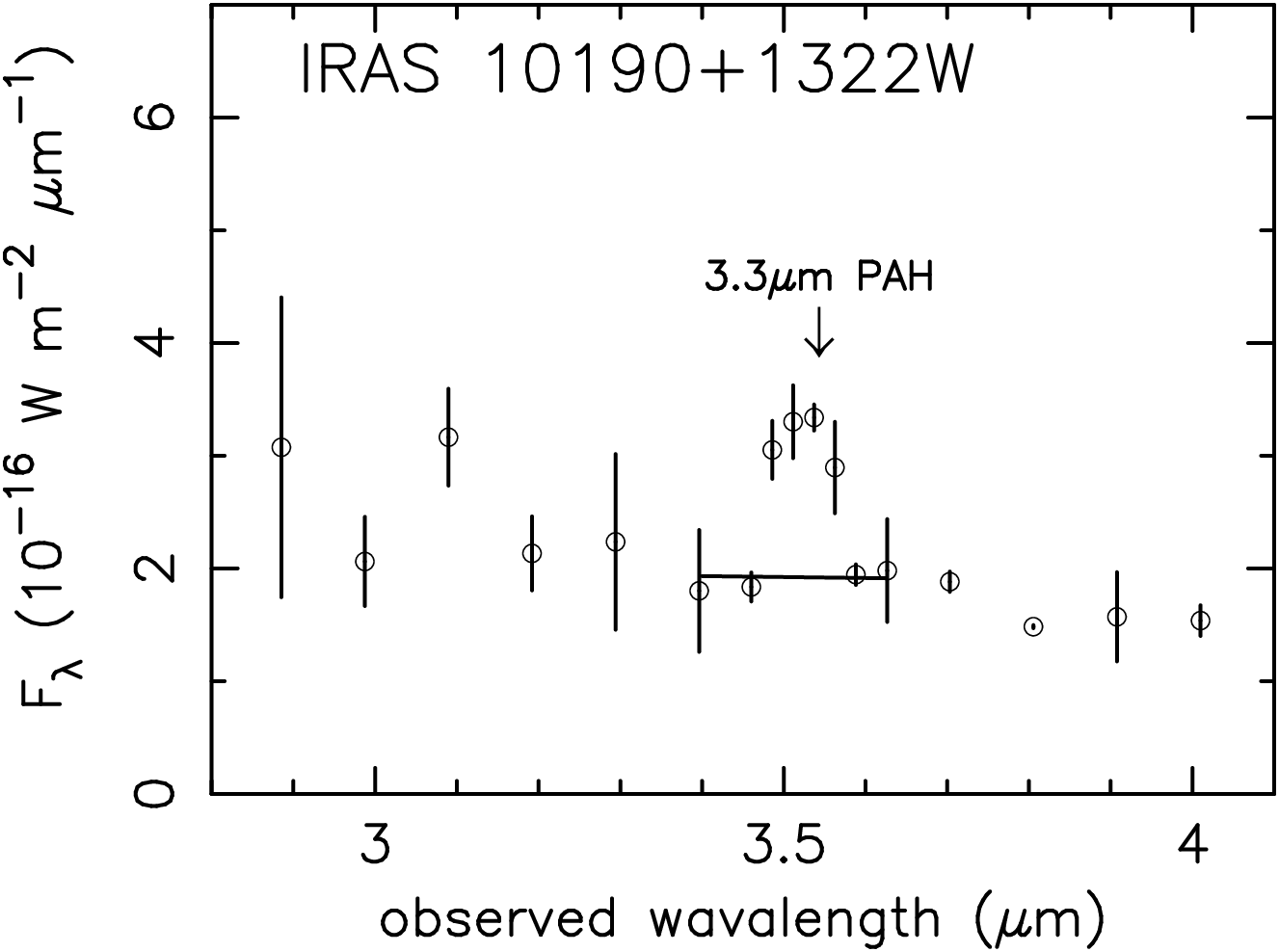} \\ 
\includegraphics[angle=-90,scale=.35]{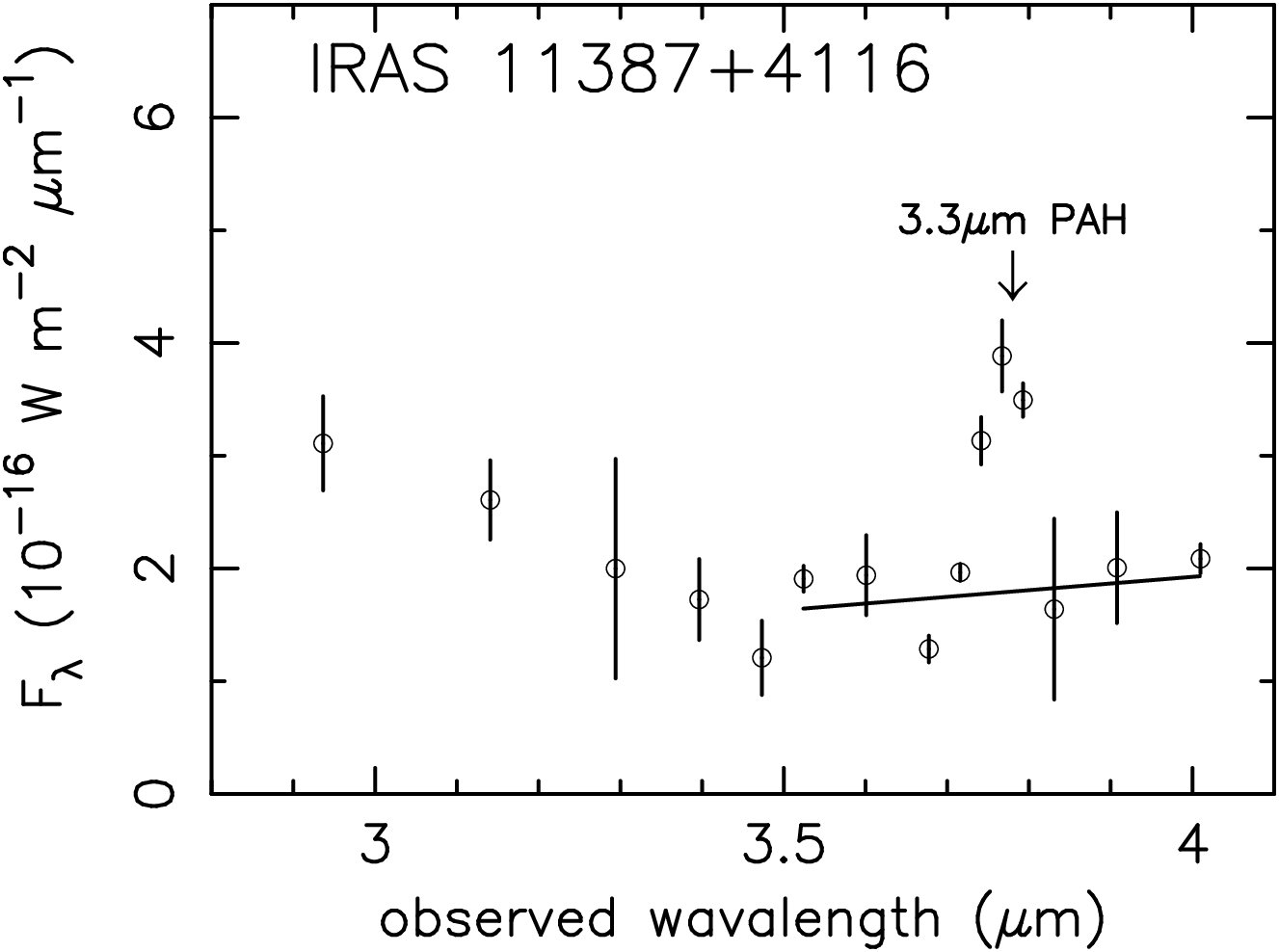} \hspace{0.3cm}
\includegraphics[angle=-90,scale=.35]{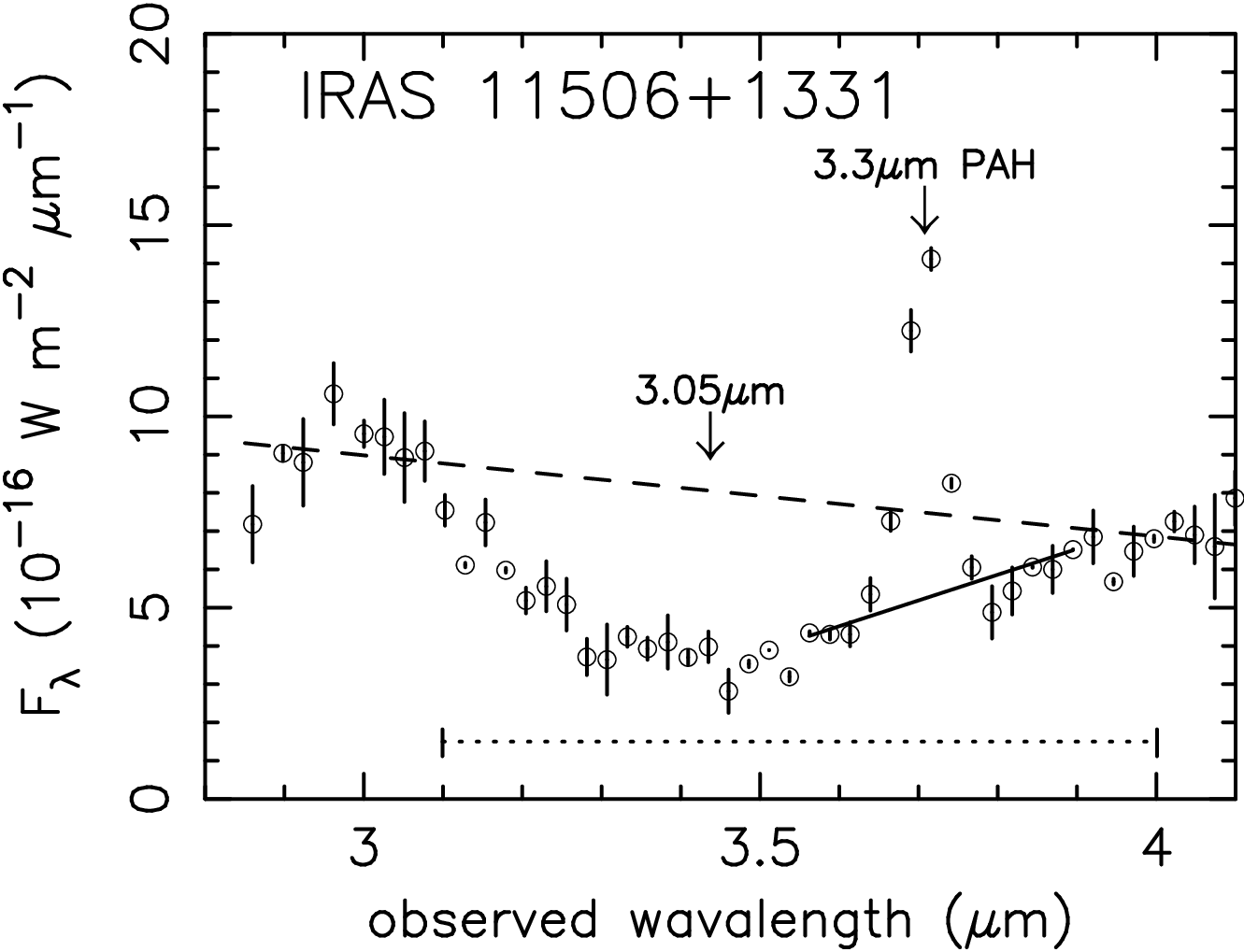} \\ 
\includegraphics[angle=-90,scale=.35]{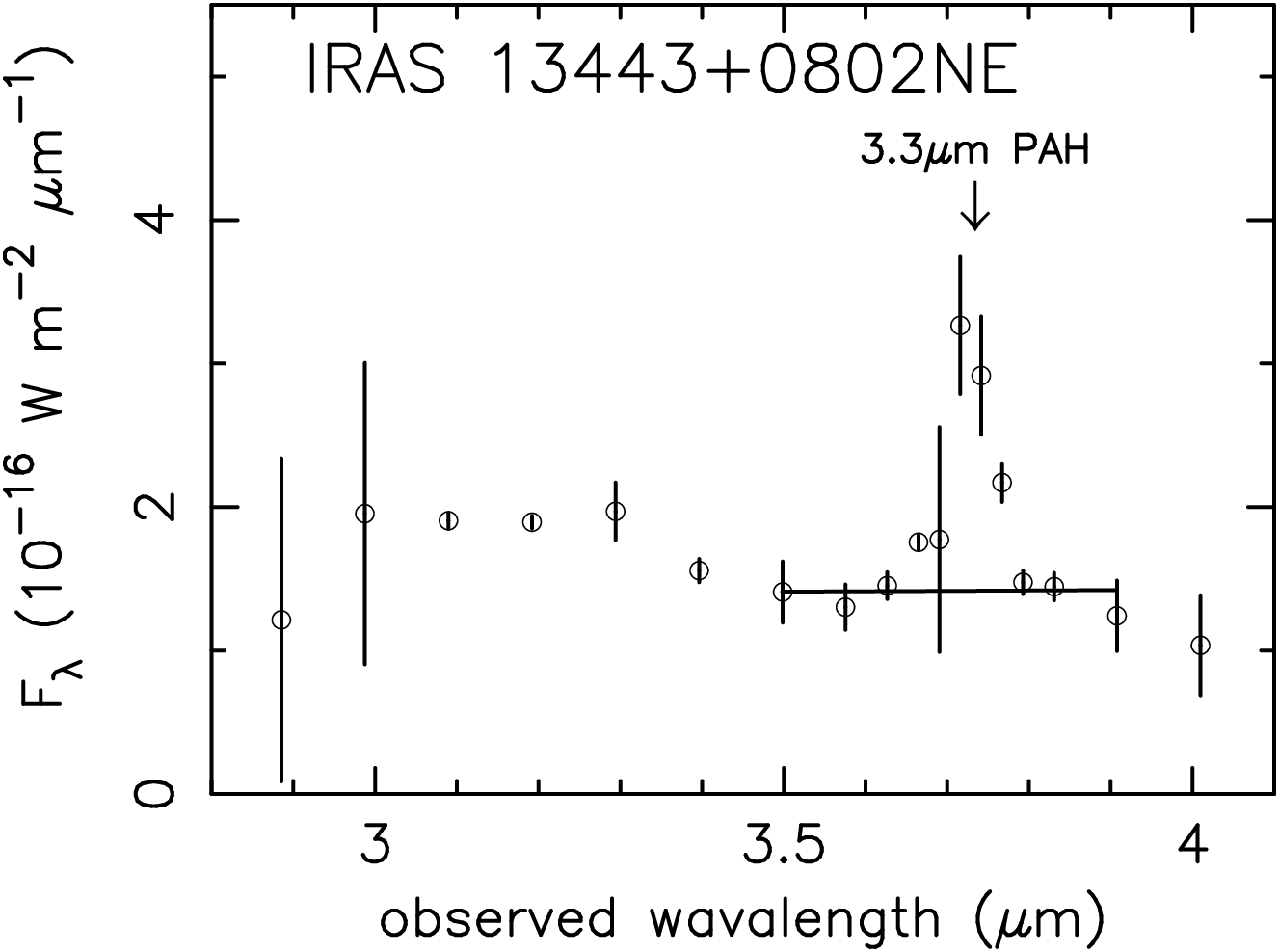} \hspace{0.3cm}
\includegraphics[angle=-90,scale=.35]{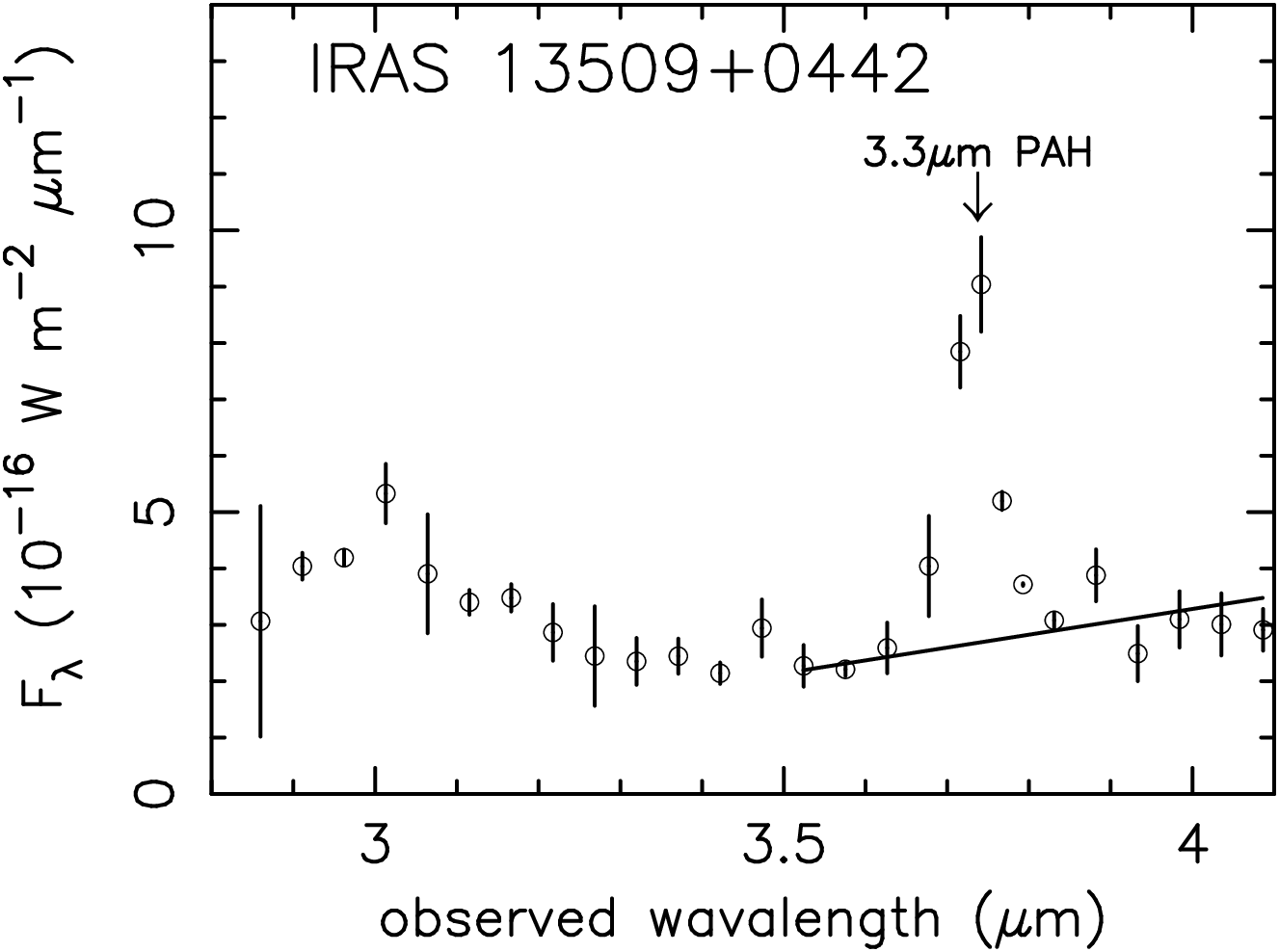} \\
\end{figure}

\clearpage

\begin{figure}
\begin{center} {\bf \Large HII (2)} \end{center}
\includegraphics[angle=-90,scale=.35]{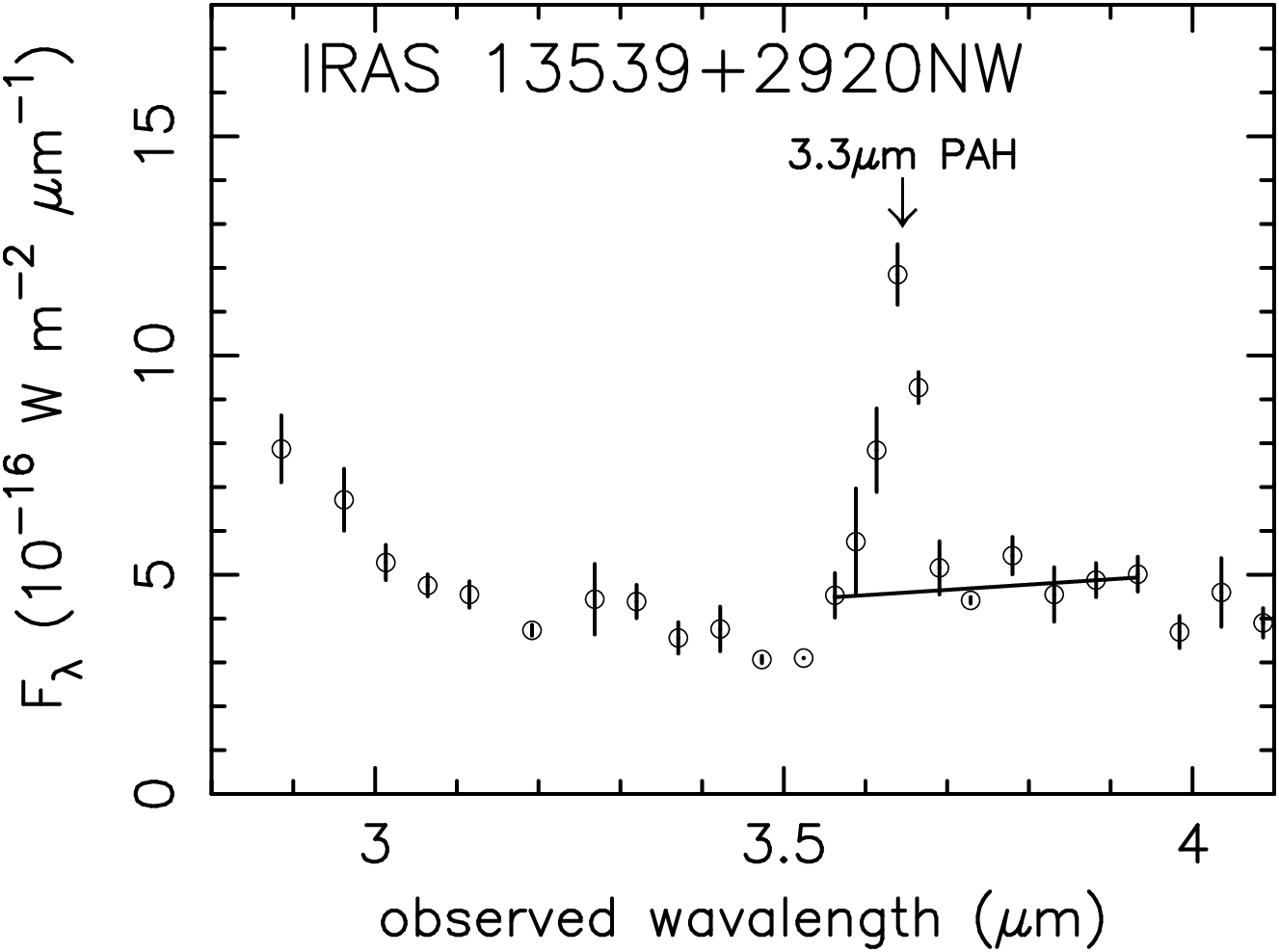} \hspace{0.3cm}
\includegraphics[angle=-90,scale=.35]{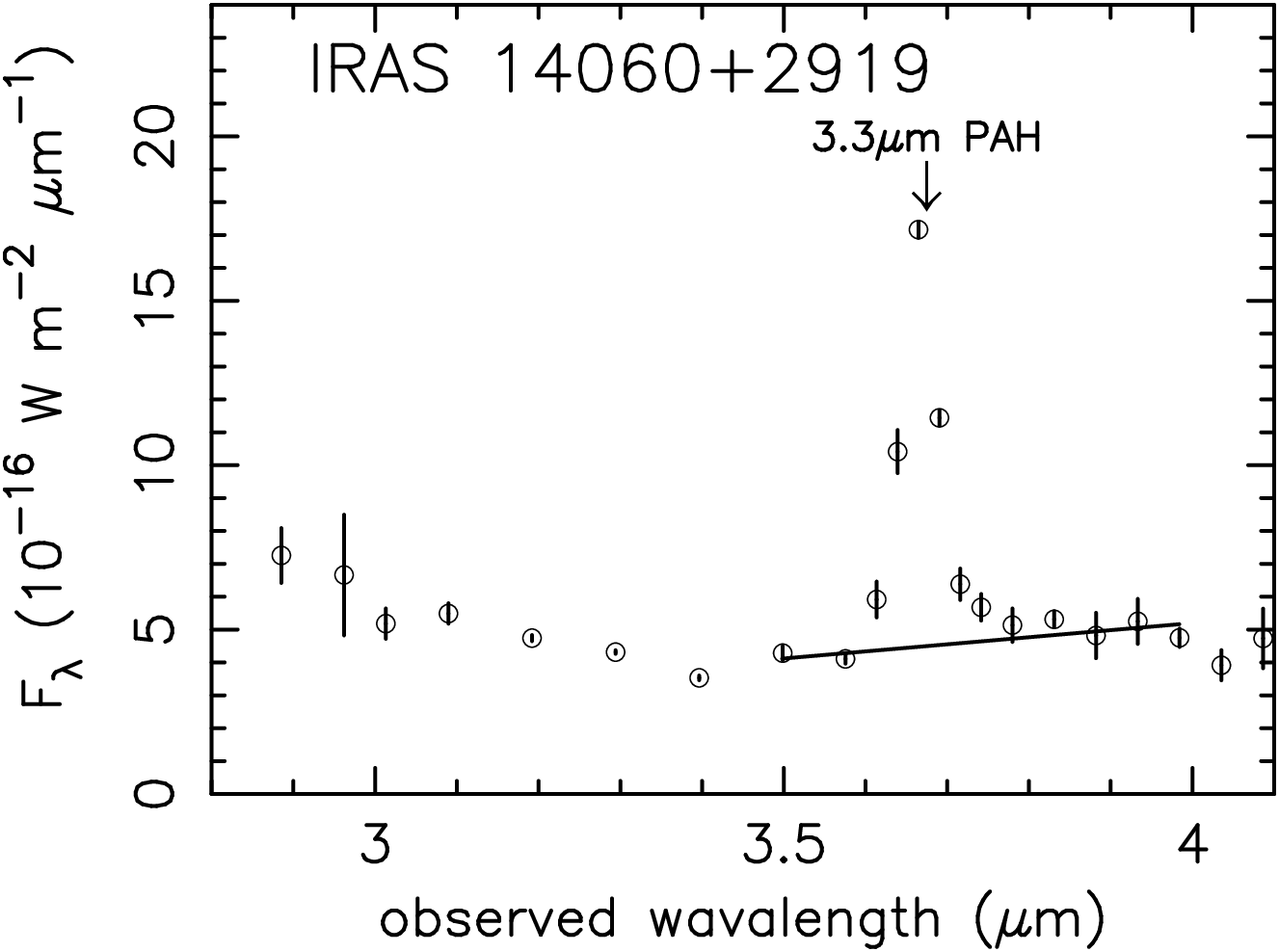} \\ 
\includegraphics[angle=-90,scale=.35]{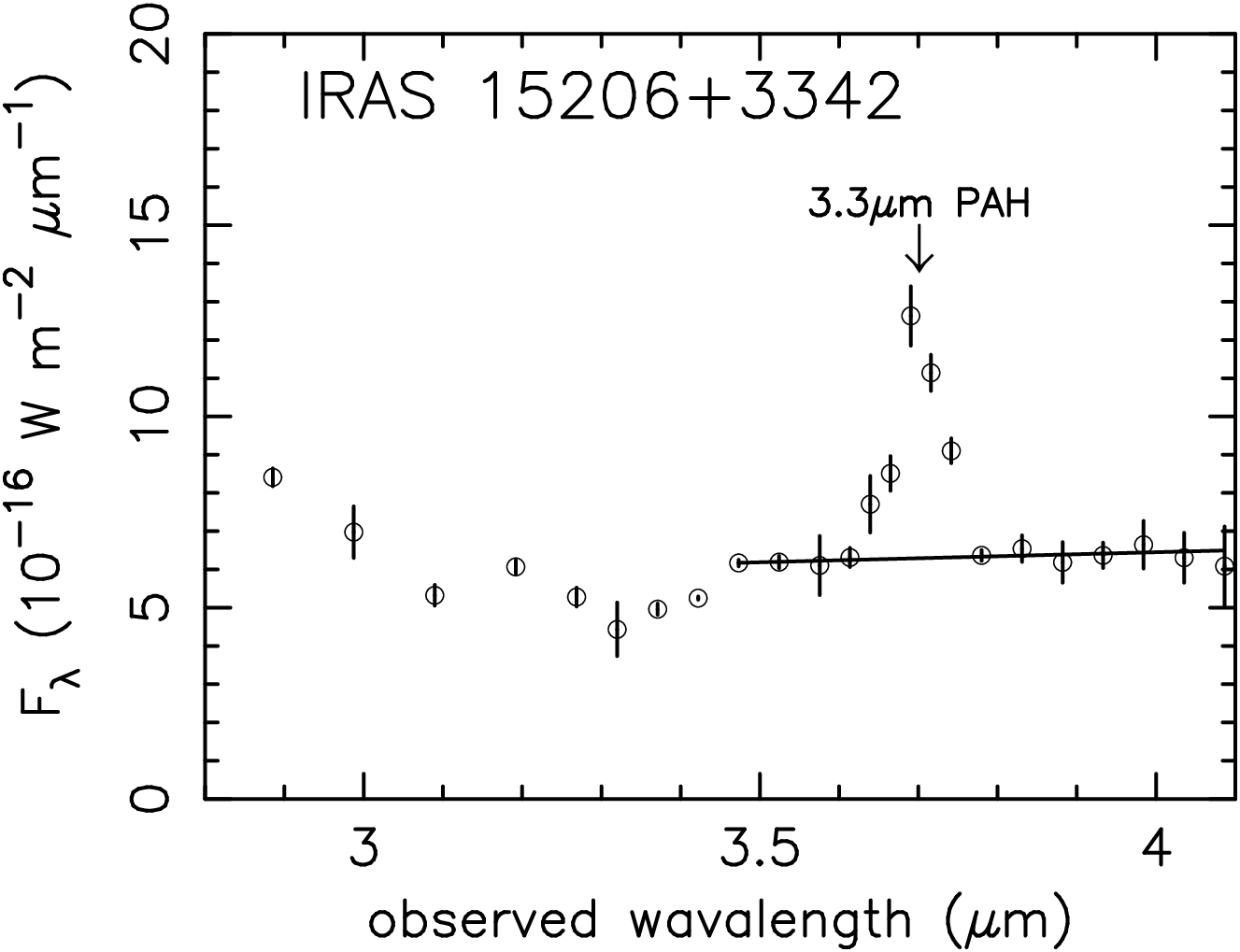} \hspace{0.3cm} 
\includegraphics[angle=-90,scale=.35]{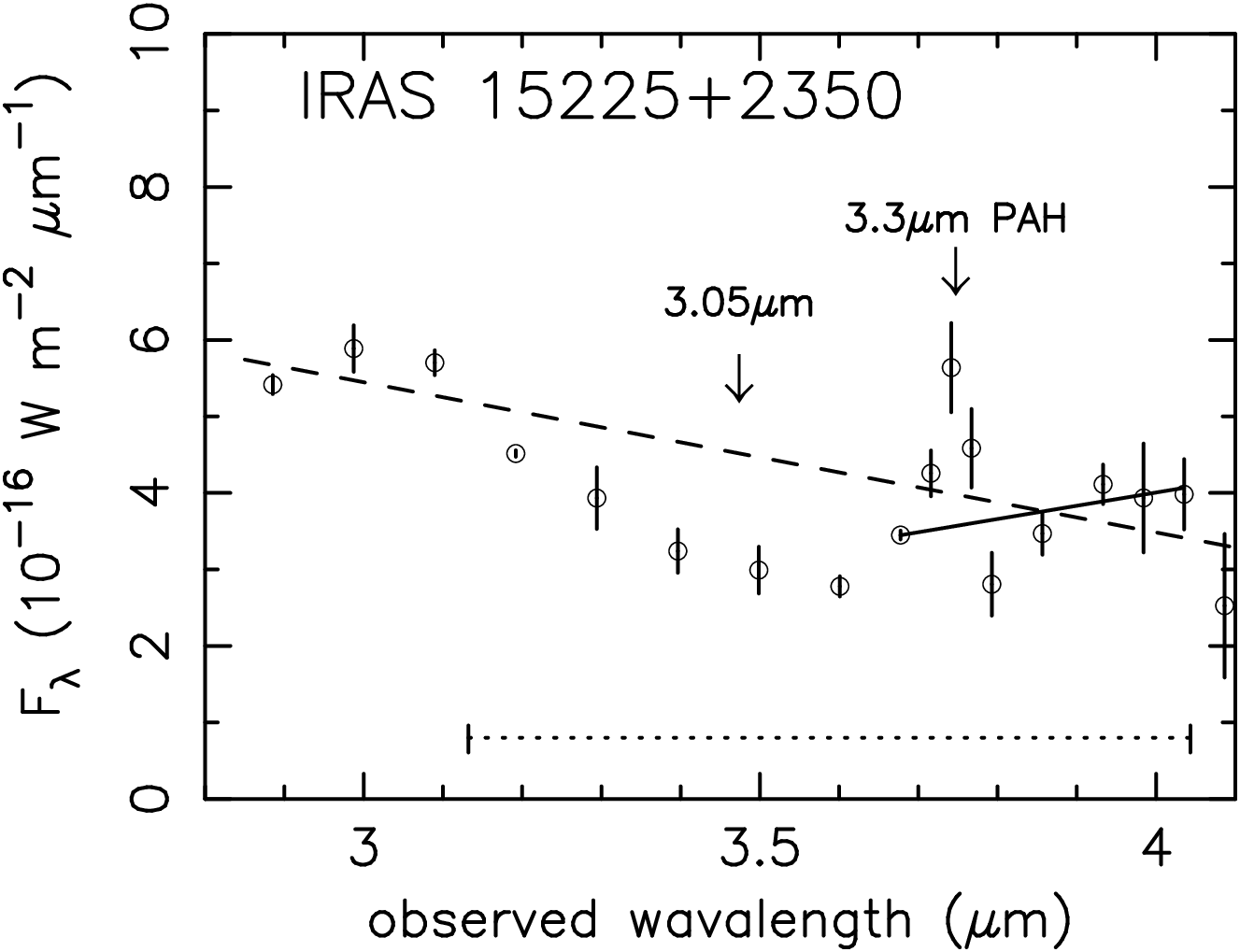} \\ 
\includegraphics[angle=-90,scale=.35]{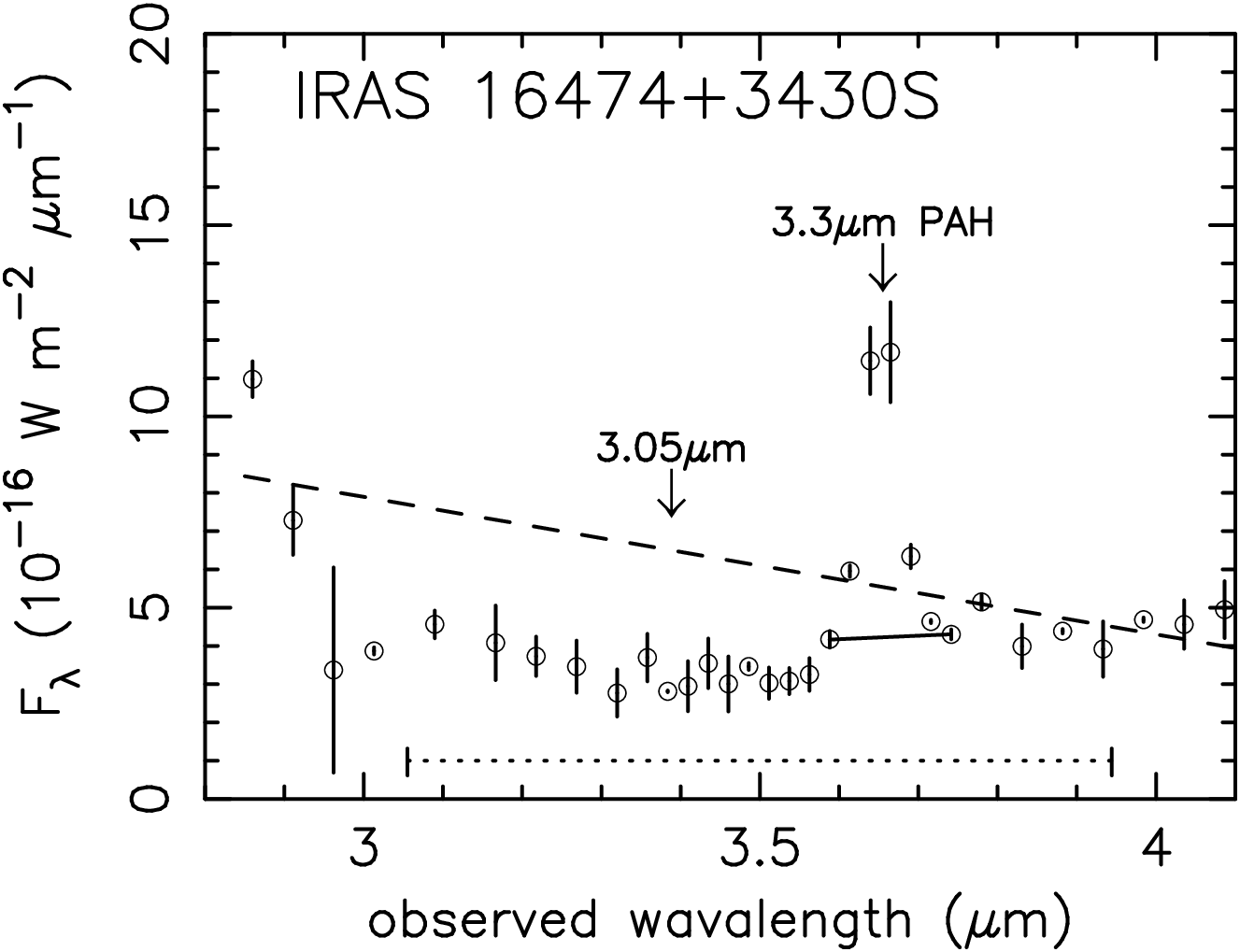} \hspace{0.3cm} 
\includegraphics[angle=-90,scale=.35]{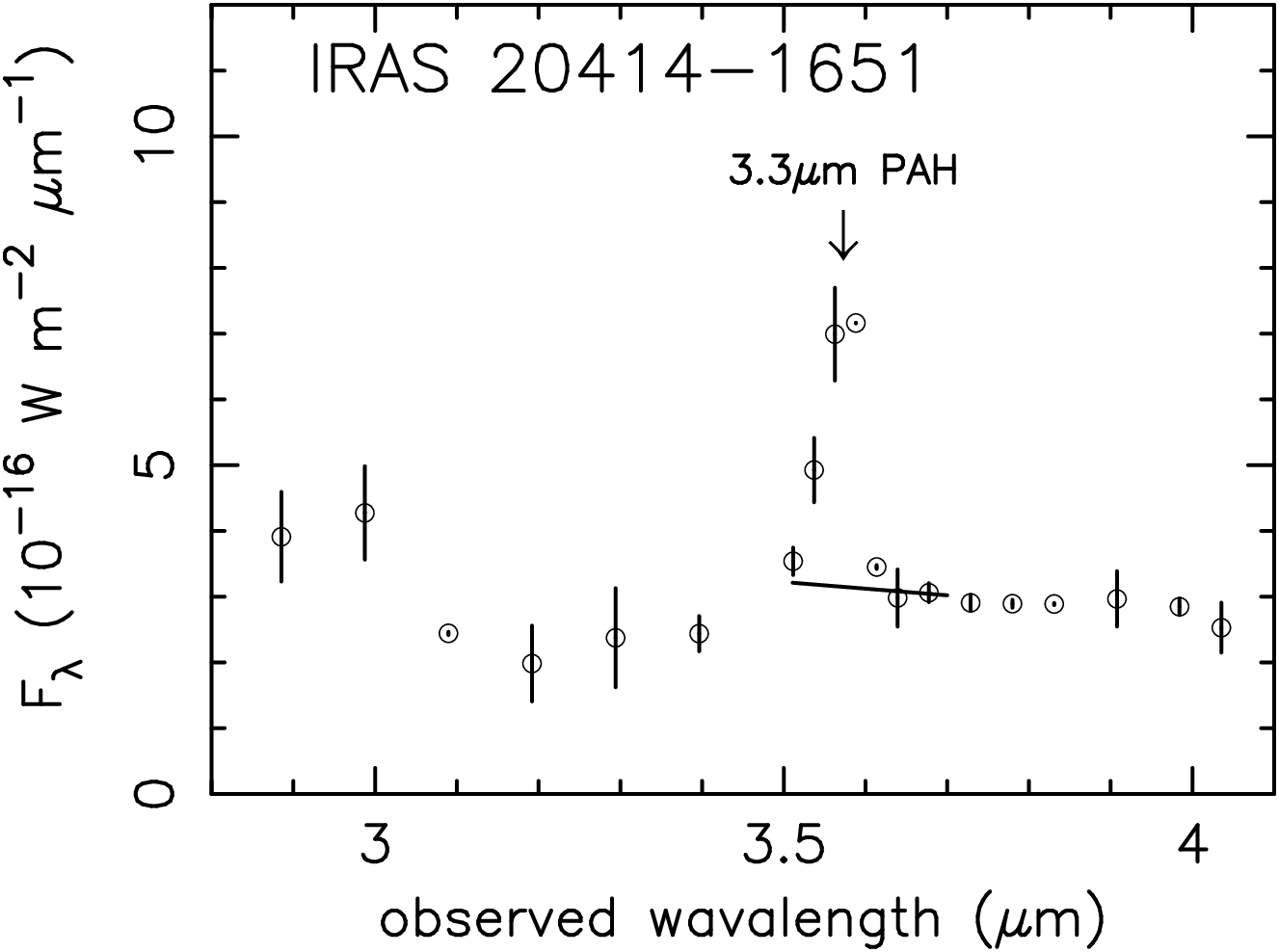} \\
\end{figure}

\clearpage

\begin{figure}
\begin{center} {\bf \Large HII (3)} \end{center}
\includegraphics[angle=-90,scale=.35]{f2m.eps} \hspace{0.3cm} 
\caption{
Infrared $L$-band ($\lambda_{\rm obs}$ = 2.8--4.1 $\mu$m) spectra of ULIRGs
classified optically as HII-regions. 
Units and symbols are the same as Figure 1.
}
\end{figure}


\begin{figure}
\begin{center} {\bf \Large Unclassified} \end{center}
\includegraphics[angle=-90,scale=.35]{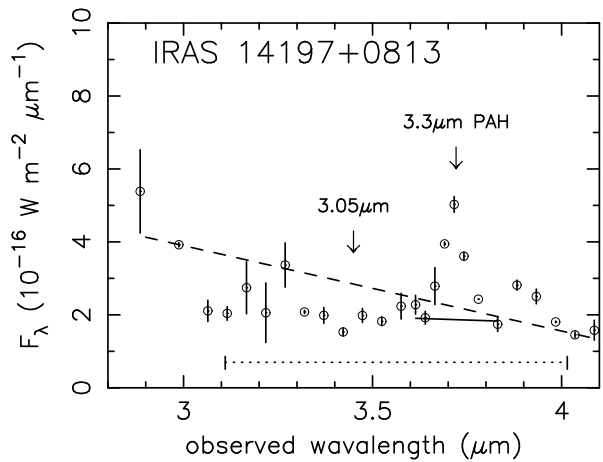} \hspace{0.3cm}
\includegraphics[angle=-90,scale=.35]{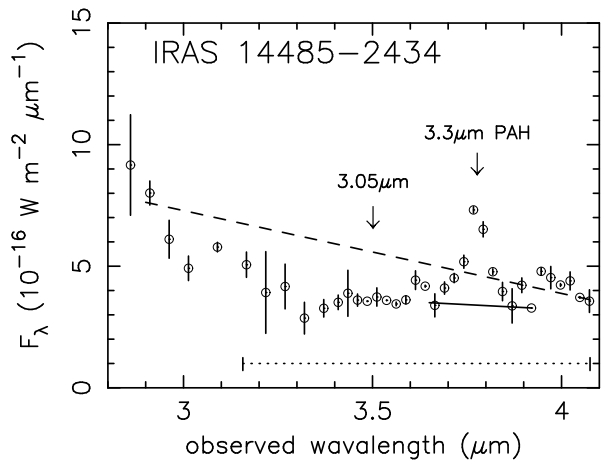} \\ 
\caption{
Infrared $L$-band ($\lambda_{\rm obs}$ = 2.8--4.1 $\mu$m) spectra of 
optically unclassified ULIRGs.
Units and symbols are the same as Figure 1.
}
\end{figure}

\clearpage

\begin{figure}
\begin{center} {\bf \Large Sy2} \end{center}
\includegraphics[angle=-90,scale=.35]{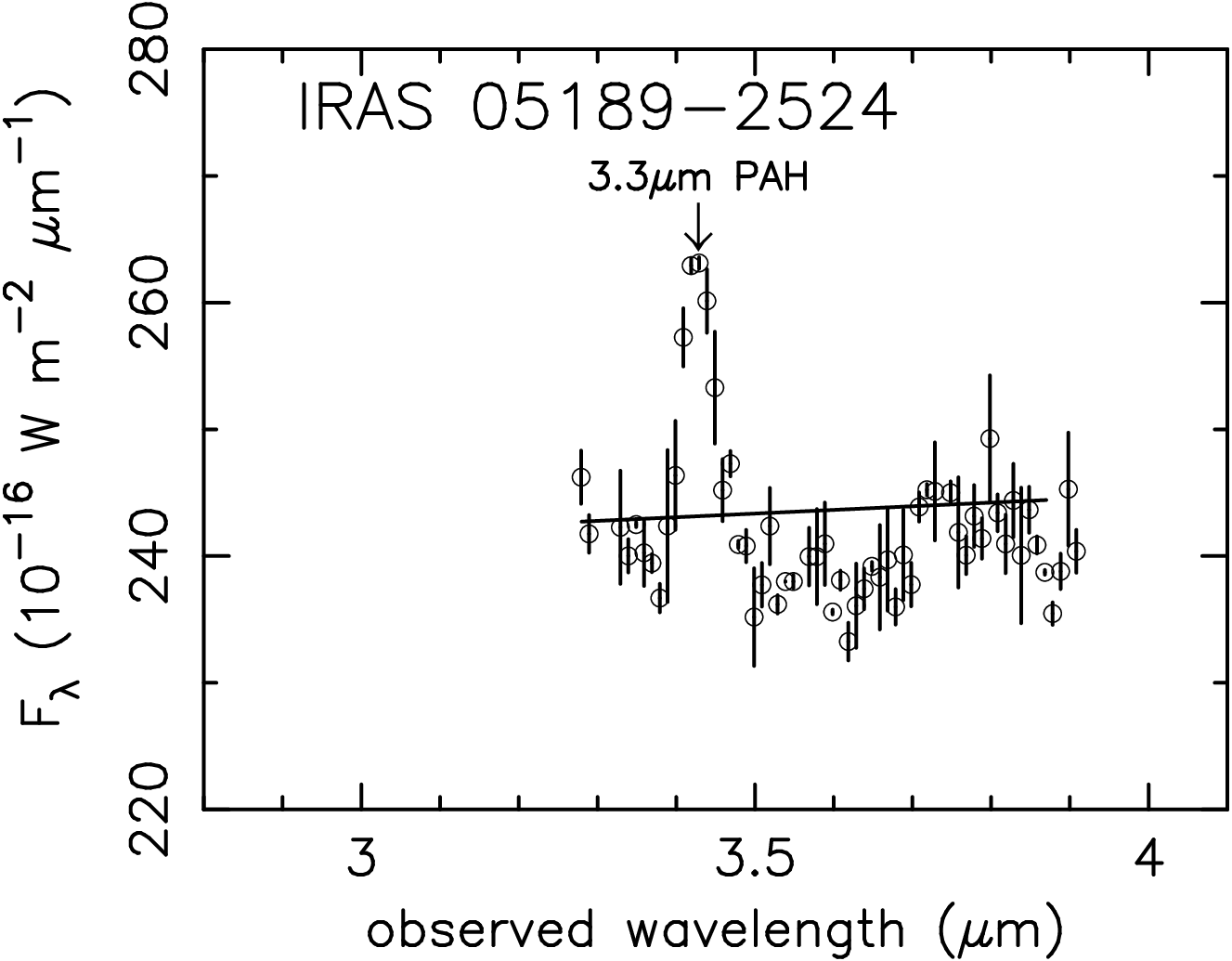} \hspace{0.3cm}
\includegraphics[angle=-90,scale=.35]{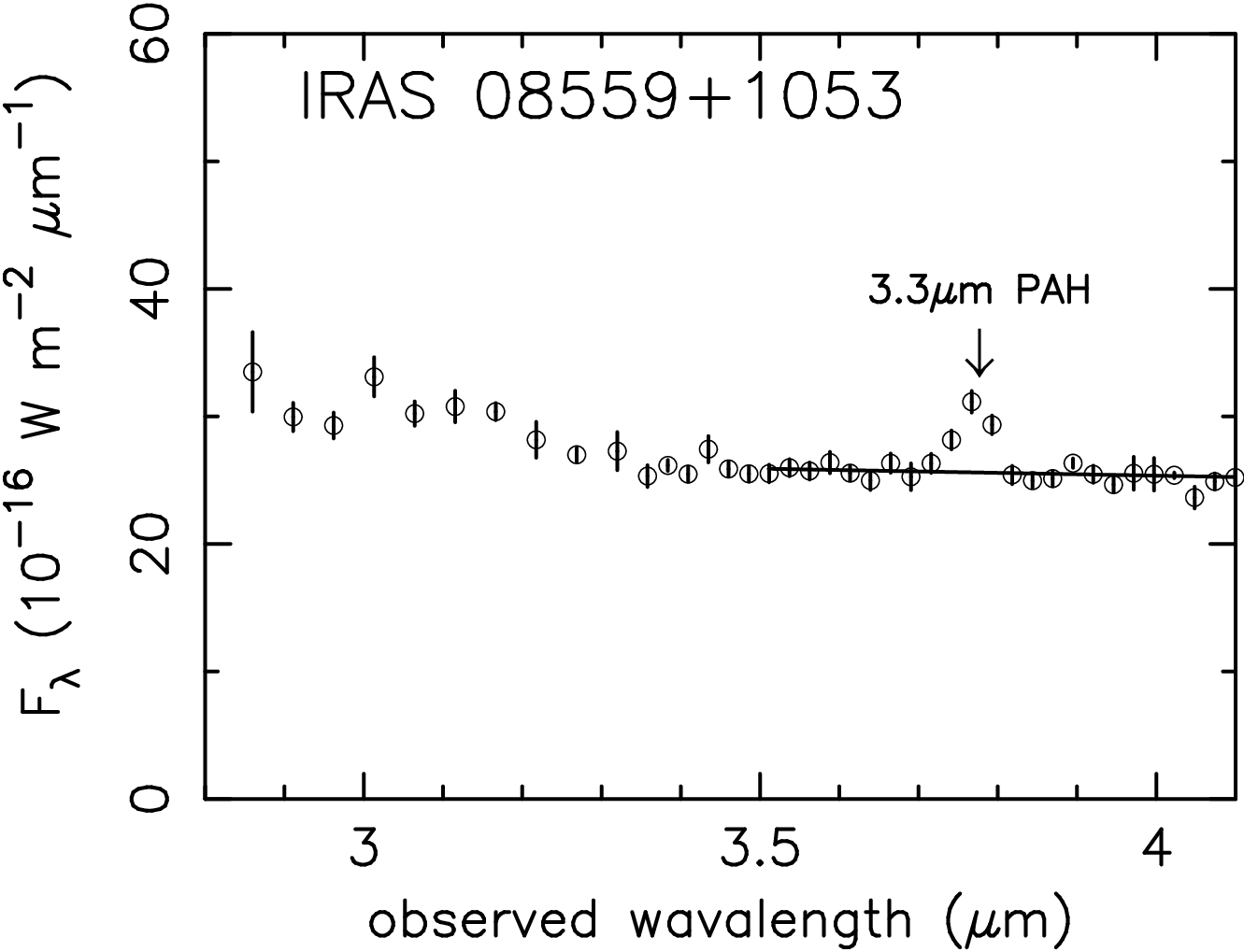} \\
\includegraphics[angle=-90,scale=.35]{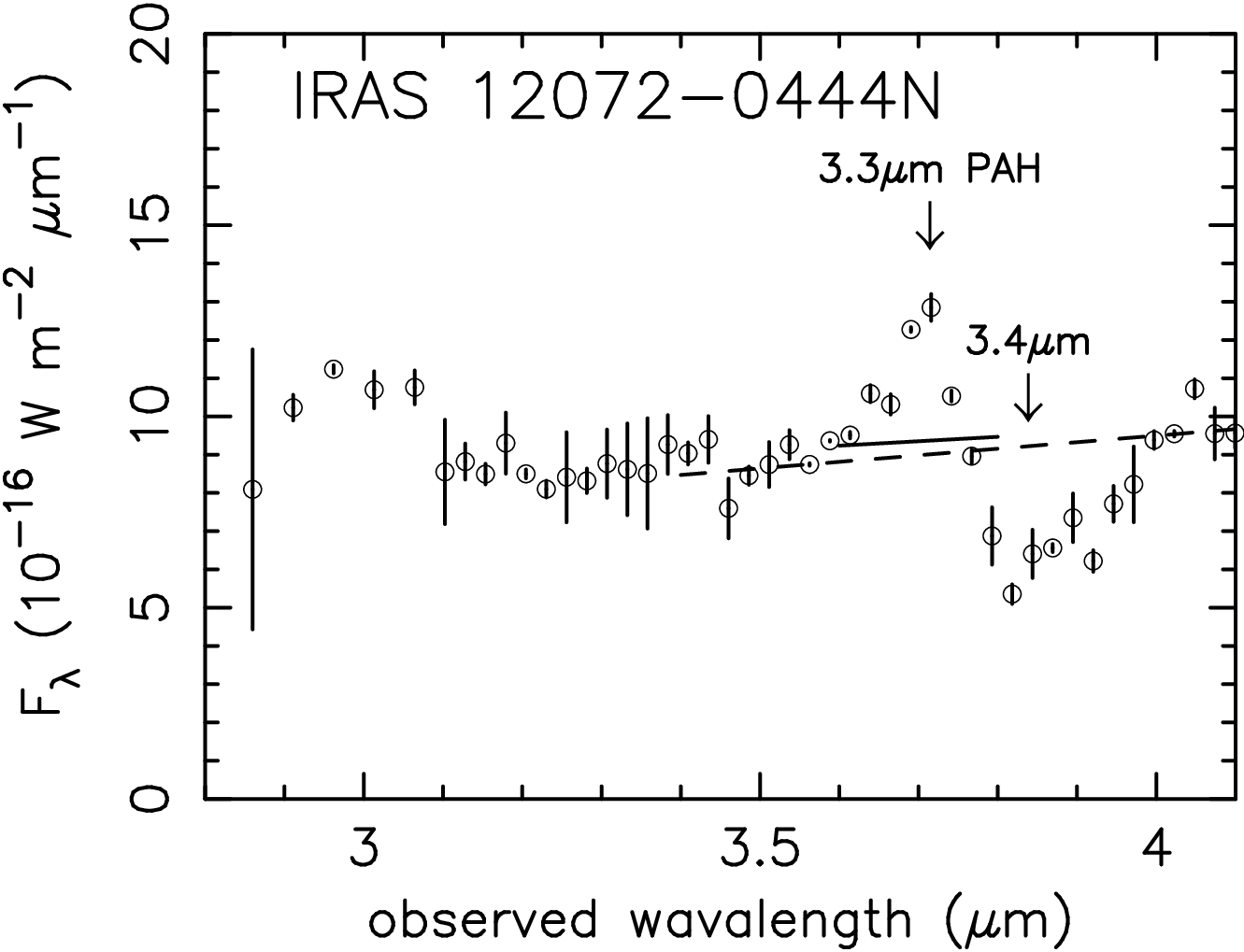} \hspace{0.3cm}
\includegraphics[angle=-90,scale=.35]{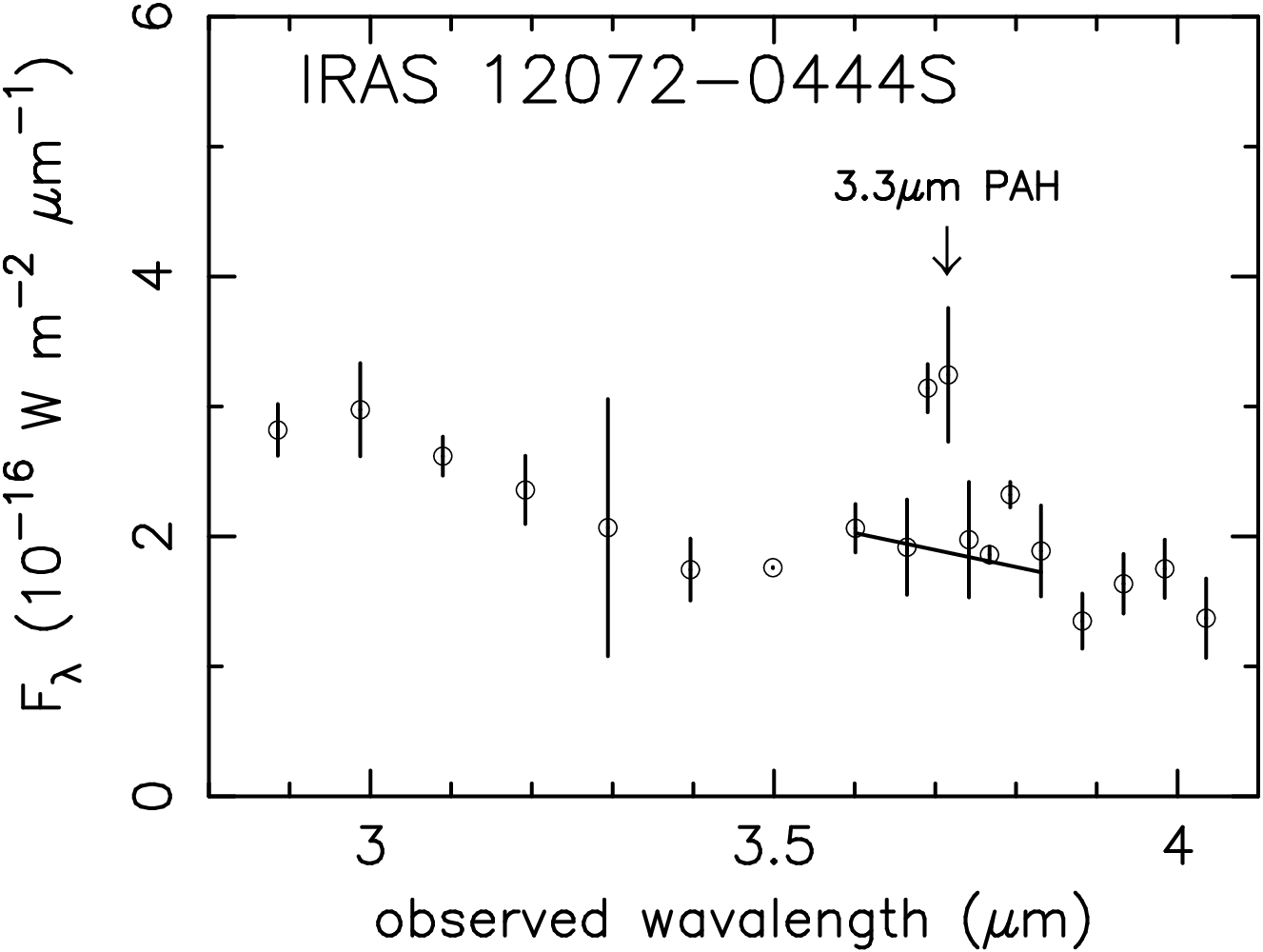} \\
\includegraphics[angle=-90,scale=.35]{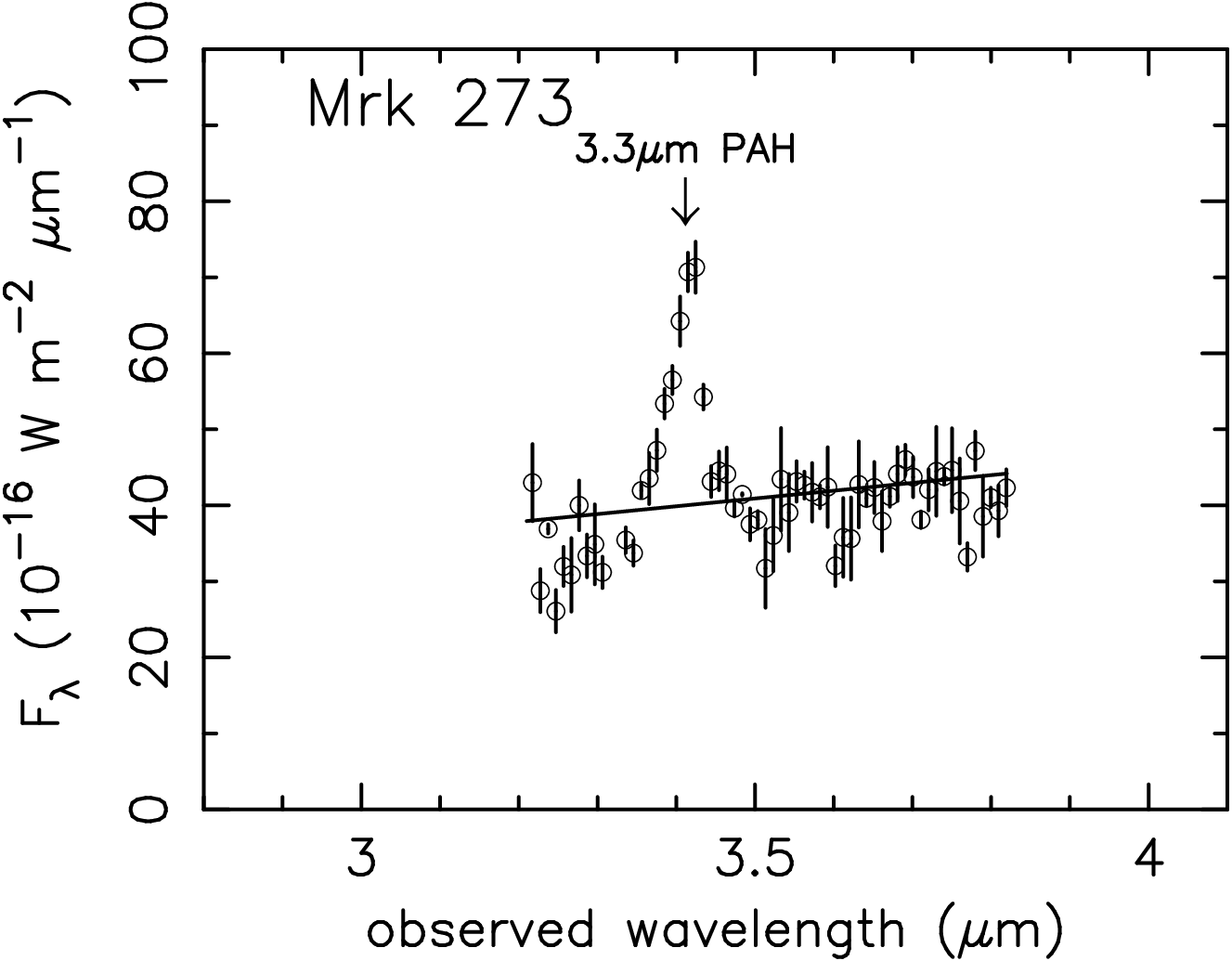} \hspace{0.3cm}
\includegraphics[angle=-90,scale=.35]{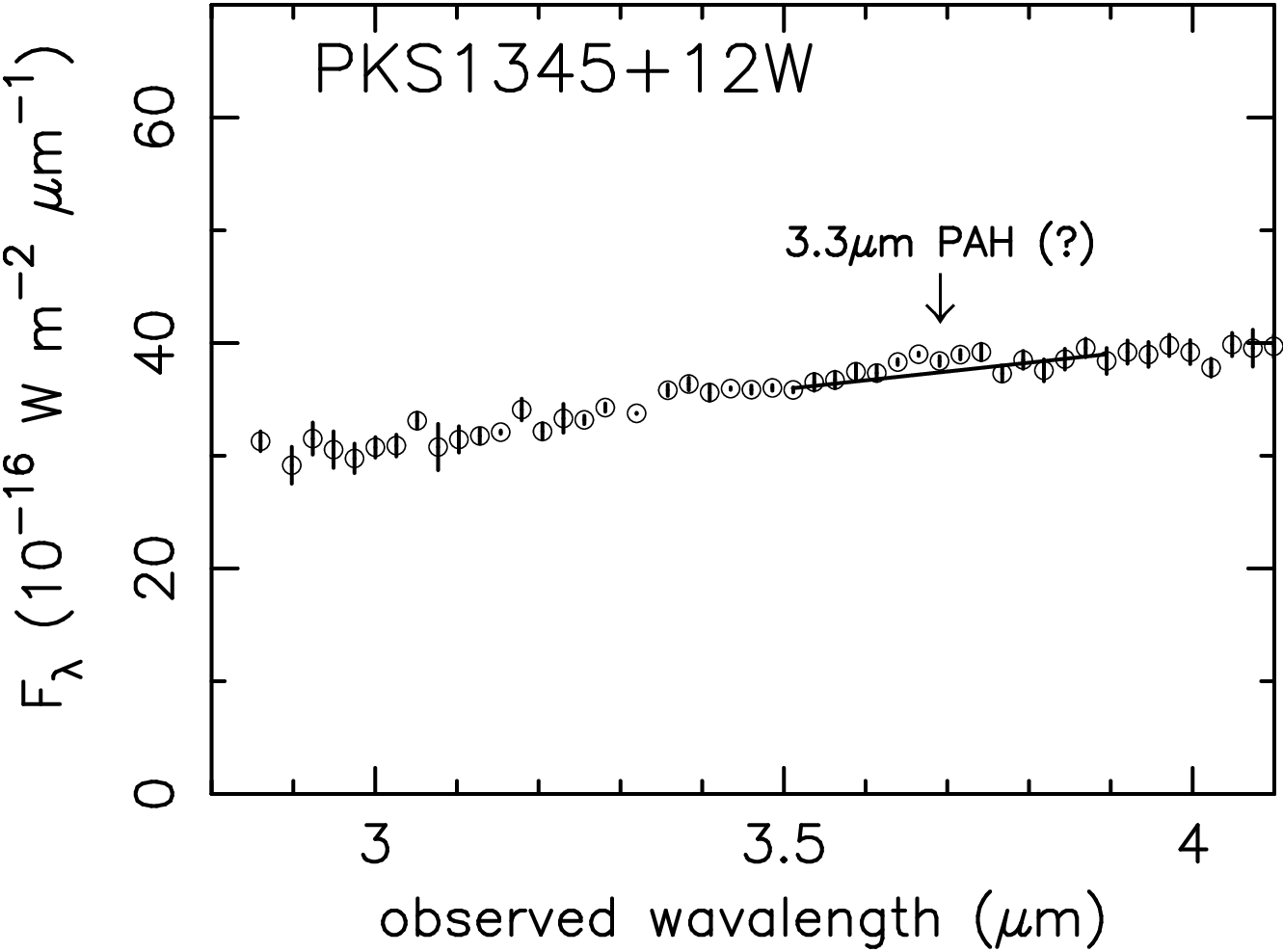} \\

\end{figure}

\clearpage

\begin{figure}
\begin{center} {\bf \Large Sy2 (II)} \end{center}
\includegraphics[angle=-90,scale=.35]{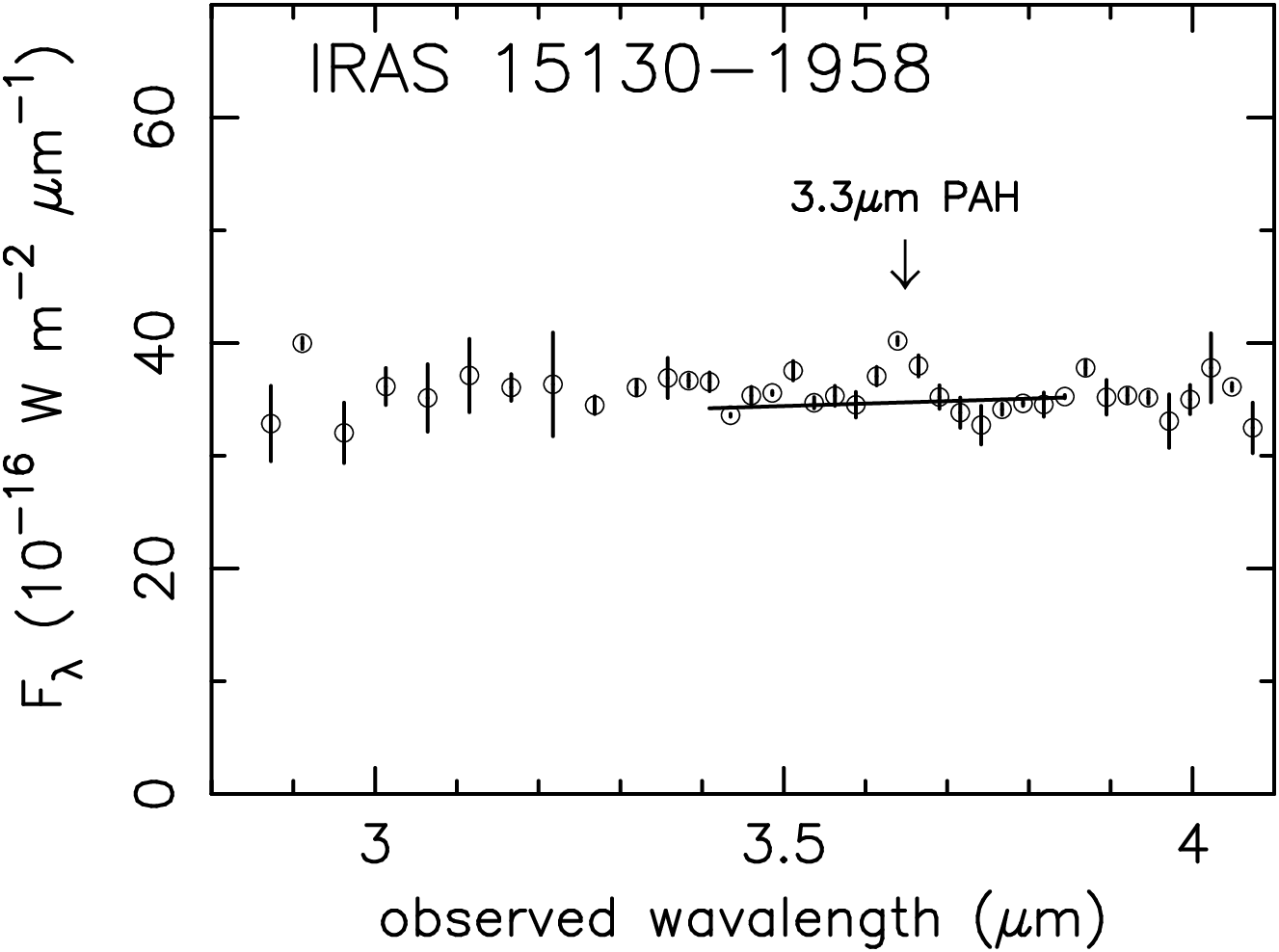} \hspace{0.3cm}
\includegraphics[angle=-90,scale=.35]{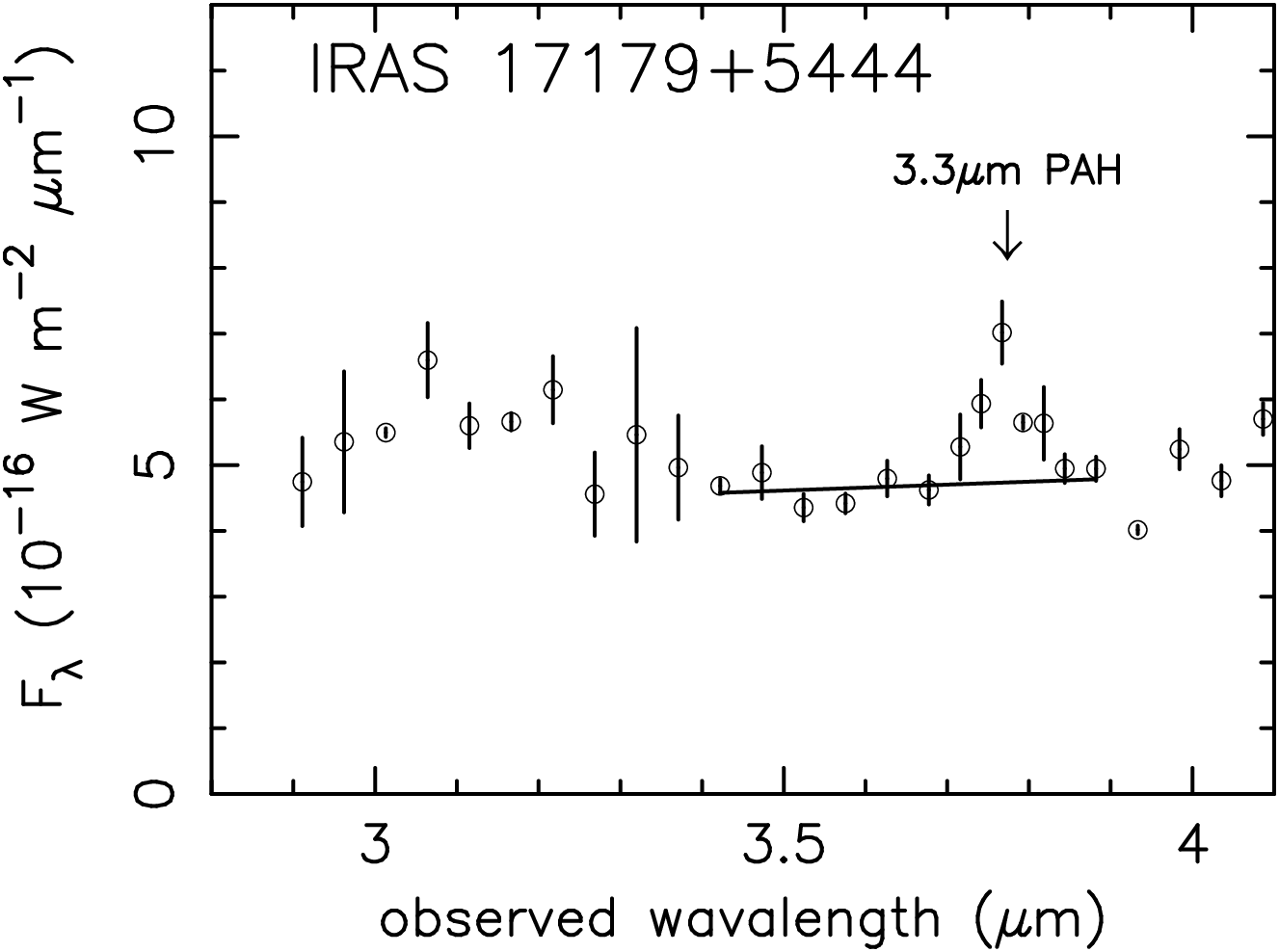} \\
\caption{
Infrared $L$-band ($\lambda_{\rm obs}$ = 2.8--4.1 $\mu$m) spectra of ULIRGs
classified optically as Seyfert-2s. 
Units and symbols are the same as Figure 1.
}
\end{figure}

\clearpage

\begin{figure}
\begin{center} {\bf \Large Sy1} \end{center}
\includegraphics[angle=-90,scale=.35]{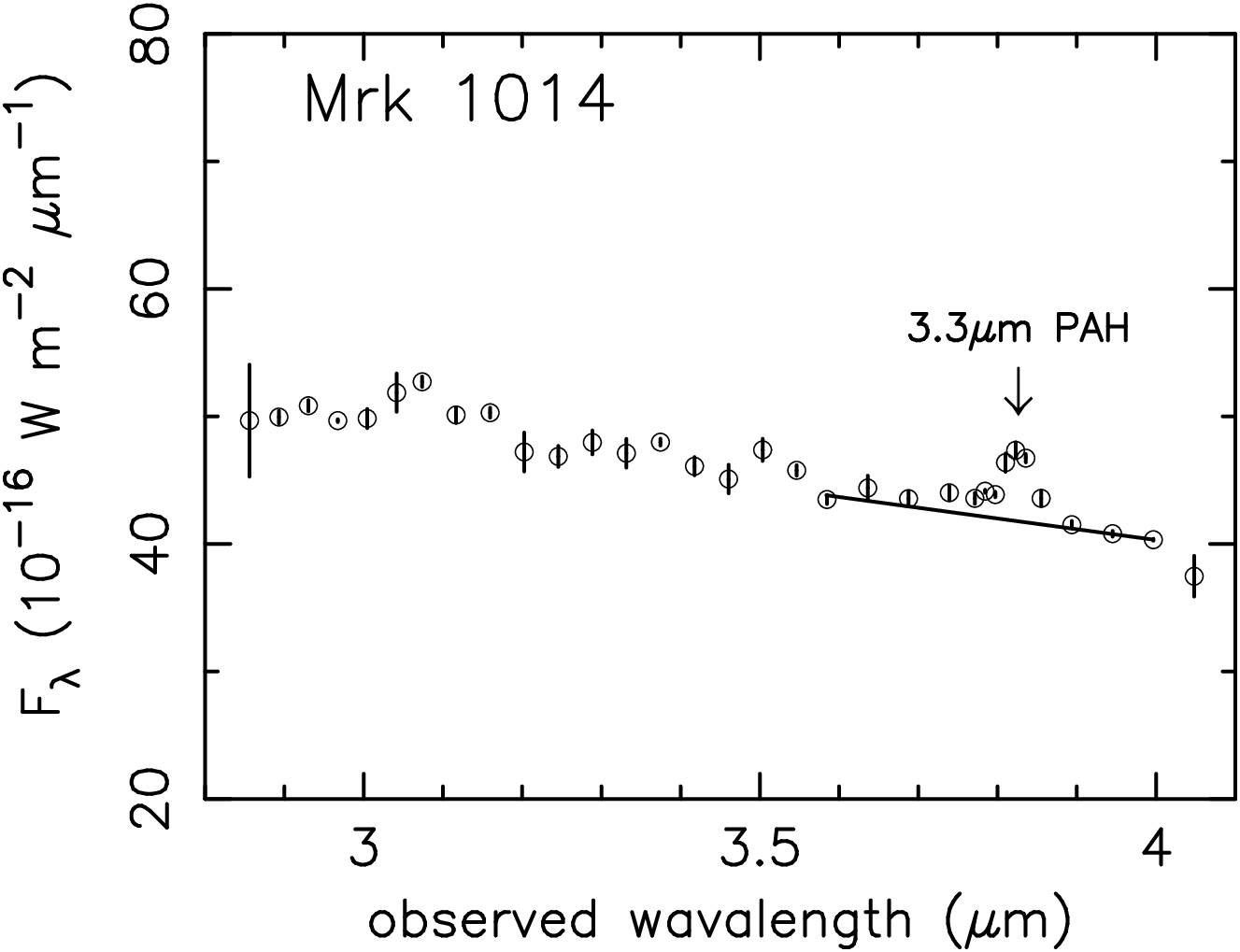} \hspace{0.3cm}
\includegraphics[angle=-90,scale=.35]{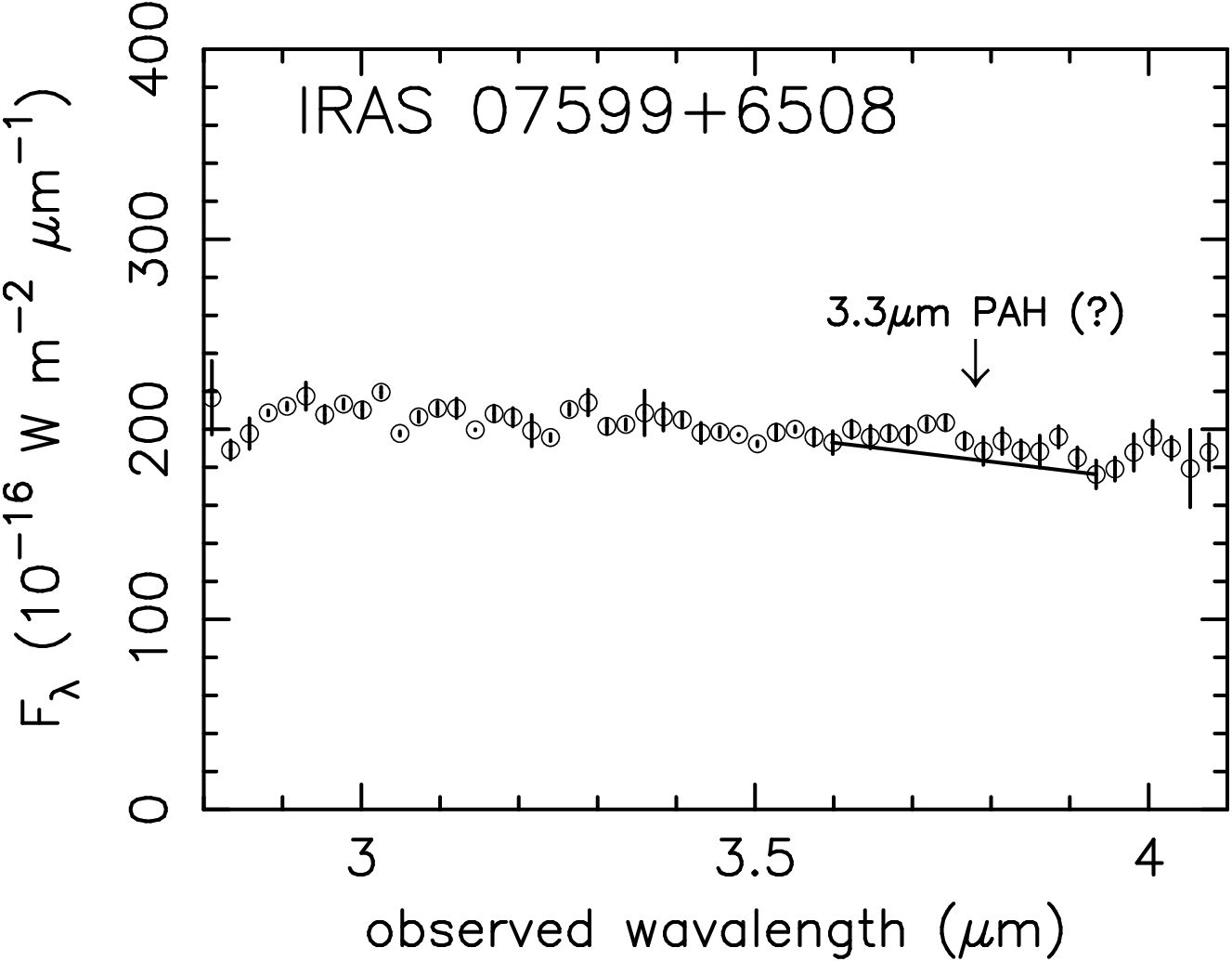}  \\ 
\includegraphics[angle=-90,scale=.35]{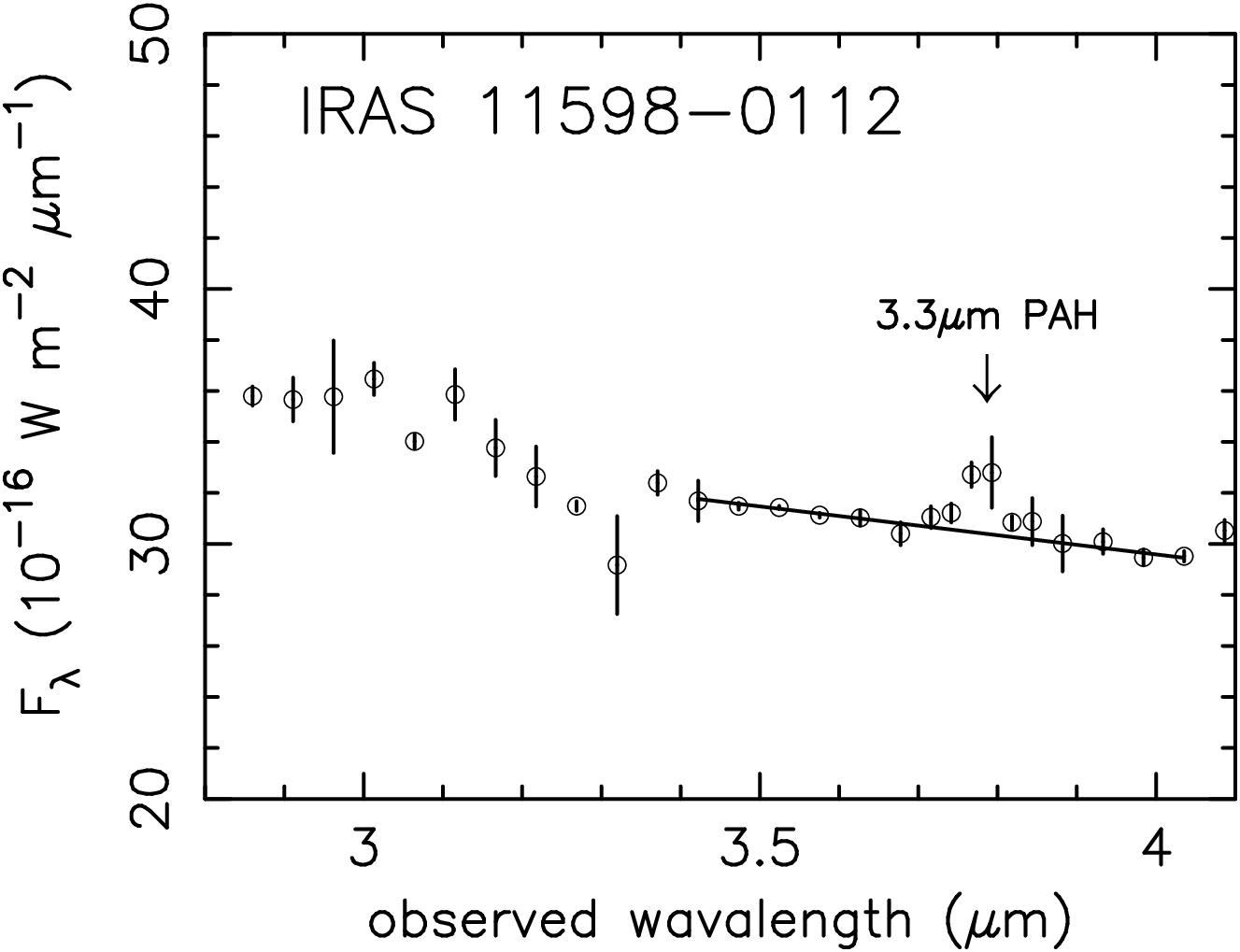} \hspace{0.3cm}
\includegraphics[angle=-90,scale=.35]{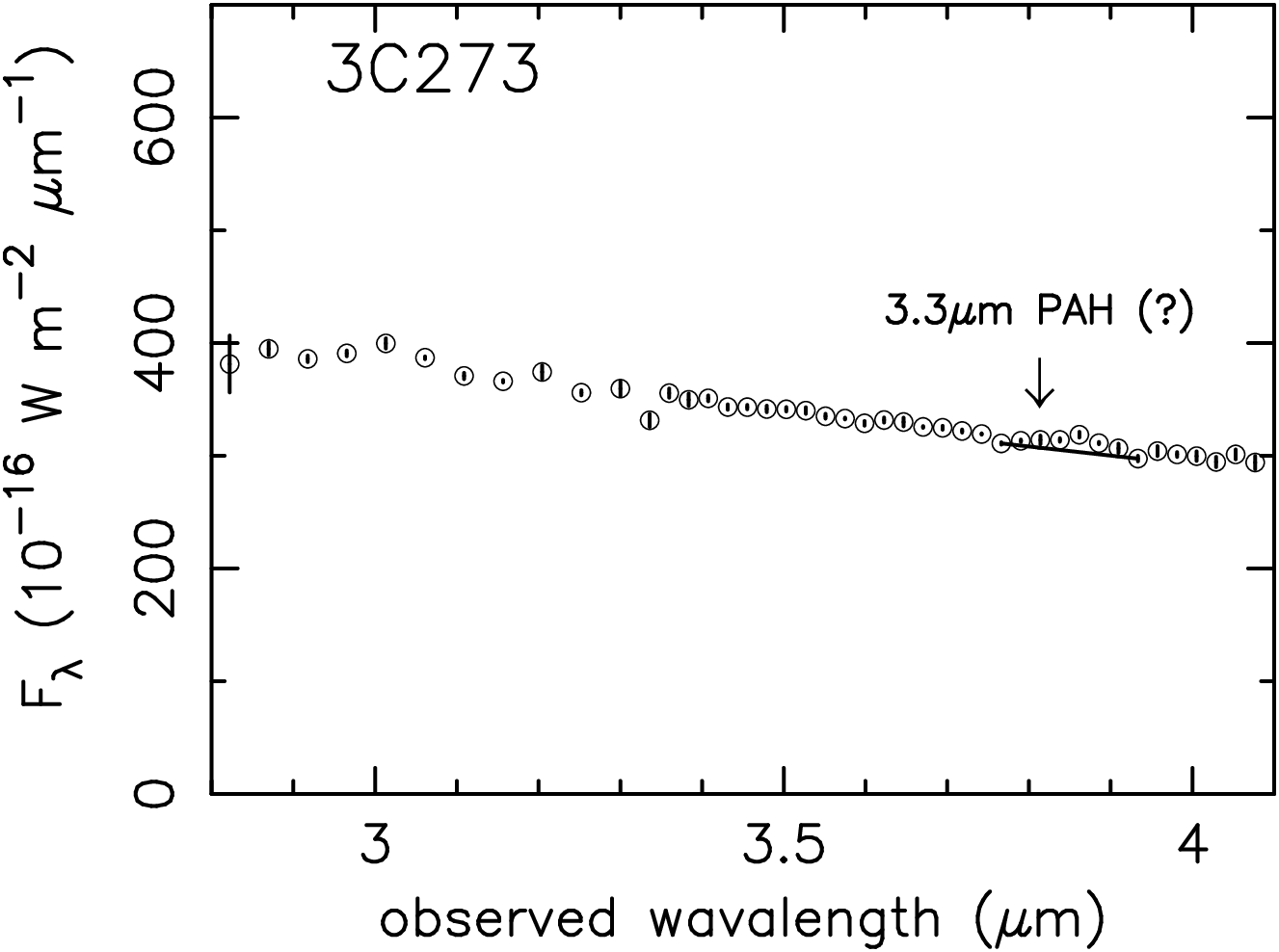}  \\ 
\includegraphics[angle=-90,scale=.35]{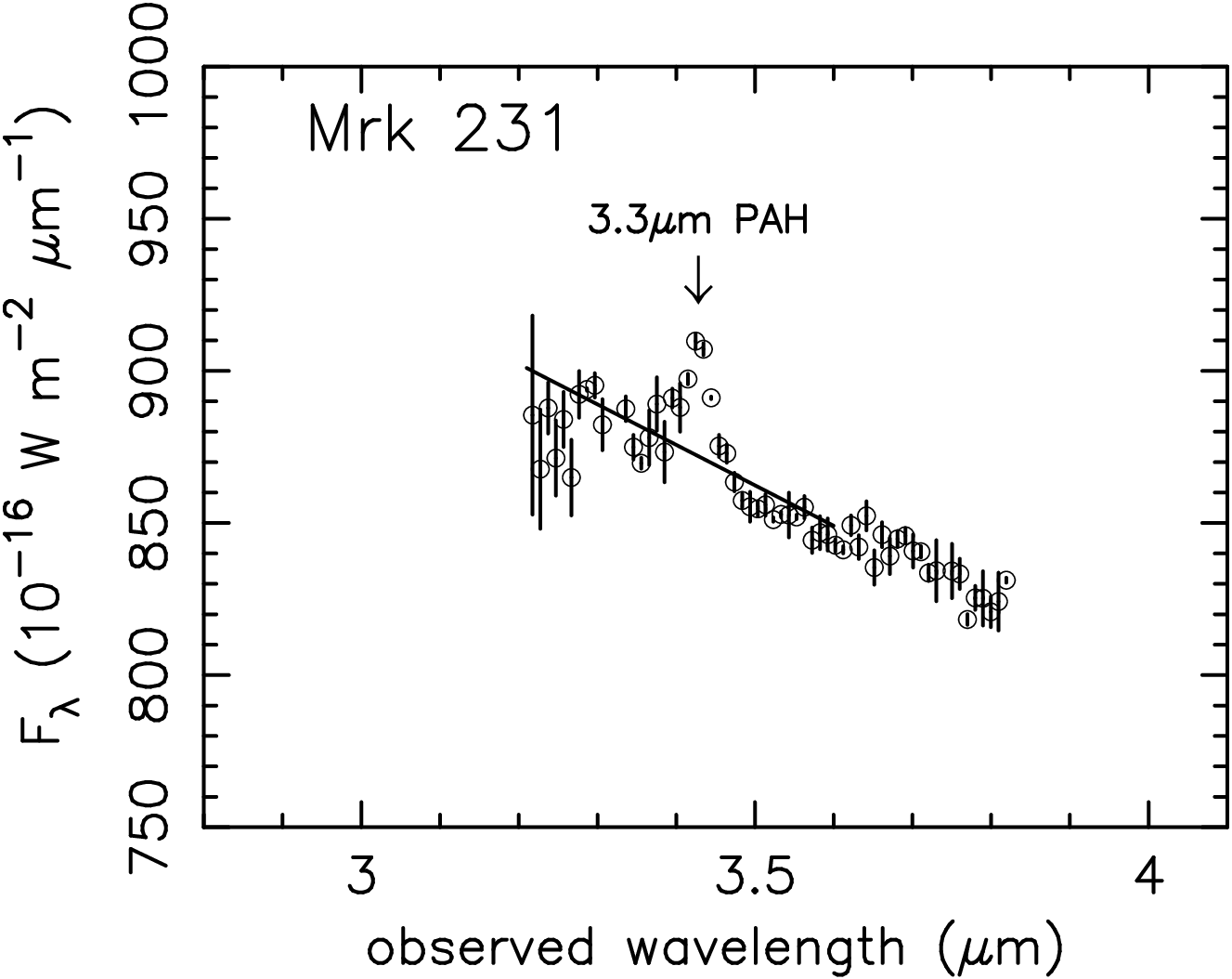}  \hspace{0.3cm}
\includegraphics[angle=-90,scale=.35]{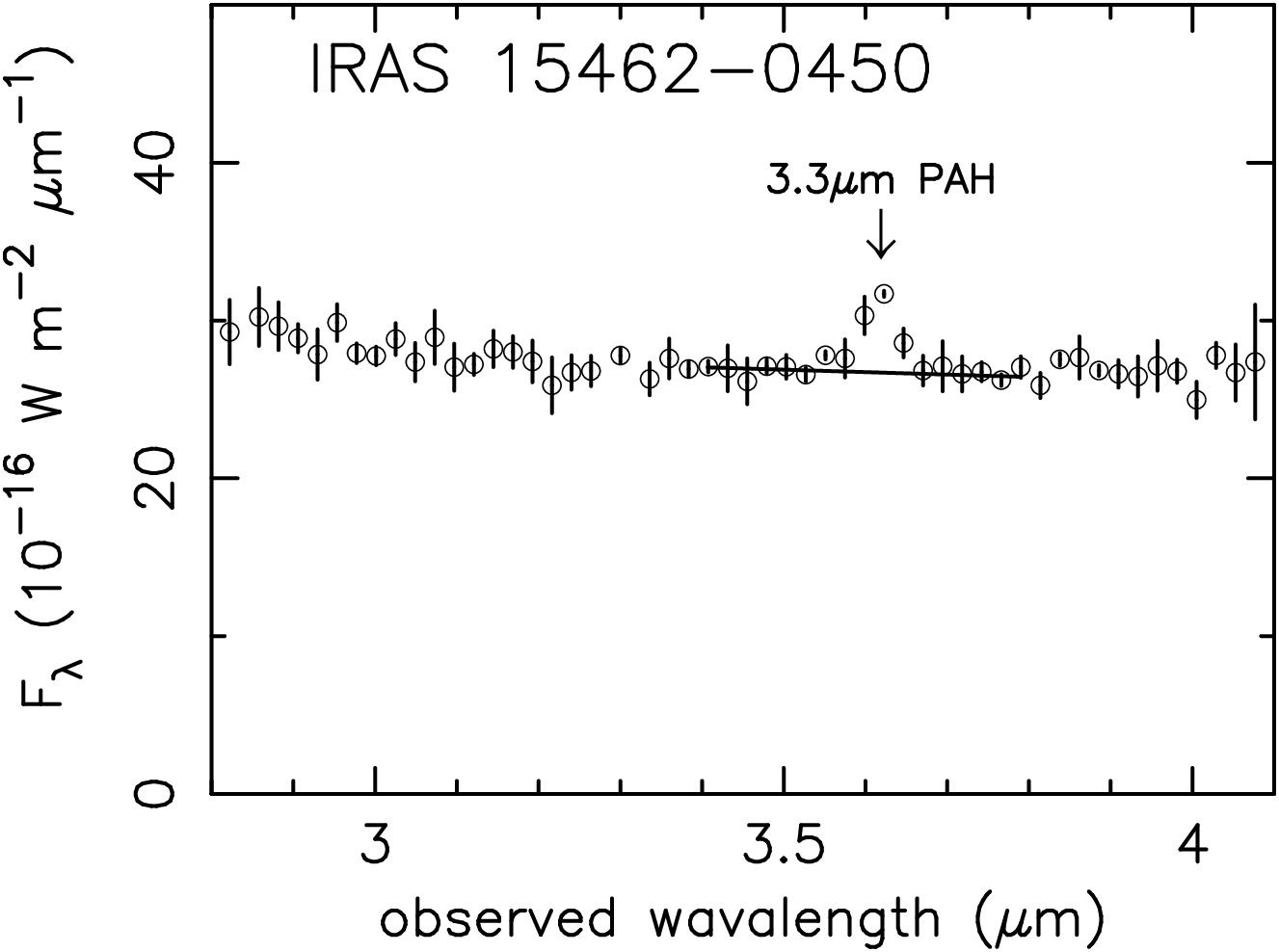} \\
\end{figure}

\clearpage

\begin{figure}
\begin{center} {\bf \Large Sy1 (II)} \end{center}
\includegraphics[angle=-90,scale=.35]{f5g.eps} \hspace{0.3cm}
\caption{
Infrared $L$-band ($\lambda_{\rm obs}$ = 2.8--4.1 $\mu$m) spectra of ULIRGs
classified optically as Seyfert-1s. 
Units and symbols are the same as Figure 1.
}
\end{figure}


\begin{figure}
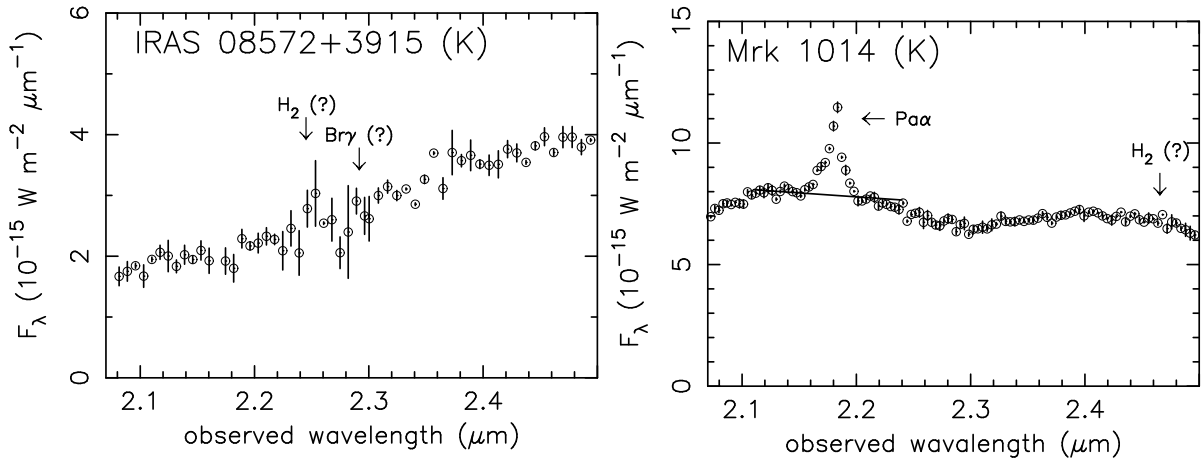

\begin{center} {\bf \Large K-band} \end{center}
\includegraphics[angle=-90,scale=.35]{f6a.eps} 
\includegraphics[angle=-90,scale=.35]{f6b.eps} \\
\caption{
Infrared $K$-band ($\lambda_{\rm obs}$ = 2.07--2.5 $\mu$m) spectra
of IRAS 08572+3915 and Mrk 1014, obtained with IRTF SpeX.
The abscissa is the observed wavelength in $\mu$m. 
The ordinate is F$_{\lambda}$ in 10$^{-15}$ W m$^{-2}$ $\mu$m$^{-1}$.
The solid line in the Mrk 1014 spectrum is the adopted continuum level
to estimate the Pa$\alpha$ emission flux.
}
\end{figure}

\clearpage

\begin{figure}
\includegraphics[angle=0,scale=.65]{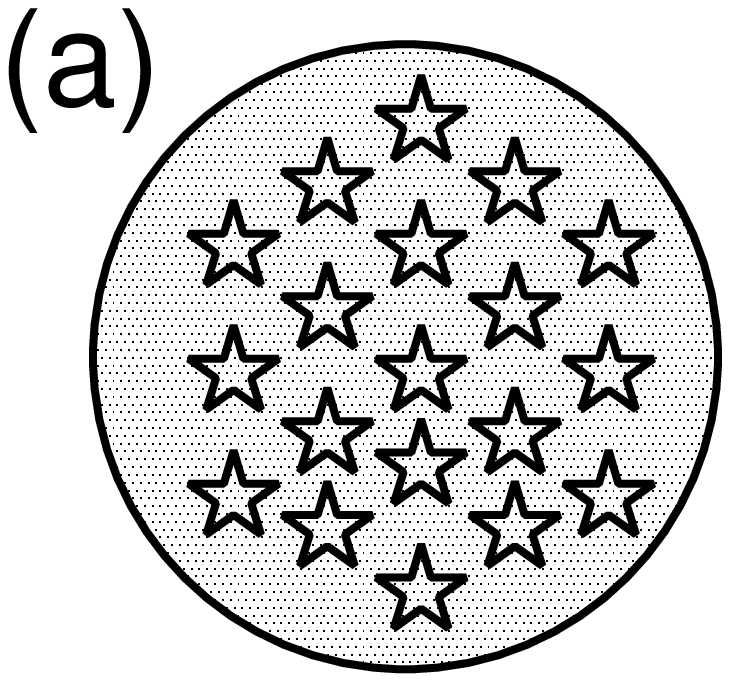} \hspace{0.7cm}
\includegraphics[angle=0,scale=.65]{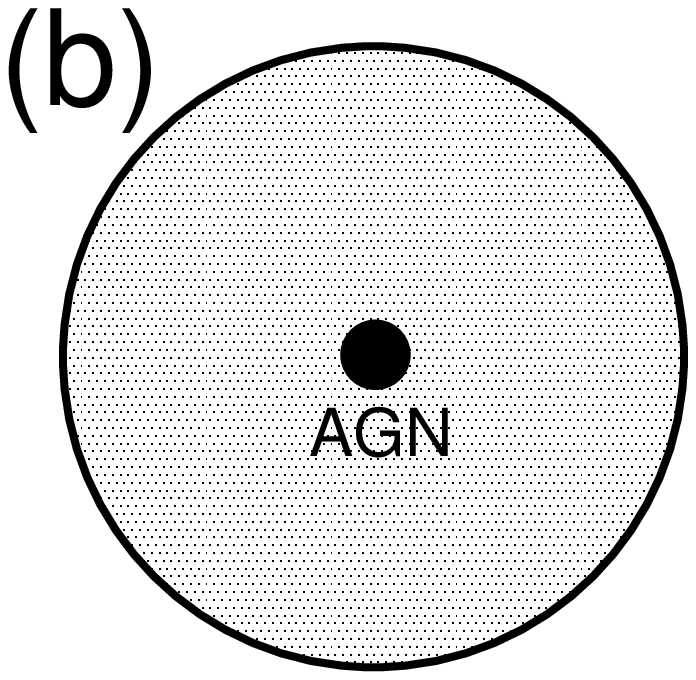} \hspace{0.2cm} 
\includegraphics[angle=0,scale=.35]{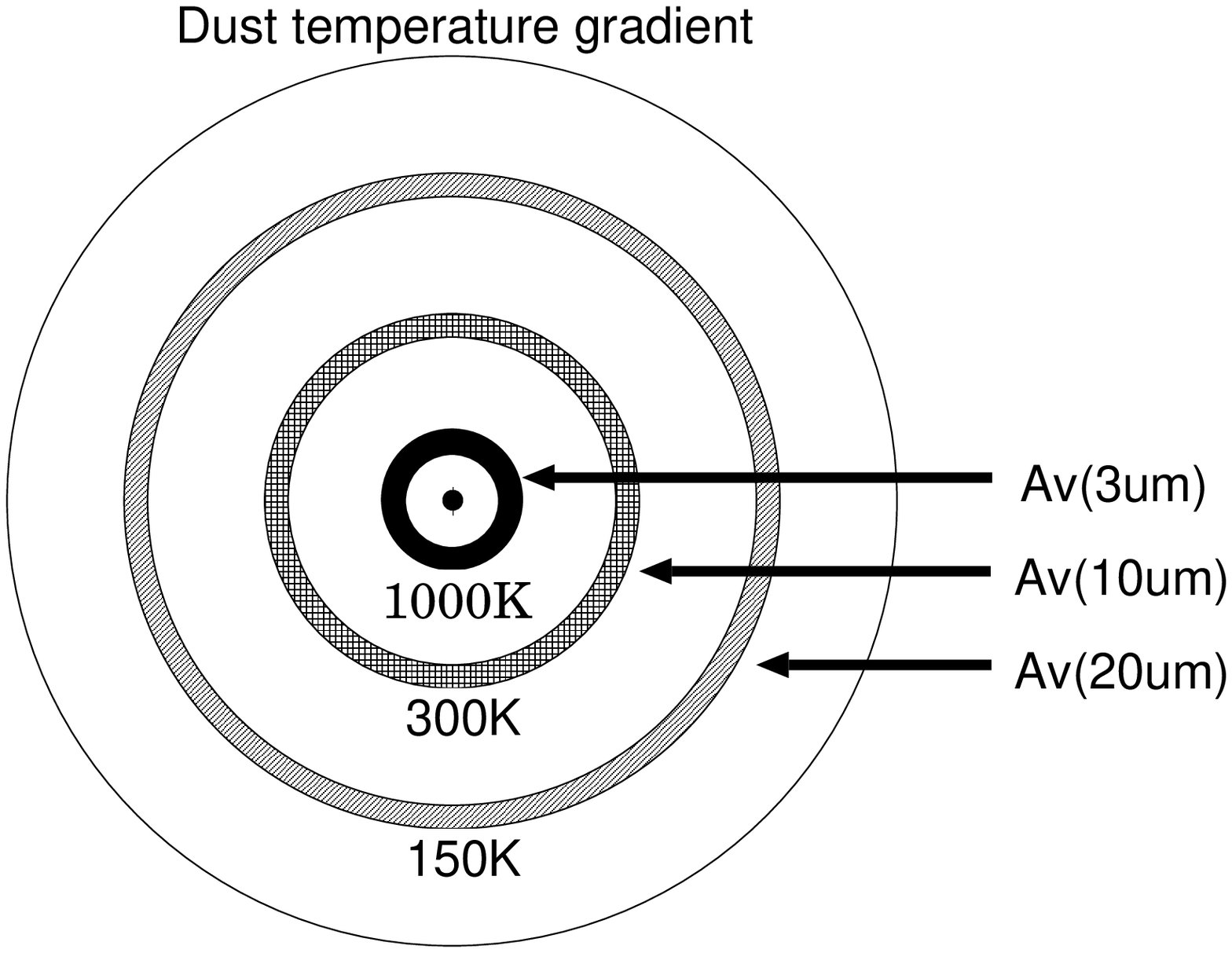} 
\caption{
{\it (a)}: Geometry of energy sources and dust in a normal
starburst. The energy sources (stars) and dust are spatially well
mixed. 
{\it (b)} ({\it Left}): Geometry of an energy source and dust in a
buried AGN. The energy source is more centrally concentrated than the
surrounding dust. 
{\it (b)} ({\it Right}): Schematic diagram of a strong dust temperature
gradient around a centrally-concentrated energy source. 
Inner dust has higher temperature. 
The blackbody radiation from inner 1000K dust, close to the dust
sublimation radius, has a peak at $\lambda \sim$ 3 $\mu$m. 
That of outer 300K dust and that of even outer 150K dust have peaks at
$\lambda \sim$ 10 $\mu$m and $\sim$20 $\mu$m, respectively. 
This is a very simplified picture, and in an actual case, 
dust with other temperatures can also contribute to the observed flux at
each wavelength of 3 $\mu$m, 10 $\mu$m, and 20 $\mu$m. 
The contribution is highly dependent on the dust radial density
distribution, but regardless of any dust radial density distribution, it
is always true that a contribution from inner, hotter dust emission is
larger at shorter wavelengths in this wavelength range.
Thus, the relation $A_V$(3$\mu$m) $>$ $A_V$(10$\mu$m) $>$
$A_V$(20$\mu$m) should hold, and dust extinction estimated using 3
$\mu$m data is a better probe for dust column density toward the AGN
itself than longer wavelengths.   
Dust extinction toward 3 $\mu$m continuum emitting region, 
A$_{\rm V}$(3$\mu$m), can be estimated from $\tau_{3.4}$ (bare dust) and
$\tau_{3.1}$ (ice-covered dust) values. 
Dust extinction toward 10 $\mu$m and 20 $\mu$m continuum emitting
regions, A$_{\rm V}$(10$\mu$m) and A$_{\rm V}$(20$\mu$m), can be
estimated from the optical depths of 9.7 $\mu$m and 18 $\mu$m silicate
dust absorption features, respectively.
These A$_{\rm V}$(10$\mu$m) and A$_{\rm V}$(20$\mu$m) data will be
available for all the non-Seyfert ULIRGs studied in this paper from
Spitzer IRS spectra of guaranteed time observations and Cycle-1
programs, including ours.  
Comparison of these values can be used to further investigate whether
the energy source is more centrally concentrated than the surrounding
dust \citep{dud97,ima00}.
}
\end{figure}

\begin{figure}
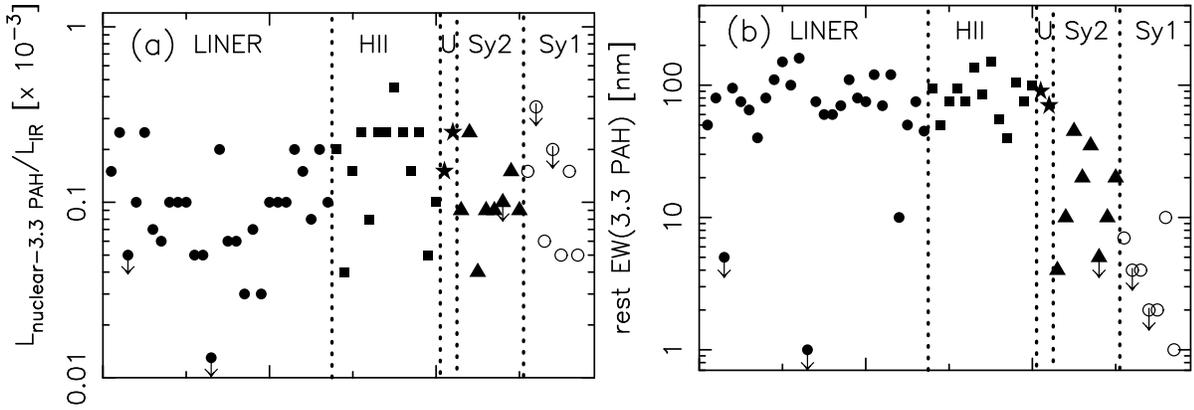

\includegraphics[angle=-90,scale=.35]{f8a.eps} 
\includegraphics[angle=-90,scale=.35]{f8b.eps} 
\caption{
{\it (a)}: Observed nuclear 3.3 $\mu$m PAH to infrared luminosity
ratio, separated by optical spectral types.
ULIRGs' nuclei classified optically as LINERs, HIIs, unclassified,
Seyfert 2s, Seyfert 1s (from the left to the right) are separated by the
vertical dashed lines. 
Filled circles: LINER. Filled squares: HII. 
Filled stars: optically unclassified. 
Filled triangles: Seyfert-2.
Open circles: Seyfert-1.
Objects' order is the same as that shown in Table 3.
The data point of IRAS 12127$-$1412 (EW$_{\rm 3.3PAH}$ = 0 nm) is ploted
at the lowest visible part in this Figure.
{\it (b)}: Rest frame equivalent widths of the 3.3 $\mu$m PAH
emission, separated by optical spectral types. 
Symbols are the same as {\it (a)}.
}
\end{figure}

\begin{figure}
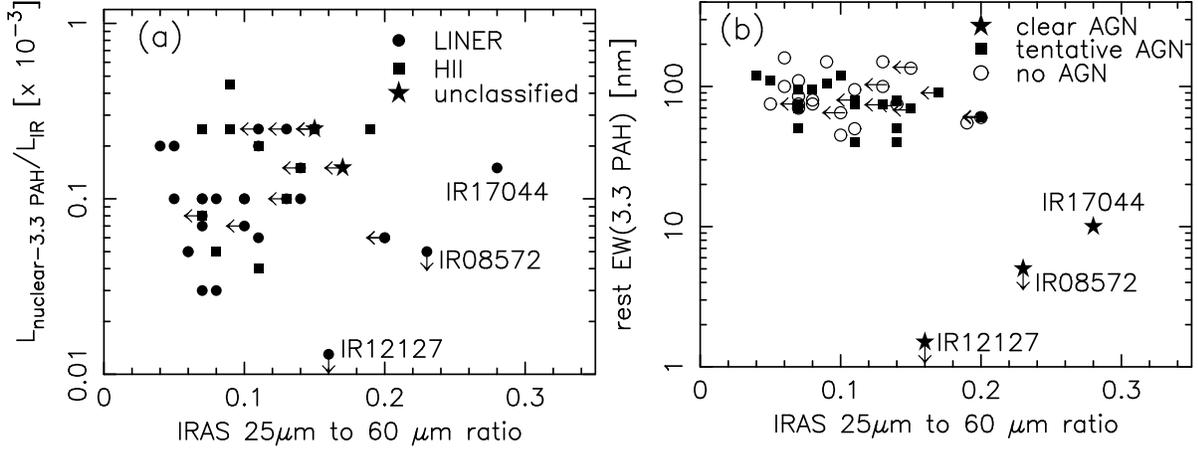

\includegraphics[angle=-90,scale=.35]{f9a.eps} 
\includegraphics[angle=-90,scale=.35]{f9b.eps} \\
\caption{
{\it (a)}: Observed nuclear 3.3 $\mu$m PAH to infrared luminosity
ratio (ordinate) and {\it IRAS} 25 $\mu$m-to-60 $\mu$m flux ratio
(abscissa) for non-Seyfert ULIRGs' nuclei. 
Filled circles: LINER ULIRGs' nuclei. Filled squares: HII ULIRGs' nuclei. 
Filled stars: optically unclassified ULIRGs' nuclei.
For several double nuclei ULIRGs, emission from individual nuclei is
resolved in our $L$-band spectra, but not in the {\it IRAS} data. 
For these sources, we assume that both nuclei have the same far-infrared
colors as measured with {\it IRAS}.
The data points of three LINER ULIRGs powered by almost pure buried
AGNs, IRAS 08572+3915NW, 12127$-$1412NE, and 17044+6720, are indicated
as ``IR08572'', ``IR12127'', and ``IR17044'', respectively. 
The data point of IRAS 12127$-$1412 (EW$_{\rm 3.3PAH}$ = 0 nm) is ploted
at the lowest visible part in this Figure.
{\it (b)}: Rest frame equivalent widths of the 3.3 $\mu$m PAH
emission (ordinate) and {\it IRAS} 25 $\mu$m-to-60 $\mu$m flux ratio 
(abscissa) for non-Seyfert ULIRGs.
Filled stars: ULIRGs with clear buried AGN signatures.
Filled squares: ULIRGs with tentative buried AGN signatures.
Open circles: ULIRGs with no obvious buried AGN signatures in our
$L$-band spectra.
The data points of IRAS 08572+3915NW, 12127$-$1412NE, and 17044+6720 are
indicated in the similar way to {\it (a)}.
}
\end{figure}

\end{document}